\documentclass{iopart}

\usepackage[comma,numbers,sort&compress]{natbib}
\usepackage{iopams}
\usepackage{setstack}
\usepackage{amsfonts}
\usepackage{amssymb}
\usepackage{multirow}
\usepackage{color}
\usepackage{colortbl}

\usepackage{epsfig}

\def\newblock{\hskip .11em plus .33em minus .07em}

\def\aj{Astron. J.}
\def\apj{Astrophys. J.}
\def\apjl{Astrophys. J. Lett.}
\def\apjs{Astrophys. J. Supp. Ser. }
\def\aa{Astron. Astrophys. }
\def\aap{Astron. Astrophys. }
\def\araa{Ann.\ Rev. Astron. Astroph. }
\def\physrep{Phys. Rep. }
\def\mnras{Mon. Not. Roy. Astron. Soc. }
\def\mmsun{M_\odot}
\def\prl{Phys. Rev. Lett.}
\def\prd{Phys. Rev. D.}
\def\azh{Soviet Astron.}

\begin{document}

\topical[The GW Signature of Core-Collapse SNe]
{The Gravitational Wave Signature\\  of Core-Collapse Supernovae}

\author{Christian D Ott}
\ead{cott@tapir.caltech.edu}

\address{\scriptsize 
  Theoretical Astrophysics, Mailcode 130-33,\\
           California Institute of Technology,
           Pasadena, California 91125, USA \\
  and \\ 
  Niels Bohr International Academy, Niels Bohr Institute,\\
  Blegdamsvej 17, 2100 Copenhagen, Denmark}

\begin{abstract}
We review the ensemble of anticipated gravitational-wave (GW) emission
processes in stellar core collapse and postbounce core-collapse
supernova evolution.  We discuss recent progress in the modeling of
these processes and summarize most recent GW signal estimates. In
addition, we present new results on the GW emission from postbounce
convective overturn and protoneutron star $g$-mode pulsations based on
axisymmetric radiation-hydrodynamic calculations. Galactic
core-collapse supernovae are very rare events, but within
$3-5\;\mathrm{Mpc}$ from Earth, the rate jumps to 1 in $\sim 2$
years. Using the set of currently available theoretical gravitational
waveforms, we compute upper-limit optimal signal-to-noise ratios based
on current and advanced LIGO/GEO600/VIRGO noise curves for the recent
SN 2008bk which exploded at $\sim$3.9 Mpc.  While initial LIGOs cannot
detect GWs emitted by core-collapse events at such a distance, we find
that advanced LIGO-class detectors could put significant upper limits
on the GW emission strength for such events.  We study the potential
occurrence of the various GW emission processes in particular
supernova explosion scenarios and argue that the GW signatures of
neutrino-driven, magneto-rotational, and acoustically-driven
core-collapse SNe may be mutually exclusive. We suggest that even
initial LIGOs could distinguish these explosion mechanisms based on
the detection (or non-detection) of GWs from a galactic core-collapse
supernova.
\end{abstract}



\section{Introduction}
\label{section:intro}

Ever since the very first experimental efforts to detect gravitational
waves (GWs), core-collapse supernovae (SNe) have been considered as
potential astrophysical emission sites.  There are very strong
indications from theory and observation that multi-D dynamics play a
prominent and probably decisive role in core-collapse SNe (see, e.g.,
\cite{burrows:00nature,janka:07}).  GWs are emitted at lowest order by
an accelerated mass-energy quadrupole moment. Hence, by their
intrinsic multi-D nature, GWs, if detected from a core-collapse event,
will very likely prove powerful messengers that can provide detailed
and live dynamical information on the intricate multi-D dynamics
occurring deep inside collapsing massive stars.

Massive stars ($8$--$10\;\mmsun \lesssim M \lesssim 100\;\mmsun$ at
zero-age main sequence [ZAMS]) form electron-degenerate cores composed
primarily of iron-group nuclei in the final stages of their exoergic
nuclear burning. Once such an iron core exceeds its effective
Chandrasekhar mass (see, e.g., \cite{baron:90,bethe:90}) it grows
gravitationally unstable.  Collapse ensues, leading to dynamical
compression of the inner core material to nuclear densities. There,
the nuclear equation of state (EOS) stiffens, resulting in the rebound
of the inner core (``core bounce''). A hydrodynamic shock wave is
launched at the outer edge of the inner core and propagates outward in
mass and radius, slamming into the still infalling outer core. Owing
to the dissociation of heavy nuclei and to energy losses to neutrinos
that stream away from the postshock region, the shock quickly loses
energy, stalls and must be \emph{revived} to plow through the stellar
envelope, blow up the star, and produce a SN explosion, leaving behind
a neutron star.  Without shock revival, black-hole (BH) formation is
inevitable and even with a successful explosion, a BH may still form
via fall-back accretion.

Iron core collapse is the most energetic process in the modern
universe, liberating some $10^{53}\;\mathrm{erg} = 100\;\mathrm{B}$
(Bethe) of gravitational energy.  Most of this energy, $\sim 99\,\%$,
is emitted in neutrinos as the protoneutron star (PNS) contracts and
cools over a timescale of $\sim 100\,\mathrm{s}$. Only $\sim 1\,\%$
goes into the asymptotic energy of the SN explosion and becomes
visible in the electromagnetic spectrum. The fundamental question that
core-collapse SN theory has been facing for the past $\sim 45$ years
is how exactly the necessary fraction of gravitational energy is
transferred to revive the shock and ultimately unbind the stellar
envelope. Shock revival must occur sooner than $1$--$1.5\;\mathrm{s}$
after bounce (depending on progenitor star structure setting the rate
of mass accretion) in order to produce a compact remnant that obeys
observational and theoretical neutron star upper baryonic mass limits
around $\sim 1.5$--$2.5\;\mmsun$ (see \cite{lattimer:07} and
references therein).

The SN \emph{explosion mechanism} may involve (a combination of)
heating of the postshock region by neutrinos, multi-dimensional
hydrodynamic instabilities of the accretion shock, in the postshock
region, and in the PNS, rotation, PNS pulsations, magnetic fields, and
nuclear burning (for a recent review, see \cite{janka:07}, but also
\cite{burrows:07bethe}). Three SN mechanisms are presently discussed
in the literature. The \emph{neutrino mechanism} has the longest
pedigree \cite{bethewilson:85,bethe:90,janka:07}, is based on
postbounce neutrino energy deposition behind the stalled shock, and
appears to require \cite{buras:06b,marek:07,murphy:08} convection and
the standing-accretion-shock instability (SASI, see, e.g.,
\cite{scheck:08} and references therein) to function in all but the
very lowest-mass massive stars which may explode even in spherical
symmetry~\cite{burrows:07c,kitaura:06}. Recent
detailed 2D neutrino-radiation hydrodynamics simulations by the
Garching group produced explosions in particular $11.2$-$\mmsun$ and
$15$-$\mmsun$ progenitor models \cite{buras:06b,marek:07}. However,
it is not yet clear how the neutrino mechanism's efficacy varies
with progenitor ZAMS mass and structure and what its detailed
dependence on the high-density nuclear EOS may be.

The \emph{magneto-rotational (or MHD) mechanism}, probably operating only
in the context of rapid progenitor rotation, depends on magnetic-field
amplification during collapse and at postbounce times. It leads to
explosions that develop in jet-like fashion along the axis of rotation
\cite{leblanc:70,symbalisty:84,yamada:04,burrows:07b,sawai:08} and may
reach hypernova energies of $\sim 10\,\mathrm{B}$~\cite{burrows:07b}.
The MHD mechanism may also be relevant in the context of long-soft
gamma-ray bursts~(GRBs, see, e.g., \cite{wb:06}) and could be a precursor,
setting the stage for a later GRB~(e.g., \cite{burrows:07b}, but see
\cite{dessart:08a}).

The \emph{acoustic mechanism} for core-collapse SNe, as recently
proposed by Burrows~et~al.~\cite{burrows:06,burrows:07a,ott:06prl,
  burrows:07bethe}, requires the excitation of large-amplitude PNS
pulsations (primarily $g$-modes) by turbulence and SASI-modulated
accretion downstreams. These pulsations damp by the emission of strong
sound waves that steepen to shocks and deposit energy in the postshock
region, eventually leading to late explosions at $\gtrsim
1\;\mathrm{s}$ after bounce. This mechanism appears to be
sufficiently robust to blow up even the most massive and extended
progenitors, but has so far not been confirmed by other groups (see
also \cite{yoshida:07,weinberg:08}).

Constraining the core-collapse SN mechanism via astronomical
observations is difficult. The intricate pre-explosion dynamics of the
SN core deep inside the supergiant presupernova star are inaccessible
by the traditional means of astronomy. Theoretical models of the SN
mechanism can currently be tested via secondary observables only,
including the asymptotic explosion energy, ejecta morphology,
nucleosynthesis products, compact remnant mass and proper motion, and
pulsar spin/magnetic fields. 

GWs and neutrinos are the only messengers with the potential of
delivering first-hand information on the physical processes leading to
explosion: Both are emitted deep inside the SN core and travel to
observers on Earth practically without interaction with intervening
material.

A small number of neutrinos were detected from SN 1987A in the
Large Magellanic (distance $D \approx 50\;\mathrm{kpc}$; see, e.g.,
\cite{bethe:90} and references therein). GWs have not yet been
observed directly, but the advent of GW astronomy has begun. An
international network of broad-band light-interferometric GW
observatories is active, encompassing the US LIGOs \cite{ligo}, the
British-German GEO600 \cite{geo600}, the French-Italian
VIRGO~\cite{virgo} and the Japanese TAMA 300 \cite{tama300}.
The three LIGO interferometers have recently
reached their design sensitivities, and, in their S5 science run, have
taken a year worth of data, partly in coincidence with GEO600 and
VIRGO. In addition, there are a number of active resonant bar/sphere
GW detectors in operation, including the four bar detectors of the
International Gravitational Event Collaboration (IGEC-2),
ALLEGRO, AURIGA, EXPLORER, and NAUTILUS~\cite{astone:07}, and the
resonant spheres MiniGrail~\cite{minigrail} and Schenberg~\cite{schenberg}.
The current status of ground-based GW detection was recently
summarized by Whitcomb~\cite{whitcomb:08}.

GWs from astrophysical sources are weak and notoriously difficult to
detect (e.g., \cite{thorne:87}).  Hence, in order to disentangle an
astrophysical GW signal from the mostly overwhelming detector noise, GW
astronomy does not only require sensitive detectors, but also
extensive processing and analysis of the detector output on the basis
of reliable theoretical estimates for the GW signals presently expected from
astrophysical sources. The latter must, in most cases, be obtained via
detailed numerical modeling of the dynamics responsible for the GW
emission in a given source.

In iron core collapse and postbounce SN evolution, the emission of GWs
is expected primarily from rotating collapse and bounce,
nonaxi\-symmetric rotational instabilities, postbounce convective
overturn/SASI, and PNS pulsations. In addition, anisotropic
neutrino emission, global precollapse asymmetries in the iron core and
surrounding burning shells, aspherical mass ejection, magnetic
stresses, and the late-time formation of a black hole may contribute
to the overall GW signature.

The aim of this topical review is to summarize the recent significant
progress in the modeling of the various GW emission processes in
core-collapse SNe with a focus on the early postbounce, pre-explosion
SN evolution up to $\sim1-2\;\mathrm{seconds}$ after bounce. We do not
cover the GW emission from the collapse of supermassive primordial or
very massive population III stars, dynamical fission processes,
late-time fall-back accretion on black holes, nuclear
phase-transitions in PNSs, or from late-time postbounce secular
instabilities such as $r$-modes. Other reviews of the GW signature of
core-collapse SNe are those of Kotake~et~al.~\cite{kotake:06a} and
Fryer~and~New~\cite{new:03}.

In \sref{section:history}, we provide a concise historical overview on
early work and go on to discuss computational core-collapse SN modeling and
the various ways in which GR and GW extraction are treated 
in \sref{section:snmodel}. Section~\ref{section:rotcoll} covers the most
extensively modeled GW emission mechanism, rotating core collapse and
bounce. The potential for and the GW emission from nonaxisymmetric
rotational instabilities is the topic of
\sref{section:rotinst}. Section~\ref{section:convsasi} is devoted to
the emission of GWs from convective overturn and SASI, while we
discuss the GW signal from PNS core pulsations in \sref{section:puls}.
To both of these sections we add new, previously unpublished results
obtained via 2D Newtonian radiation-hydrodynamics calculations.  In
\sref{section:neutrinos}, we discuss the emission of GWs from
anisotropic neutrino radiation fields and in \sref{section:others} we
summarize the GW signals associated with rapid aspherical outflows,
precollapse global asymmetries, strong magnetic stresses, and PNS
collapse to a black hole.

While the SN rate in the Milky Way and the local group of galaxies is
rather low and probably less than 1 SN per two decades (e.g.,
\cite{vdb:91}), there may be 1 SN occurring about every other year
between $3-5\;\mathrm{Mpc}$ from Earth~\cite{ando:05}. The recent SN
2008bk, which exploded roughly $3.9\;\mathrm{Mpc}$ away, is an example
SN from this region of space. Thus, in \sref{section:2008bk}, we
present optimal single-detector matched-filtering signal-to-noise
ratios for LIGO/GEO600/VIRGO and advanced LIGOs for a subset of the
gravitational waveforms reviewed in this article and with an assumed
source distance corresponding to that of SN 2008bk. We find that initial
LIGO-class detectors had no chance of detecting GWs from SN 2008bk.
Advanced LIGOs, however, could put some constraints on the GW emission
strength, but still would probably not allow detailed GW observations.
We wrap up our review in section~\ref{section:conclusions} with a
critical summary of the subject matter and discuss in which way the
various GW emission processes can be linked to particular SN explosion
mechanisms. We argue that the GW signatures of the neutrino,
MHD, and acoustic SN mechanisms may be mutually exclusive and that the
mere detection, or, in fact non-detection, of GWs from a nearby
core-collapse SN can constrain significantly the core-collapse SN
explosion mechanism.

In this review article, all values of the dimensionless GW signal
amplitudes $h_+$ and $h_\times$ are given for optimal source-observer
orientation and a source distance of $10\;\mathrm{kpc}$ is typically
assumed. In most figures showing GW signals, the waveforms are
plotted as $h_{+,\times}\, D$, rescaled by distance $D$ and in units
of centimeters. Summaries of GW extraction methods can be found in
\cite{thorne:87,moenchmeyer:91,zwerger:97,ott:04,ott:06phd} and a
comparison of Newtonian and general relativistic methods were presented
in \cite{shibatasekiguchi:03,baiotti:08b}.

\section{A Short Overview on Early Work}
\label{section:history}
Although gravitational waves were proposed by Einstein himself 
in 1918 \cite{einstein:18}, there
was general disagreement on the waves' reality and their ability to
carry off energy from their source until a 1957 gedankenexperiment by
Bondi \cite{bondi:57} restored trust in the fundamental theory of GWs
(see, e.g., \cite{thorne:87} and references therein for more
details). This lead to a renaissance of GW physics in the
early to mid 1960s, resulting in much theoretical work on
astrophysical sites of GW emission (see, e.g., \cite{thorne:87,eardley:83}) 
and to the first experimental searches for GWs led
by Weber with resonant bar detectors~\cite{weber:60,whitcomb:08}. 

GW emission in the core-collapse SN context first appeared in a
1966 paper by Weber \cite{weber:66} in which he referred to
unpublished work from 1962 by Dyson \cite{dyson:62} on astrophysical
sources of GWs, including binary neutron star systems and
non-spherical stellar collapse\footnote{ Upon being asked by the
  author to comment on his early work on GW emission from stellar
  collapse, Dyson wrote: \it I have no idea who first calculated the
  emission of gravitational radiation from a collapsing star with
  rotation.  Ruffini and Wheeler \cite{ruffini:71} may have been the
  first. It was certainly obvious to everyone who thought about it
  that a collapsing star with rotation would give rise to a strong
  pulse of gravitational waves. I make no claim to have thought of
  this first.  \rm F.\ Dyson, priv. comm., Oct. 2006.}. Also in 1966,
Wheeler, based on unpublished work by Zee and himself, published first
quantitative estimates of the emission of a GW burst by quadrupole
oscillations of a PNS associated with core bounce~\cite{wheeler:66}.
In the following years, the theory of NS pulsations and their GW
emission characteristics was worked out by Thorne, Bardeen, Meltzer,
and collaborators \cite{bardeen:66,meltzer:66,
  thorne:67,price:69,thorne:69a,thorne:69b,thorne:70}, leading to many
subsequent studies and opening up a rich independent branch of
research (see, e.g., \cite{kokkotas:99,andersson:03} for reviews).

In their 1971 review \cite{ruffini:71}, Ruffini and Wheeler provided a
first comprehensive summary of scenarios for GW emission in
the stellar collapse context and provided quantitative estimates for
the GW signal from rotating core collapse and bounce, (early postbounce)
neutron star pulsations, and nonaxisymmetric neutron star
deformations.

A number of subsequent studies in the 1970s and early 1980s focussed
on aspherical collapse and relied upon semi-analytical descriptions of
collapsing and/or oscillating homogeneous spheroids and
ellipsoids. Key studies were those carried out by Thuan and
Ostriker~\cite{thuan:74}, Novikov~\cite{novikov:75},
Shapiro~\cite{shapiro:77}, Saenz and Shapiro~\cite{saenzshapiro:78,
saenzshapiro:79,saenzshapiro:81}, Epstein~\cite{epstein:76},
Moncrief~\cite{moncrief:79}, and Detweiler and
Lindblom~\cite{detweiler:81}.  Most of these early and ground-breaking
investigations severely overestimated the strength (amplitude/energy)
of the GW burst from core collapse and bounce and failed to capture
the qualitative features of the core bounce GW burst observed later in
numerical simulations. Some of the former and
other authors using similar techniques also studied the GW emission
from nonaxisymmetric rotational deformations of
rapidly rotating postbounce neutron stars
\cite{chin:65,chau:67,chandrasekhar:70b,miller:74,
ipsermanagan:84,saenzshapiro:81}.

In 1979, Turner and Wagoner~\cite{turnerwagoner:79} followed a
different approach and used perturbation theory to calculate the GW
emission produced by rotationally-induced perturbations of spherically
symmetric numerical iron core collapse models of Van
Riper~\cite{vanriper:78} and Wilson \cite{wilson:77}. Seidel and
collaborators followed along those lines and perturbatively analyzed
detailed spherically-symmetric general-relativistic (GR) core collapse
simulations in the 1980s~\cite{seidelmoore:87,seideletal:88}. Although
the perturbative approach is in principle appropriate only for small
deviations from sphericity and, hence, for very slowly rotating
stellar cores, the GW signals from core bounce predicted by
Seidel~et~al.\ is in rough qualitative agreement with the signals
obtained from more recent multi-D simulations of
even rapidly rotating stellar cores.

Epstein \cite{epstein:78} and independently Turner~\cite{turner:78}
considered in their 1978 studies the GW emission from anisotropic
neutrino emission in stellar core collapse and developed the
linear-theory formalism that has since been applied to extract GWs
from aspherical neutrino radiation fields in multi-D
radiation-hydrodynamic simulations.

The early epoch of work on the GW emission in stellar core collapse
was dominated by analytic, semi-analytic, and perturbative studies,
largely because of a lack of computers sufficiently powerful to carry
out multi-D numerical non-linear hydrodynamics simulations. The end of
this epoch was marked by the emergence of the first supercomputers and
their general availability to the astrophysics community beginning in
the late 1970s and early 1980s. 

\section{Modeling Core-Collapse Supernovae}
\label{section:snmodel}

The Einstein equations as well as the equations of
radiation-magnetohydrodynamics are families of coupled non-linear
partial differential equations for which solutions can be found
analytically only in very few limiting cases, leaving numerical
modeling as the only viable route to their solution in complex
scenarios such as stellar collapse and core-collapse SNe. Hence, for an
accurated understanding of the core-collapse SN phenomenon and the GW
signals emitted by the involved multi-D dynamics, detailed numerical
models are required.

Modeling stellar core collapse and the postbounce evolution
of the SN core is a multi-scale multi-physics problem that involves
lengthscales from the extended pre-SN stellar core (thousands of
kilometers) down to small-scale turbulence in the postbounce flow (on
the order of meters) and timescales from $\lesssim 10^{-6}\,\mathrm{s}$ 
(the typical computational timestep) to up to $1-2\,\mathrm{s}$
(the time for the development of a full explosion or for BH formation
to occur).

An ideal and complete (``realistic'') model should resolve all
lengthscales, should be free of artificially imposed symmetries (i.e.,
be 3D) and should fully include general relativity, GR MHD, multi-D GR
Boltzmann neutrino transport, and a microphysical treatment of EOS and
nuclear and neutrino interactions. In addition, since the high-density
EOS, the pre-SN stellar structure and angular momentum configuration
are not well constrained, extensive parameter studies are necessary to
plow the parameter space of possible EOS and pre-SN configurations.
Present-day 2D core-collapse SN simulations, e.g., those carried out
by \cite{marek:07,burrows:07b,burrows:07a,ott:08}, scale to $\sim
48$-$128$ compute cores and typically require multiple months to
complete on current supercomputers.  A single 3D full GR
radiation-(magneto)hydrodynamics simulation will have an at least a
100 times (and more realistically, 1000 to 10000 times) greater
computational complexity and will need to scale to thousands of
compute cores on next-generation supercomputers to complete within
reasonable time\footnote{See \cite{ES-Schnetter2007a} for a more
  quantitative discussion of the computational complexity of the
  core-collapse SN problem.}.

None of the currently published numerical models of core collapse and
postbounce SN evolution live up to the above standards and all studies
employ approximations for multiple of the physical ingredients listed. 
A detailed discussion of the various approximations cannot be
provided here, but since the focus of this review is on GWs, a
consequence of GR and, in principle, requiring a full GR treatment, we
present in the following the different ways in which GR gravity,
relativistic dynamics, and GW extraction from matter and spacetime
dynamics are presently handled in multi-D core-collapse simulations.
\renewcommand{\labelenumi}{(\roman{enumi})}
\begin{enumerate}
\item {\bf Newtonian gravity and dynamics}. Many recent multi-D
  simulations
  (e.g.,\cite{burrows:06,burrows:07a,burrows:07b,ott:04,ott:08,ott:06prl,
    kotake:07a,kotake:04,kotake:03,fryer:02,fh:00,obergaulinger:06a})
  employ a solution of the Newtonian Poisson equation for computing
  the gravitational acceleration terms and treat 
  (magneto)-hydrodynamics and radiation transport (if included) in
  Newtonian fashion. GWs are extracted from the fluid motions via the
  slow-motion, weak-field quadrupole formalism (e.g.,
  \cite{thorne:87,finnevans:90,blanchet:90}).

\item {\bf Approximate GR gravity}. Motivated by the work of the
  Garching SN group~\cite{rampp:02,marek:06,bmueller:08}, a set of recent
  core-collapse SN simulations \cite{mueller:04,buras:06b,marek:07,
    marek:08b,scheidegger:08} take GR into account via replacing the
  spherical component of the multipole decomposition of the Newtonian
  gravitational potential with an effective GR potential modeled after
  the Tolman-Oppenheimer-Volkoff (TOV) potential~(e.g.,
  \cite{shapteu:83}). The hydrodynamics, however, are treated in
  standard Newtonian fashion.  Redshift and time-dilation
  effects are taken into account in the radiation transport sector in
  the simulations of the Garching SN group \cite{rampp:02,buras:06b}. 
  GWs are extracted via the slow-motion,
  weak-field quadrupole formalism \cite{thorne:87,finnevans:90,blanchet:90}.

\item {\bf Conformally-flat (CFC) GR}.  In the conformally-flat (or
  conformal-flatness condition [CFC]) approximation to GR introduced
  by Isenberg~\cite{isenberg:08} and first used by Wilson et
  al.~\cite{wilson:96}, the general 3-metric of the $3+1$
  decomposition (e.g., \cite{baumgarte:02}) is replaced by the
  flat-space Minkowski metric scaled with a conformal-factor $\phi^4$.
  The CFC approximation is exact in spherical symmetry. In multi-D, a
  CFC spacetime behaves as an approximation of GR at first
  post-Newtonian order \cite{kley:99} and may be regarded as GR minus
  the dynamical degrees of freedom of the gravitational field that
  correspond to the GW content at infinity. However, even stationary
  spacetimes without GWs can be non-CFC, e.g., rotating NSs or Kerr
  BHs~(e.g., \cite{garat:00,cook:96}). An extension of the CFC
  approach with terms of second post-Newtonian order, CFC+, was
  introduced by Cerd\'a-Dur\'an and collaborators~\cite{cerda:05}.
  The CFC approximation permits the use of the full GR (M)HD equations
  (e.g., \cite{font:08}) and is employed in simulations of
  axisymmetric rotating core collapse by Dimmelmeier and collaborators
  \cite{dimmelmeier:02a,dimmelmeier:02,dimmelmeier:05,dimmelmeier:07,
    dimmelmeier:08,cerda:07,cerda:08}. At least for this scenario it
  has been shown to be an excellent approximation to full GR
  \cite{cerda:05,ott:07cqg}.  Since a CFC spacetime does not contain
  GWs, they must be extracted from the fluid dynamics via variants of
  the slow-motion, weak-field quadrupole formalism, making use of the
  conserved variables of GR (M)HD (e.g.,
  \cite{dimmelmeier:02,dimmelmeier:08}).

\item {\bf Full GR}. Multi-D simulations in full $2+1$ and $3+1$ GR
  evolve the Einstein equations without approximations and with
  appropriate choices of temporal and spatial gauge~(e.g.,
  \cite{baumgarte:02}).  Full GR simulations of rotating core collapse
  and early postbounce evolution were carried out by
  Shibata~et~al.~\cite{shibata:04,shibata:05,shibata:06} and by
  Ott~et~al.~\cite{ott:07prl,ott:07cqg,ott:06phd} who employed the
  BSSN spacetime evolution system
  \cite{Shibata95,Baumgarte99,baumgarte:02}. In principle, GWs can be
  extracted directly from the spacetime in regions sufficiently far
  away from the emission region (e.g., \cite{nagar:05} and references
  therein).  Due to the relative weakness of the GWs emitted in core
  collapse, the direct extraction has proven to be numerically
  difficult and most studies resort to the same slow-motion,
  weak-field quadrupole formalism used in the CFC 
  context\footnote{See \cite{shibatasekiguchi:03,baiotti:08b} for
  comparisons of gauge-invariant extraction techniques with variants
  of the quadrupole formalism.}.

\end{enumerate}

\noindent Studies addressing the GW signature of stellar core collapse
and core-collapse SNe historically followed two general approaches. 

One approach focussed on the detailed treatment of EOS, microphysics
and radiation transport as necessary for physically accurate models of
the collapse and postbounce phase. Studies following this approach were
either Newtonian or employed approximate GR, neglecting the additional
complication of CFC or full GR (e.g., \cite{moenchmeyer:91,
  bh:96,jm:97,kotake:03,kotake:04,mueller:04,ott:06prl}). Due to the
large computational complexity (dominated by neutrino transport) of
such detailed simulations, these studies were limited to small model
sets and, hence, were unable to cover the parameter space of possible
initial configurations and EOS.

Studies following the second approach were primarily concerned with
the GW signal of rotating core collapse and bounce. Their emphasis was
on performing large parameter studies of precollapse rotational
configurations in Newtonian gravity, CFC, or full GR and, in some cases,
included magnetic fields. Their schemes typically
did not take into account a 
microphysical EOS, detailed neutrino transport and microphysics, and
presupernova models from stellar evolutionary calculations 
(see, e.g., \cite{zwerger:97,dimmelmeier:02,shibata:04,ott:04,
obergaulinger:06a,obergaulinger:06b}, but note that \cite{ott:04}
employed a microphysical EOS, but neglected deleptonization).

Only recently, these two approaches have begun to come together in the
first extensive CFC and full GR parameter studies of rotating core
collapse with a microphysicsl EOS and an approximate deleptonization
scheme carried out by
Dimmelmeier~et~al.~\cite{dimmelmeier:07,dimmelmeier:08} and
Ott~et~al.~\cite{ott:06phd,ott:07prl,ott:07cqg}. The results of these
studies will be discussed extensively in the following section.

\section{Rotating Core Collapse and Bounce}
\label{section:rotcoll}

\begin{figure}
\centering
\includegraphics[width=8.7cm]{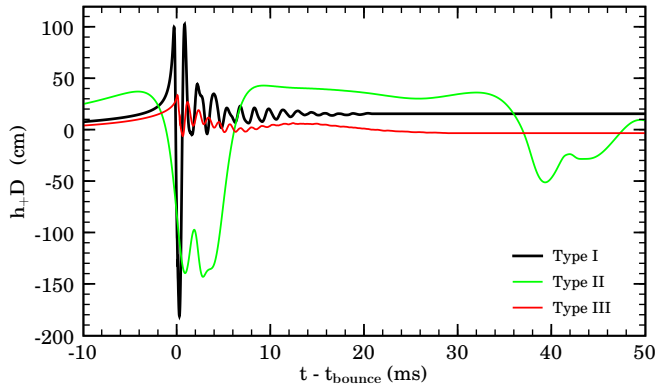}
\caption{Axisymmetric GW burst signal ($h_+\,D$ in units of cm, where
$D$ is the distance of the source) as a function of time after core
bounce for models A1B3G3 (type I), A2B4G1 (type II), and A1B3G5 (type
III) of the 2002 GR study using polytropic initial models and a simple
analytic hybrid polytropic/ideal-fluid EOS carried out by Dimmelmeier
et al.~\cite{dimmelmeier:02}. The waveforms were obtained from 
\cite{garchingcat}.}
\label{fig:gw_dimmelmeier}
\end{figure}

Rapid precollapse rotation, in combination with angular momentum
conservation during collapse, leads to significant asphericity 
in the form of a GW emitting, rapidly time-varying $\ell = 2$
oblate (quadrupole) deformation of the collapsing and bouncing
core.

Rotating collapse and core bounce is the most extensively studied GW
emission process in the massive star collapse context.  In 1982,
M\"uller published in \cite{mueller:82} the first GW signals from
rotating core collapse and bounce that were based on the axisymmetric
(2D) Newtonian simulations of M\"uller and
Hillebrandt~\cite{muellerhille:81}. A large number of 2D studies
followed with varying degrees of microphysical detail, inclusion of GR,
and precollapse model sets (see, e.g.,
\cite{moenchmeyer:91,yamadasato:95,zwerger:97,
dimmelmeier:02,kotake:03,ott:04,shibata:04} and references therein).
These computational parameter studies demonstrated that the general
analytic picture of stellar core collapse derived by Goldreich \&
Weber \cite{goldreich:80} and Yahil \cite{yahil:83} for
spherically-symmetric collapse also holds to good approximation for
rotating cores: From the beginning of collapse, the collapsing core
separates into a subsonically homologously ($v \propto r$) contracting
\emph{inner core} and a supersonically infalling \emph{outer
core}. The mass of the inner core at core bounce sets the mass cut for
the matter that is dynamically relevant in bounce (see, e.g.,
\cite{bethe:90}) and responsible for the GW burst. It also determines
the initial size of the PNS.

Furthermore, these studies identified at least three GW signal
``types'' that can be associated with distinctly different types of
collapse and bounce dynamics (\fref{fig:gw_dimmelmeier} displays
representative examples). Type~I models undergo core bounce governed
by the stiffening of the nuclear EOS at nuclear density and ``ring
down'' quickly into postbounce equilibrium. Their waveforms exhibit
one pronounced large spike at bounce and then show a gradually damped
ring down. Type~II models, on the other hand, are affected
significantly by rotation and undergo core bounce dominated by
centrifugal forces at densities below nuclear. Their dynamics exhibits
multiple slow harmonic-oscillator-like damped
bounce--re-expansion--collapse cycles (``multiple bounces''), which is
reflected in the waveform by distinct signal peaks associated with
every bounce. It is interesting to note that type-II models are
related to {\it fizzlers}, proposed (e.g.,
\cite{shapiro:76,tohline:84,imamura:03}) collapse events of
self-gravitating fluid bodies that become temporally or permanently
stabilized by centrifugal forces before core bounce or black-hole
formation.

Type~III models are characterized by fast collapse
(owing to a very soft subnuclear EOS or very efficient electron
capture), extremely small masses of the homologously collapsing inner
core, and low-amplitude GW emission and a subdominant negative spike
in the waveform associated with bounce.

GWs from collapse with magnetic fields were studied by
\cite{kotake:04,obergaulinger:06a,obergaulinger:06b,shibata:06,scheidegger:08}
who found an additional dynamics/signal type~IV which occurs only in the
case of very strong precollapse core magnetization ($B \gtrsim
10^{12}$ G). Such strong precollapse magnetic fields are unlikely to
occur in nature and the aforementioned studies showed that weaker magnetic
fields have little dynamical consequence during collapse and bounce
(see also the discussion in \cite{burrows:07b}). However, MHD-driven
jet-like bipolar outflows and time-changing magnetic fields themselves
can lead to the emission of GWs (see the discussion in
section~\ref{section:others}) 

\begin{figure}
\centering
\includegraphics[width=8.7cm]{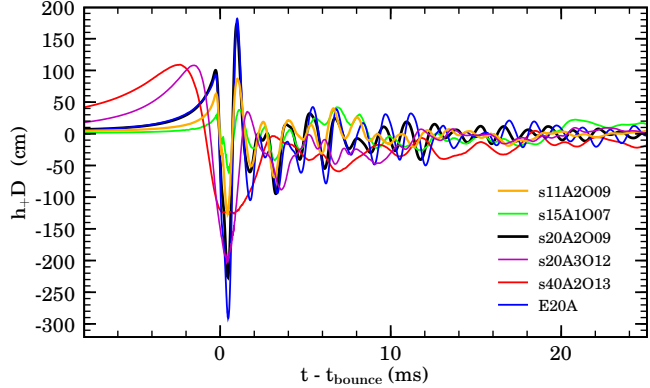}
\caption{GW signals ($h_+\,D$ in units of cm,
where $D$ is the distance of the source) for a few examples from the
2D GR model set of Dimmelmeier~et~al.~\cite{dimmelmeier:08}. The models
shown here were computed with the Shen EOS \cite{shen:98a,shen:98b}
and employ 1D presupernova models of \cite{whw:02}, spanning the
progenitor mass range from $11.2\;\mmsun$ (s11) to $40\;\mmsun$
(s40). The models were set up with precollapse central angular
velocities $\Omega_\mathrm{c,i}$ from $\sim
1.5\;\mathrm{rad}\,\mathrm{s}^{-1}$ to $\sim
11\;\mathrm{rad}\,\mathrm{s}^{-1}$. For details of the rotational
setup, see \cite{dimmelmeier:08}. Model E20A uses a $20$-$\mmsun$
presupernova model that was evolved by \cite{heger:00} with a 1D prescription
for rotation. Note the generic shape of the waveforms, exhibiting one
pronounced spike at core bounce and a subsequent ring down. Very rapid
precollapse rotation ($\Omega_\mathrm{c,i} \gtrsim
6\;\mathrm{rad}\,\mathrm{s}^{-1}$; models s20A3O12 and s40A2O13
in this plot) results in a significant slow-down
of core bounce, leading to a lower-amplitude and lower-frequency GW
burst. The GW signal data are available for download from
\cite{garchingcat}.}
\label{fig:gw_rotbounce}
\end{figure}

Recently, Ott~et~al.~\cite{ott:07cqg,ott:07prl} and
Dimmelmeier~et~al.~\cite{dimmelmeier:07,dimmelmeier:08} presented the
first 3D full GR and 2D CFC GR simulations of rotating stellar
collapse that incorporate GR as well as a microphysical
finite-temperature nuclear EOS and an approximate scheme by
Liebend\"orfer (first presented in \cite{liebendoerfer:05fakenu}) to
account for electron capture and neutrino losses during collapse.
These calculations for the first time included all the known most
relevant physics in the collapse and bounce phase and, in the case of
\cite{dimmelmeier:08}, also considered two different nuclear EOS, the
Lattimer-Swesty EOS~\cite{lseos:91} and the Shen
EOS~\cite{shen:98a,shen:98b}.

 The results of Dimmelmeier~et~al.~\cite{dimmelmeier:08}  show
that the combined effects of deleptonization and GR decrease the mass
and extent of the homologous inner core and reduce significantly the
relevance of centrifugal support for a large set of progenitor models
and a wide range of precollapse rotational configurations. This set
encompasses and goes beyond what is deemed realistic in the context of
collapsing massive stars (see, e.g., \cite{ott:06spin,heger:05}).  In
particular, they find that the GW signal of rotating collapse and core
bounce is of generic Type-I shape. Type-II dynamics with 
the associated ``multiple bounce''
GW signals do not obtain when GR and deleptonization are included.
Type-III dynamics are also absent, since they require very efficient
electron capture and resulting very small masses of the inner core
$\lesssim 0.3\; \mmsun$ that do not occur in the iron core
collapse context where the smallest inner core mass is 
$\gtrsim 0.45\; \mmsun$~\cite{dimmelmeier:08}.

\begin{table*}
\caption{Summary of the GW signal characteristics of rotating iron
core collapse and core bounce based on the waveforms of Dimmelmeier et
al.~\cite{dimmelmeier:08}. All models exhibit type-I dynamics and
waveform morphology and can be organized into three
distinct groups based primarily on their precollapse central angular
velocity $\Omega_\mathrm{c,i}$. $|h_\mathrm{max}|$ is the maximum
gravitational wave strain amplitude (scaled to 10~kpc) at bounce,
$E_\mathrm{GW}$ is the energy radiated away in gravitational waves,
$f_\mathrm{peak}$ is the frequency at which the GW energy spectrum
$dE_\mathrm{GW}/df$ peaks, and $\Delta f_{50}$ is the frequency
interval centered about $f_\mathrm{peak}$ that contains $50\%$ of
$E_\mathrm{GW}$.  Note that $f_\mathrm{max}$ used by Dimmelmeier~et~al.\ is
the peak of the GW signal spectrum and not the peak of
$dE_\mathrm{GW}/df$.  Also note that for the slowly rotating group
prompt postbounce convective overturn contributes significantly to the
overall GW signal.  The convective contribution was removed from the
waveforms before analysis, since the deleptonization scheme employed
by Dimmelmeier~et~al.\ and Ott et~al.\ is ineffective at postbounce times and
is likely to overestimate the strength and duration of prompt
convection after bounce \cite{dimmelmeier:08}. } {\footnotesize
\begin{tabular}{clllll}
\multicolumn{1}{c}{Group}
&\multicolumn{1}{c}{$\Omega_\mathrm{c,i}$}
& \multicolumn{1}{c}{$|h_\mathrm{max}|$}
& \multicolumn{1}{c}{$E_\mathrm{GW}$}
& \multicolumn{1}{c}{$f_\mathrm{peak}$} 
& \multicolumn{1}{c}{$\Delta f_{50}$}\\
&\multicolumn{1}{c}{($\mathrm{rad}\;\mathrm{s}^{-1}$)}
& \multicolumn{1}{c}{($10^{-21}$ at $10\;\mathrm{kpc}$)}
& \multicolumn{1}{c}{($10^{-8} \mmsun\, c^2$)}
& \multicolumn{1}{c}{(Hz)}
& \multicolumn{1}{c}{(Hz)}\\
\hline
1
&$\lesssim 1$--$1.5$
&$\lesssim 0.5$
&$\lesssim 0.1$
&$\sim 700$--$800$
&$\sim 400$\\
\hline
2 
&$1$--$2$ to $6$--$13$
&$0.5$ to $10$
&$0.1$ to $5$
&$\sim 400$--$800$
&$100$ to $400$\\
&&&&most models: $700$-$800$
\\
\hline
3
&$\gtrsim 6$--$13$
&$3.5$ to $7.5$
&$0.07$ to $0.5$
&$70$ to $200$
&$80$ to $250$\\
\end{tabular}}
\label{table:rotcollapse}
\end{table*}

Furthermore, Dimmelmeier~et~al.~\cite{dimmelmeier:08} demonstrate
that the rotating collapse and bounce dynamics and the resulting GW
burst signal depend primarily on the precollapse central angular
velocity $\Omega_\mathrm{c,i}$ and secondarily on the progenitor mass,
which influences precollapse iron core entropy and mass, which, in
turn affects the mass of the inner core at bounce
(see, e.g., \cite{baron:90,burrows:83}).  
The degree of differential rotation and the stiffness of the
particular choice of nuclear EOS appear to have little influence on
the general characteristics of the GW signal.

Figure~\ref{fig:gw_rotbounce} shows example waveforms of a subset of
the models considered by Dimmelmeier~et~al.~\cite{dimmelmeier:08}
whose waveform data are available from \cite{garchingcat}.
Although all models exhibit generic type-I dynamics and signal
morphology, the dynamics and GW signals can be ordered into three
groups (for more details, see \cite{dimmelmeier:08} and the summary in
table~\ref{table:rotcollapse}): 
\renewcommand{\labelenumi}{(\arabic{enumi})}
\begin{enumerate}
\item Slowly rotating iron cores with $\Omega_\mathrm{c,i} \lesssim
1$--$1.5\;\mathrm{rad}\,\mathrm{s}^{-1}$ undergo core bounce dominated
by nuclear pressure effects, develop only a small quadrupole deformation and
yield small peak GW amplitudes $|h_\mathrm{max}|$ at core bounce that
are generally below $\sim 5\times10^{-22}$ for a galactic
core-collapse event at $10\;\mathrm{kpc}$ distance. All slowly
rotating models exhibit strong prompt postbounce convective overturn
in the postshock region owing to the negative entropy gradient left
behind by the stalling the shock. Because of the lack of postbounce
neutrino transport, the strength and duration of the prompt convection
is probably overestimated in these models \cite{dimmelmeier:08}
(see also section \ref{section:promptconv} in this article).

\item Moderately rapidly to rapidly rotating iron cores with
  $1$--$2\;\mathrm{rad}\,\mathrm{s}^{-1} \lesssim \Omega_\mathrm{c,i}
  \lesssim 6$--$13\;\mathrm{rad}\,\mathrm{s}^{-1}$ still experience
  pressure-dominated bounce, but develop larger quadrupole
  deformations and have a rotationally increased mass of the inner
  core at bounce.  This results in sizeable peak GW amplitudes and
  energy emissions (see table \ref{table:rotcollapse}). Prompt
  postbounce convection is suppressed by positive specific angular
  momentum gradients in the postshock region and does not contribute
  to the GW signal (see, e.g., \cite{ott:06spin}). The upper end of
  the range in $\Omega_\mathrm{c,i}$ for this group is dependent on
  progenitor characteristics. Progenitors with massive (and
  higher-entropy) iron cores tend to transition to the next group at
  lower $\Omega_\mathrm{c,i}$. Dimmelmeier~et~al.\ find approximate
  transition $\Omega_\mathrm{c,i}$ of $\gtrsim 13$, $\sim 9$, $\sim
  11$, and $\sim 7\;\mathrm{rad\;s}^{-1}$ for their $11.2$, $15$,
  $20$, and $40\;\mmsun$ models, respectively.

\item Very rapid precollapse rotation ($\Omega_\mathrm{c,i} \gtrsim
6$--$13\;\mathrm{rad}\,\mathrm{s}^{-1}$; the actual value depending on
the progenitor model) results in slow core bounce governed exclusively
by centrifugal forces at significantly subnuclear densities. This
results in a decrease of $|h_\mathrm{max}|$ and a shift of the peak of
the GW energy spectrum to frequencies below $\sim 400\;\mathrm{Hz}$.
\end{enumerate}

\begin{figure}
\centering
\includegraphics[width=8.7cm]{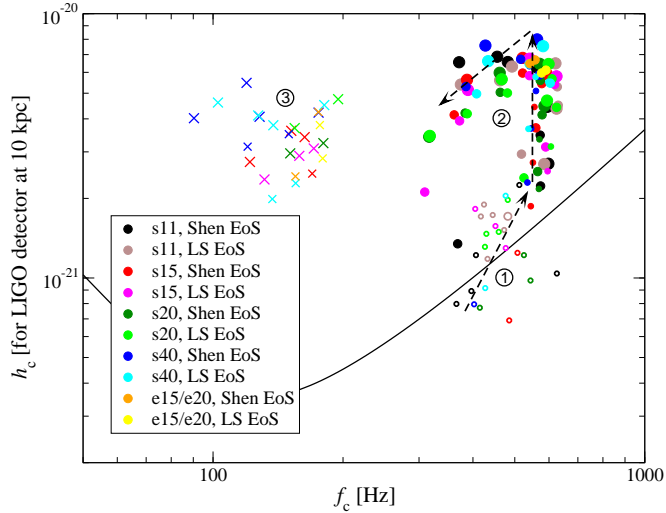}
\caption{Location of the GW signals of Dimmelmeier et
al.~\cite{dimmelmeier:08} in the $h_c$-$f_c$ plane (using the
detector-dependent definitions of $h_c$ and $f_c$ given in
\cite{thorne:87}) relative to the $h_\mathrm{rms} = \sqrt{f S(f)}$
sensitivity of initial LIGO and for a source distance of
$10\;\mathrm{kpc}$. The three groups of signals discussed in the text
and summarized in table~\ref{table:rotcollapse} are marked.  This plot
is a similar to figure~16 of \cite{dimmelmeier:08} and based on
freely available data~\cite{garchingcat}.}
\label{fig:gw_rotbounce_ligo}
\end{figure}

\Fref{fig:gw_rotbounce_ligo} contrasts the strengths of the GW bursts
of the models of Dimmelmeier~et~al.~\cite{dimmelmeier:08} with initial
LIGO sensitivity~\cite{ligo}. Plotted is the initial LIGO
dimensionless rms strain sensitivity $h_\mathrm{rms} = \sqrt{f S(f)}$
as a function of frequency $f$, where $S(f)$ is the noise spectral
density in units of $(\mathrm{Hz})^{-1/2}$. Each dot represents a
particular model and is plotted at that model's detector-dependent
characteristic GW frequency $f_c$ and characteristic strain $h_c$,
computed by Dimmelmeier~et~al.\ according to the prescription given in
\cite{thorne:87}. The models cluster in the $h_c$--$f_c$ plane
according to their membership in one of the three groups discussed
above. Slowly rotating models (group 1) have small $h_c$ and
moderately high $f_c$. With increasing $\Omega_\mathrm{i,c}$, the
model move upward and slightly to the right (higher $h_c$ and slightly
higher $f_c$). Moderately rapid models (group 2) reach high $h_c$ and
cluster in $f_c$ between $600$ and $800\;\mathrm{Hz}$. The most
rapidly rotating models of that group move somewhat to the left in
$f_c$, while very rapidly rotating and centrifugally bouncing models
(group 3) cluster at low $f_c$ and lower $h_c$, but still exhibit a
high signal to noise ratio (SNR) with initial LIGO and at a distance of
$10\;\mathrm{kpc}$. Hence, according to \fref{fig:gw_rotbounce_ligo},
GWs from moderately rapidly to rapidly rotating core collapse and
bounce within the Milky Way should be detectable by current
LIGO-class detectors. In addition, it may be possible, based on a
determination of the dominant emission frequency (ideally in
combination with a determination of the progenitor mass by other
astronomical means), to constrain the precollapse iron core rotation.

To conclude this section, we point out that according to the most
recent stellar evolutionary models and pulsar birth spin estimates
\cite{heger:05,ott:06spin}, garden-variety massive stars at solar
metallicity are probably spinning with central angular velocities
significantly below $1\;\mathrm{rad}\,\mathrm{s}^{-1}$ (group 1).  GW
bursts from core bounce of such models are unlikely to be detectable
by current detectors and may be marginally detectable by advanced
LIGO.  More rapid rotation may be relevant in the context of gamma-ray
burst progenitors (see, e.g., \cite{wb:06}) which could make up
perhaps $\sim 1\%$ of the single massive star population, but may
occur primarily at low metallicities \cite{woosley:06}.

\section{Rotational Nonaxisymmetric Instabilities}
\label{section:rotinst}
\setcounter{footnote}{1} From the classical theory of Newtonian
MacLaurin spheroids (see, e.g.,
\cite{chandrasekhar69c,stergioulas:03}) one finds that rotating
axisymmetric fluid bodies become unstable to nonaxisymmetric
deformations due to a dynamical instability at rotation rates $\beta =
T/|W| \gtrsim 0.27 \equiv \beta_\mathrm{dyn}$, where $T$ is the
rotational kinetic and $|W|$ the gravitational energy of the
spheroid. A secular (dissipation driven\footnote{Possible mechanisms
are physical viscosity and gravitational-wave reaction. See, e.g.,
\cite{stergioulas:03,andersson:03}.})  instability may set in at
$\beta = T/|W| \gtrsim 0.14 \equiv \beta_\mathrm{sec}$, though, due to
its secular nature, has longer growth times than the dynamical
instability.

Stars that go unstable to the classical MacLaurin-type dynamical or
secular instability develop global azimuthal (nonaxisymmetric)
structure that, at least in the linear regime, can be characterized in
terms of modes $m$ with spatial structure proportional to
$\exp(im\varphi)$, where $\varphi$ is the azimuthal angle. In most
cases the bar-like $m=2$ mode is dominant and one frequently speaks of
a ``barmode instability''.  Direct numerical simulations have
demonstrated (e.g., \cite{saijo:01,baiotti:07,manca:07}) that
$\beta_\mathrm{dyn}$ holds approximately even in the case of
differentially rotating compressible fluid configurations in GR. For
the secular instability, perturbative studies suggest an onset at
significantly lower $\beta$ in GR (see, e.g., \cite{stergioulas:03,
  andersson:03} for discussions). However, GR
hydrodynamic studies of the secular instability remain yet to be
carried out.

A spinning bar is a simple system that emits GWs at twice its
rotational frequency (due to its $\pi$-symmetry) with amplitudes
$|h_\mathrm{bar}| \propto MR^2 \Omega^2/D$, where $M$ is the bar's
mass, $2R$ its length, and $\Omega = 2\pi f$ its angular
velocity. Using the Newtonian quadrupole approximation, one can derive
an estimate for the GW amplitude \cite{fryer:02},
\begin{eqnarray}
\label{eq:bar}
|h_\mathrm{bar}| &\approx 4.5\times10^{-21} 
\bigg(\frac{\epsilon}{0.1}\bigg)
\bigg(\frac{f}{500\,\mathrm{Hz}}\bigg)^2 \times \nonumber\\
&\;\;\;\;\;\;\bigg(\frac{D}{10\, \mathrm{kpc}}\bigg)^{-1}
\bigg(\frac{M}{0.7\, \mmsun}\bigg)
\bigg(\frac{R}{12\, \mathrm{km}}\bigg)^2\,\,,
\end{eqnarray}
where $\epsilon$ is the ellipticity of the bar. The scaling
of \eref{eq:bar} is set up to reflect a PNS core of $0.7\;\mmsun$
with a $2\;\mathrm{ms}$ period and a radius of $12\;\mathrm{km}$ that
is deformed only moderately ($\epsilon = 0.1$) and located at a distance
of $10\;\mathrm{kpc}$.

\subsection{Instability at High-$\beta$}

The literature on nonaxisymmetric instabilities in rotating fluid
bodies is extensive and cannot be reviewed here in the necessary
detail. Relatively recent reviews and relevant references can be found
in \cite{stergioulas:03,andersson:03,baiotti:07}.  In the context of
massive star collapse and PNSs, the first Newtonian 3D simulations
were carried out by Rampp~et~al.~\cite{rmr:98} and
Brown~\cite{brown:01} who followed the postbounce development of
nonaxisymmetric dynamics in extremely differentially and rapidly
rotating simplified stellar models that reached up to $\beta \approx
0.35$ at bounce in 2D and were mapped to 3D shortly before or after
bounce.  With a similar approach, though in full GR, Shibata \&
Sekiguchi~\cite{shibata:05} studied the development of nonaxisymmetric
dynamics in models that reached $\beta$ near $\beta_\mathrm{dyn}$ and
reported the development of $m=2$ and $m=1$ dynamics in their models.

Ott~et~al.~\cite{ott:06spin,ott:07prl,ott:07cqg} and very recently
Dimmelmeier~et~al.~\cite{dimmelmeier:08} studied via their 3D full GR 
and 2D CFC GR core-collapse simulations the prospects for nonaxisymmetric
rotational instabilities in massive star collapse and postbounce
evolution. Their models included microphysical details and they
considered a large number of precollapse rotational configurations
that covered a parameter space, encompassing and going beyond current
predictions from stellar evolution theory and constraints from pulsar
birth spin estimates \cite{ott:06spin,heger:05,woosley:06}.

Ott~et~al.\ and Dimmelmeier~et~al.\ found that even their most extreme
models do not reach values of $\beta$ close to
$\beta_\mathrm{dyn}$ during collapse, bounce, and during early postbounce
times. Their work confirmed results obtained in studies
considering evolutionary sequences of rotating PNSs \cite{villain:04}
and 2D Newtonian simulations \cite{ott:04,ott:06spin}.

On the basis of the current understanding, it appears rather unlikely
that a PNS in nature develops a high-$\beta$ rotational instability
before, at, or early after bounce. As the PNS contracts and cools on a
timescale of seconds to minutes and if angular momentum is conserved
and not redistributed by other means (see, e.g.,
\cite{shibata:06,burrows:07b,ott:06spin}), it spins up and may reach
$\beta_\mathrm{dyn}$.  It is, however, more likely that the secular
instability driven by viscous dissipation or GW back reaction, which
in PNSs has a growth timescale of the order of or larger than 
$\sim 1\;\mathrm{s}$ \cite{lai:95,lai:01}, may set in first.

\subsection{Instability at Low-$\beta$}

\begin{figure}
\centering
\includegraphics[width=5.4cm]{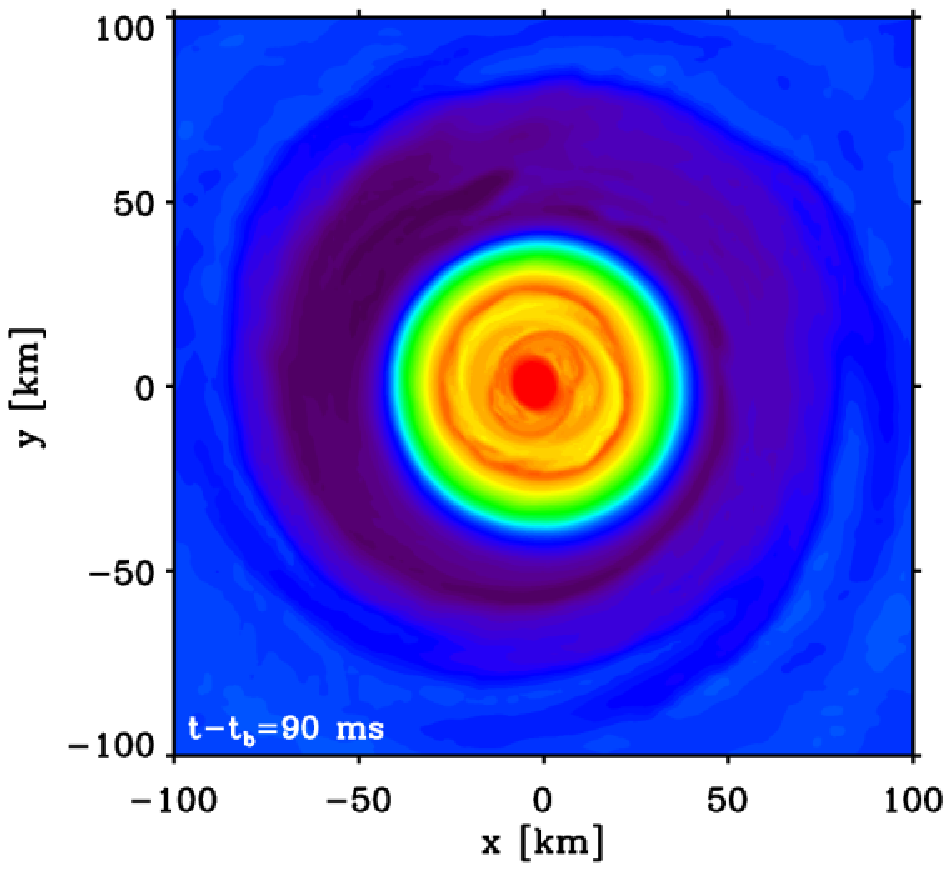}
\includegraphics[width=6.3cm]{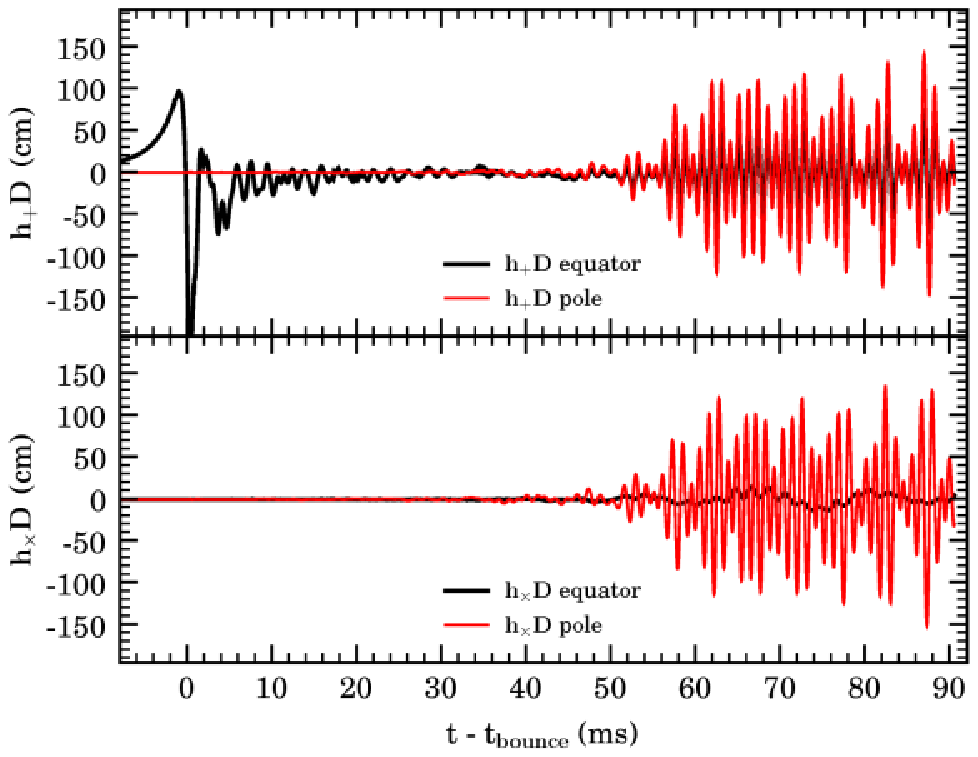}
\caption{{\bf Left}: Colormap depicting the specific entropy
distribution in the equatorial plane of the $20$-$\mmsun$ model
s20A2B4 of Ott~et~al.~\cite{ott:06phd,ott:07prl,ott:07cqg} at 90~ms
after core bounce. Red and yellow regions of the PNS core have low
entropy ($\sim 1$--$3\;\mathrm{k_B}$/baryon) while dark blue and black
symbolize high entropy $> 6 \;\mathrm{k_B}$/baryon. The nonaxisymmetric
structures are of primarily $m=\{1,2,3\}$ nature with radial variations
in dominance.  {\bf Right}: GW polarizations
$h_+$ (top panel) and $h_\times$ (bottom panel) multiplied by distance
$D$ as seen by observers along the equator (black) and along the pole
(red) in the same model.  Significant nonaxisymmetric dynamics with
$m=2$ components develop around $30$--$40\;\mathrm{ms}$ after core
bounce at a rotation rate $\beta \sim 0.13$. Note that the GW burst
signal from core bounce is purely axisymmetric, since an axisymmetric
system has vanishing $h_\times$ and vanishing GW emission along the
axis of symmetry. The waveforms are available from \cite{ottcatalog}.}
\label{fig:gw_nonaxi}
\end{figure}

Via 3D full GR simulations that were carried out self-consistently
from the onset of collapse,
Ott~et~al.~\cite{ott:06phd,ott:07prl,ott:07cqg} confirmed that stellar
cores with precollapse parameters in the investigated parameter space
stay axisymmetric throughout collapse, bounce and the very early
postbounce phase. Furthermore, they found the development of
nonaxisymmetric dynamics at postbounce times $\gtrsim 20\;\mathrm{ms}$
in a subset of their models.  The 3D dynamics were of predominantly
$m=\{1,2,3\}$ spatial character (see the left panel of
\fref{fig:gw_nonaxi}) and were followed to late times in only two
models, one having a postbounce $\beta$ of $0.09$, the other one
spinning with $\beta \approx 0.13$. Very recently,
Scheidegger~et~al.~\cite{scheidegger:08}, who performed a 3D Newtonian
study with a monopole gravitational potential with relativistic
corrections, found a similar instability in a model with a smaller
postbounce value of $\beta$ around $0.04$--$0.05$.

The instability at low-$\beta$ seen in simulations may be related to a
class of co-rotation dynamical shear instabilities that operates on
the shear-energy stored in differential rotation and that powers
azimuthal modes via resonant coupling with the background fluid. This
occurs at co-rotation points where the mode pattern speed coincides with the
fluid angular velocity \cite{watts:05,saijo:06}. Such a low-$\beta$
instability was first observed in a compact-star context by
Centrella~et~al.~\cite{centrella:01} in 2001 and multiple subsequent studies
confirmed its presence in differentially rotating equilibrium
stellar models below the classical high-$\beta$ instability limits
(e.g., \cite{saijo:03,shibata:04a,ou:06,cerda:07b}).  In 2005,
Ott~et~al.~\cite{rotinst:05} discovered a low-$\beta$ instability
in a 3D simulation of a simplified postbounce stellar core.

Differential rotation develops naturally in the outer core during
stellar collapse \cite{ott:06spin}. Hence, such low-$\beta$
instabilities could -- in principle -- occur in any PNS with (1)
angular velocities at the edge of the inner core that are comparable
to the pattern speed of unstable modes and (2) with sufficient total
rotational energy and strong enough differential rotation to provide
power for significant nonaxisymmetric deformation.

The right part of figure~\ref{fig:gw_nonaxi} depicts the GW signal of
a $20$-$\mmsun$ model of Ott~et~al.~\cite{ott:07prl,ott:07cqg} that
experiences the low-$\beta$ nonaxisymmetric instability. Shown are the
$+$ (top panel) and $\times$ (bottom panel) GW polarizations for
observers located along the polar axis (red) and observers
located in the equatorial plane at zero azimuth (black). The PNS and
postshock region stay axisymmetric through bounce and exhibit GW
emission only in the $+$ polarization and away from the poles.  The
nonaxisymmetric dynamics reach sizable amplitudes around
$30$--$40\;\mathrm{ms}$ and their $m=2$ components emit GWs in
bar-like fashion with correlated strong emission in $h_+$ and
$h_\times$ that is strongest for observers located perpendicular to
the equatorial plane. The simulation on which figure~\ref{fig:gw_nonaxi}
is based was tracked to $\sim 90\;\mathrm{ms}$ after core bounce, but
the instability could potentially continue for hundreds of
milliseconds until angular momentum has been redistributed by the
instability or by another process (e.g., via the magneto-rotational
instability [MRI]; see, e.g., \cite{bh:91,akiyama:03} and references
therein).

The simple and rough estimate of equation \eref{eq:bar} for a slightly deformed
PNS core rotating with a period of $2\;\mathrm{ms}$ matches the GW
signal of the low-$\beta$ instability shown in \fref{fig:gw_nonaxi}
quite well and other models of Ott~et~al.\ and Scheidegger~et~al.\ to
within a factor of a few. Concretely, these studies found maximum GW
signal amplitudes in the range of $\sim 1$--$5 \times 10^{-21}$ (at
$10\;\mathrm{kpc}$) and total emitted GW energies $E_\mathrm{GW}$ of
$\sim 5$--$15\times 10^{-8}\;\mmsun\,c^2$. Most of the energy ($>
50\%$) is emitted in a frequency interval of $\Delta f_{50} \approx
50$--$200\;\mathrm{Hz}$ about the frequency $f_\mathrm{peak}$ at which
$dE_\mathrm{GW}/df$ peaks, which is $\sim 900$--$940\;\mathrm{Hz}$ in
the models of Ott~et~al.\ and Scheidegger~et~al.  Comparing the just
stated quantitative results with those summarized for rotating
collapse and bounce in table~\ref{table:rotcollapse}, one notes that
the GW emission due to nonaxisymmetric rotational dynamics does not
necessarily dominate in signal amplitude, but does so clearly in total
emitted energy. Its $E_\mathrm{GW}$ scales with the number $n$ of 
``emission cycles,'' which, of course, depends on
the duration of the nonaxisymmetric dynamics.

As demonstrated by
\cite{ott:07prl,ott:07cqg,scheidegger:08,rotinst:05}, the potential
for a strong enhancement of the GW signature of rotating core collapse
by a low-$\beta$-type rotational instability is
great. However, the number of 3D core-collapse models in which such an
instability was observed is still very limited and the systematic
dependence on $\beta$, degree of differential rotation, and
thermodynamic and magneto-hydrodynamic configuration of the PNS
remains to be established.  Furthermore, it will be necessary to
understand the long-term behavior of the instability and its
interaction and competition with other shear instabilities such as the
MRI.

\section{Postbounce Convection and SASI}
\label{section:convsasi}

Convective instability is a central feature of the postbounce
evolution of core-collapse SN and has been discussed extensively
(e.g., \cite{herant:94,bhf:95,jankamueller:96,
  keil:96,mezzacappa:98,fh:00,buras:06b,
  dessart:06pns,burrows:06,burrows:07a,janka:07} and references
therein). According to the Schwarzschild-Ledoux criterion
\cite{ledoux:47}, convective overturn develops in the presence of
negative radial entropy or composition gradients (see, e.g.,
\cite{bhf:95,buras:06b}).

Convective overturn in postbounce SN cores is expected as entropy- and
lepton-gradient driven \emph{prompt convection} which may occur
immediately after bounce, lepton-gradient driven \emph{PNS
  convection}, and \emph{neutrino-driven convection} in the postshock
heating region.  Convection will occur at postbounce times in
virtually all core-collapse events, but can be weakened and limited to
polar regions by positive specific angular momentum gradients
in rapidly rotating cores~\cite{endal:78,fh:00,ott:06spin,ott:08}.

The standing-accretion-shock instability (SASI; see, e.g.,
\cite{foglizzo:07,burrows:06,blondin:03,blondin:06,scheck:08} and
references therein) is believed to be caused by either an
advective-acoustic \cite{foglizzo:07,scheck:08} or a purely
acoustic \cite{blondin:06,blondin:03} feedback cycle, leading
to the growth of perturbations in the stalled shock. In axisymmetry,
the dominant SASI modes are of $\ell = \{1,2\}$ character, while
in the largely unexplored 3D case, power may go into azimuthal
$m$ modes. This can reduce the saturation amplitudes in the $\ell$
modes (see \cite{iwakami:07} for
first 3D results, but also \cite{yamasaki:07} for a perturbative
analysis without symmetry constraints). 

The SASI grows to non-linear amplitudes over a timescale of typically
$200$--$300\;\mathrm{ms}$ after bounce with some dependence on initial
accretion shock radius, neutrino luminosity, and accretion rate (e.g.,
\cite{scheck:08,murphy:08}).  Rapid rotation may delay the growth of
the SASI and weaken the overall phenomenon in 2D \cite{ott:08}, while
favoring azimuthal SASI modes in 3D~\cite{yamasaki:07}.  

Once active at significant amplitudes, the SASI heavily distorts the
postshock region, affecting convection to great extent. In addition,
it enhances and modulates accretion funnels that appear as kinks in
the shock and channel low entropy material at high rates onto the PNS
core \cite{burrows:06}.

Both convection and SASI are intrinsically multi-D phenomena
and lead generically to time-varying mass quadrupole moments, hence,
emit GWs. The emission of GWs by convection was first considered by
M\"uller and Janka \cite{jm:97} in 2D and 3D postbounce models with an
approximate treatment of neutrino heating and cooling.  Subsequent
studies with more detailed neutrino transport and microphysics were
carried out by M\"uller et al.\ \cite{mueller:04},
Fryer~et~al.~\cite{fryer:04}, 
Ott~et~al.~\cite{ott:06prl,ott:06phd}, and very recently by
Marek~et~al.~\cite{marek:08b}.
Kotake~et~al.~\cite{kotake:07a} were the first to specifically address
GW emission associated with the SASI using a simple neutrino 
heating/cooling scheme.

In the following, we discuss the GW emission from convection and
SASI, present new results that were performed with the VULCAN/2D SN
code \cite{livne:04,ott:04,burrows:07a} and contrast them with
previous work.  VULCAN/2D is an axisymmetric Newtonian multi-group
flux-limited diffusion (MGFLD) radiation-hydrodynamics code.  It
employs an unsplit arbitrary Lagrangian-Eulerian (ALE) scheme that
allows for a central pseudo-Cartesian grid that transitions to a
polar-type grid at a transition radius (see figure 4 of
\cite{ott:04}).

\begin{figure}
\centering
\includegraphics[width=8.7cm]{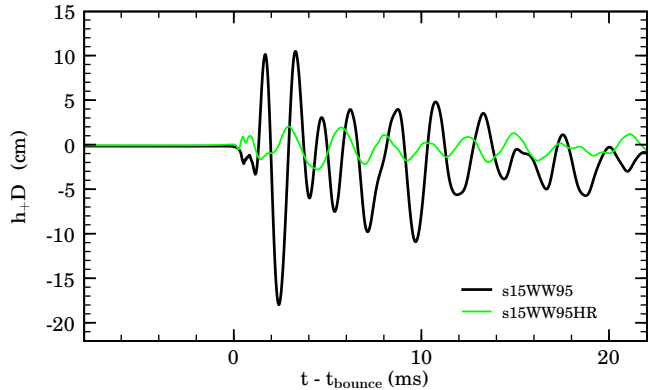}
\caption{GW burst signal ($h_+ D$ in units of cm) due
to 2D prompt postbounce convection in a nonrotating $15$-$\mmsun$ model
that employed the precollapse profile of~\cite{ww:95}.  The model
calculations were performed with the VULCAN/2D code
\cite{ott:04,livne:04,burrows:07a}. VULCAN/2D uses a central
pseudo-Cartesian grid that is matched to a polar-type grid at a
transition radius ($20\;\mathrm{km}$ in s15WW95HR and
$30\;\mathrm{km}$ in s15WW95). The transition introduces artificial
perturbations that act as seeds for prompt convection.  Model
s15WW95HR is a high-resolution variant of s15WW95 with $\sim 45\%$
more zones in the central region. The resolution -- in other words --
seed perturbation dependence of prompt convection is apparent. The GW
signal data shown here are available from \cite{ottcatalog}.}
\label{fig:promptconv}
\end{figure}

\subsection{Prompt Convection}
\label{section:promptconv}

As the stalling bounce shock passes through outer core material, it
leaves behind a negative entropy gradient. Furthermore, following
neutrino shock breakout and the associated burst of electron neutrinos
(e.g., \cite{thompson:03,buras:06b,dessart:06pns}), a negative lepton
gradient arises at the outer edge of the PNS immediately below the
neutrinosphere\footnote{The neutrinosphere can be defined as the
  surface at which the optical depth for neutrinos $\tau_\nu$, given
  by $\tau_\nu = \int_\infty^R dr{\lambda_\nu}^{-1}$, is equal to
  $2/3$. Here, $\lambda_\nu$ is the (energy-averaged) neutrino
  mean-free path.  Note that the neutrino sphere is strongly dependent
  on neutrino energy and species.}. The two negative gradients lead to
a convectively unstable region according to the Schwarzschild-Ledoux
criterion \cite{burrows:92,bruenn:94,swesty:05,marek:08b}.

Neutrino losses and, to a limited degree, neutrino energy deposition
behind the stalling shock, smooth out the large negative entropy
gradient in the immediate postshock region, but prompt 
convection can still develop rapidly and last
for $\gtrsim 10$--$20\,\mathrm{ms}$ if significant seed perturbations
are present in the immediate postbounce flow
\cite{burrows:92,bruenn:94,swesty:05,ott:04,fryer:04}.

The magnitude and distribution of seed perturbations in the central
regions of iron cores in nature is presently unknown. However, it is
not unlikely that the late burning stages of stellar evolution result
in inhomogeneities in the iron core that do not smooth out completely
between the end of core burning and core collapse (see also the
discussion of large-scale precollapse asymmetries in
section~\ref{section:global}). Such inhomogeneities are frozen in
during collapse and may act as seed perturbations for prompt
convection \cite{lai:00,murphy:04,fryer:04}.

Here, we present results from VULCAN/2D simulations using a
nonrotating $15$-$\mmsun$ progenitor model of \cite{ww:95}. Model
s15WW95 uses the standard simulation setup presented in
\cite{burrows:07a,ott:06prl}, while model s15WW95HR is set up with
$45\%$ higher radial and angular resolution inside $\sim
300\;\mathrm{km}$.  The Cartesian--polar transition of the VULCAN/2D
grid introduces artificial perturbations that converge away with
increasing resolution and decreasing transition radius.  These
numerical perturbations act as seeds for prompt convection.

Figure~\ref{fig:promptconv} depicts the GW signal of prompt convection
as seen in the VULCAN/2D simulations.  In both calculations, prompt
convection is strongest inside a radius of $\sim 60\;\mathrm{km}$ from
the origin. However, one should keep in mind that the region in which
prompt convection is most pronounced depends on the location and
strength of the negative entropy gradient \emph{and} on the location
and magnitude of the seed perturbations.  In the standard-resolution
variant, the large seed perturbations lead to strong prompt convection
and a correspondingly large GW burst setting in within $\sim
1\;\mathrm{ms}$ of core bounce.  The maximum GW amplitude is $\sim 6
\times 10^{-22}$ (at $10\;\mathrm{kpc}$) and within $\sim
20\;\mathrm{ms}$, an energy of $\sim 1.5\times10^{-10}\,\mmsun c^2$ is
emitted in GWs.  The peak of the GW energy spectrum is at
$f_\mathrm{peak} \approx 680\;\mathrm{Hz}$ and $50\%$ of the energy is
emitted at frequencies $\delta_{f,50} = 150\,\mathrm{Hz}$ centered
about $f_\mathrm{peak}$.  The higher resolution calculation s15WW95HR
has considerably smaller seed perturbations, hence prompt convection
is much weaker and and a smaller GW signal is emitted with
$|h_\mathrm{max}| \approx 9\times10^{-23}$, $E_\mathrm{GW} \approx
6\times10^{-12}\, \mmsun c^2$, $f_\mathrm{peak} \approx
430\;\mathrm{Hz}$, and $\Delta f_{50} \approx 250\,\mathrm{Hz}$.

The above numbers and the GW signals shown in
figure~\ref{fig:promptconv} demonstrate the general seed-perturbation
dependence of prompt convection and of the resulting GW burst signal.
However, they constitute only two examples and do not bracket the
parameter space of possible immediate-postbounce configurations.

Marek~et~al.~\cite{marek:08b} also observed prompt convection in their
very recent study focussing on the EOS dependence of neutrino and GW
emission in the postbounce phase of a nonrotating $15$-$\mmsun$
model. They compared a model run with a soft variant of the
Lattimer-Swesty EOS~\cite{lseos:91} with a counterpart model run with
the rather stiff Wolff EOS~\cite{hillewolff:85}. Their models produce
GW bursts associated with prompt convection that last $\sim
30-40\;\mathrm{ms}$ and produce maximum amplitudes of $\sim
1-2\times10^{-22}$ (at $10\;\mathrm{kpc}$) with most of the emission
occurring between $\sim 60$ and $\sim150\;\mathrm{Hz}$.  Furthermore,
Marek~et~al.\ found that the model employing the stiffer nuclear EOS
yields up to a factor of 2 larger maximum amplitudes and somewhat
higher frequency emission than its softer-EOS counterpart. In the VULCAN/2D
models discussed in the above, the GW emission from prompt postbounce
convection occurs primarily at $400-700\;\mathrm{Hz}$. This
discrepancy with Marek~et~al.\ is most likely due to different
numerical seed perturbations and different locations of the former and
of the convectively unstable regions.

For a more complete picture and an understanding of the possible
prompt-convection GW signals, it will be necessary to conduct a
comprehensive study of the quantitative systematics of prompt convection
with variations in the magnitude and position of seed perturbations.
Ideally such a study should be carried out with a code that uses a
nearly perturbation-free grid and with clearly quantifiable
perturbations added to the flow shortly before core bounce.

\begin{figure}
\centering
\includegraphics[width=7.5cm]{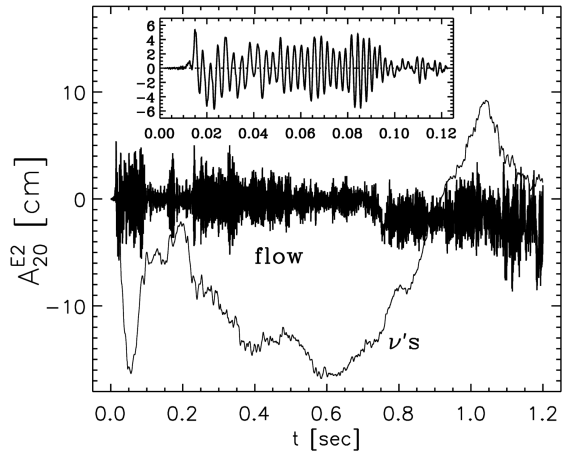}
\caption{GW signal due to PNS convection in the axisymmetric PNS model
of Keil~et~al.~\cite{keil:96,keil:97} as analyzed by
M\"uller~et~al.~\cite{mueller:04} in terms of the quadrupole pure-spin
tensor harmonic amplitude $A_{20}^{E2}$, and $h_+ D =
\frac{1}{8}\sqrt{\frac{15}{\pi}} \sin^2\theta A^{E2}_{20}$.  The
matter GW signal and the GW signal due to anisotropic neutrino
emission are shown (for a discussion of the latter, see
section~\ref{section:neutrinos}).  This figure is figure~5 of
\cite{mueller:04} and used by kind permission from the authors.}
\label{fig:mueller2004}
\end{figure}

\subsection{PNS convection} 
\label{section:pnsconv}

Owing to a negative radial lepton gradient, PNSs are unstable to
convective overturn in a radial interval from $\sim 10\;\mathrm{km}$
to $\sim 30\;\mathrm{km}$ \cite{keil:96,dessart:06pns,buras:06b}.
Convection sets in $\sim 20$--$50\;\mathrm{ms}$ after bounce and may
last for several seconds as the PNS slowly contracts and deleptonizes
\cite{keil:96} after a successful SN explosion. If the explosion
is not successful (or weak) and a black hole is formed, 
PNS convection and the associated GW emission stop abruptly.
Also, if present, large-amplitude PNS core $g$-modes can distort PNS
convection and themselves lead to GW emission much stronger than that
due to the convective motions (see section~\ref{section:gmodes}).

Figure~\ref{fig:mueller2004} displays the GW signal
of PNS convection as analyzed by M\"uller~et~al.~\cite{mueller:04} for
the axisymmetric PNS convection simulations of
Keil~et~al.~\cite{keil:96,keil:97} who followed the evolution of an
isloated nonrotating PNS core from shortly after core bounce to
$1.2\;\mathrm{s}$. M\"uller~et~al.\ reported typical strain amplitudes
$|h|$ of $2-5\times10^{-23}$ (at $10\;\mathrm{kpc}$), an energy
emission $E_\mathrm{GW}$ of $1.6\times10^{-10}\;\mmsun c^2$ over
$1.2\;\mathrm{s}$ and a broad spectrum peaking around $\sim
1300\;\mathrm{Hz}$ and with most of the energy emitted at
$700$--$1500\;\mathrm{Hz}$.

\setcounter{footnote}{1}

We analyze the GW signal emitted by PNS convection in the inner $\sim
50\;\mathrm{km}$ of the aforementioned VULCAN/2D simulations s15WW95
and s15WW95HR in the interval of $20$--$250$~ms after bounce. This
interval is chosen, (1) to exclude as much as possible prompt
postbounce convection, and, (2), because GWs from PNS core $g$-modes
begin to dominate the convective GW signal at times later than $\sim
250\;\mathrm{ms}$ in the VULCAN/2D simulations
(see section~\ref{section:gmodes}). For the two calculations, we find
comparable GW strain amplitudes $|h|$ in the range of
$1-4\times10^{-23}$ (at $10\;\mathrm{kpc}$). For the high-resolution
variant specifically, we find an emitted energy $E_\mathrm{GW}$ of
$\sim 7.0\times10^{-12}\;\mmsun c^2$ with a peak of
$dE_\mathrm{GW}/df$ at $\sim 350\;\mathrm{Hz}$ and most of the energy
being emitted within $\pm 150\;\mathrm{Hz}$ about this peak. By
scaling $E_\mathrm{GW}$ to the $1.2\;\mathrm{s}$ considered by
\cite{mueller:04}, we obtain $\sim 3.4\times 10^{-11}\;\mmsun c^2$
which is a factor of $\sim 5$ smaller than their value. 
The reason for this discrepancy is most likely the
higher average frequency of the GW emission in the model considered by
\cite{mueller:04}. Note, however, based on the inset plot shown in
\fref{fig:mueller2004} (figure 5 of \cite{mueller:04}), one can
estimate a dominant emission frequency around $\sim 300\;\mathrm{Hz}$
in the first $\sim 100\;\mathrm{ms}$ of PNS convection. Thus, the
higher-frequency emission must take place at later times. This is
consistent with the fact that the PNS becomes increasingly compact
with time, leading to convection at decreasing radii and generally on
smaller spatial scales. As a consequence, one may expect a secular
``chirp'' of the GW emission frequency from $\sim 300\;\mathrm{Hz}$ to
$\gtrsim 1000\;\mathrm{Hz}$ over $\sim 1\;\mathrm{s}$.

\subsection{Neutrino-driven Convection and SASI} 
\label{section:gwconvsasi}

Neutrino heating below the stalled shock peaks in the inner part of
the gain region\footnote{The gain region is defined as the region of
  postbounce space in which neutrino heating dominates over neutrino
  cooling.}  and decreases outward (see, e.g., figure~9 of
\cite{ott:08}). This establishes a negative radial entropy gradient
and makes the gain region unstable to convective overturn (e.g.,
\cite{buras:06b,jankamueller:96,jm:97} and references therein).
Convection develops within $30$--$50$~ms after bounce in the gain
region which extends at these times typically from $\sim
50$--$80\;\mathrm{km}$ out to almost the radius of the stalled shock
at $\sim 150$--$250\;\mathrm{km}$.

\begin{figure}[t]
\centering
\includegraphics[width=6.4cm]{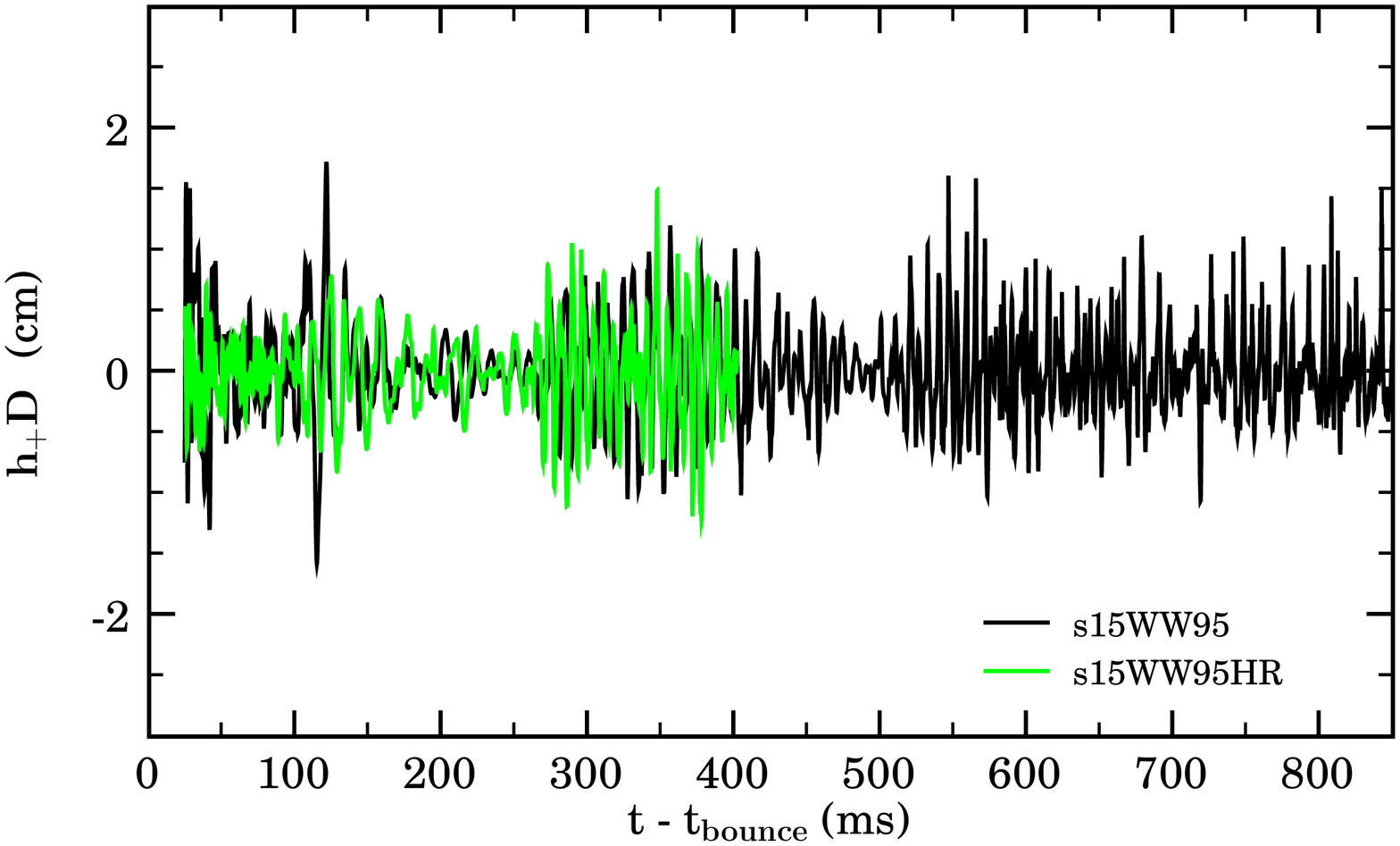}
\includegraphics[width=6.4cm]{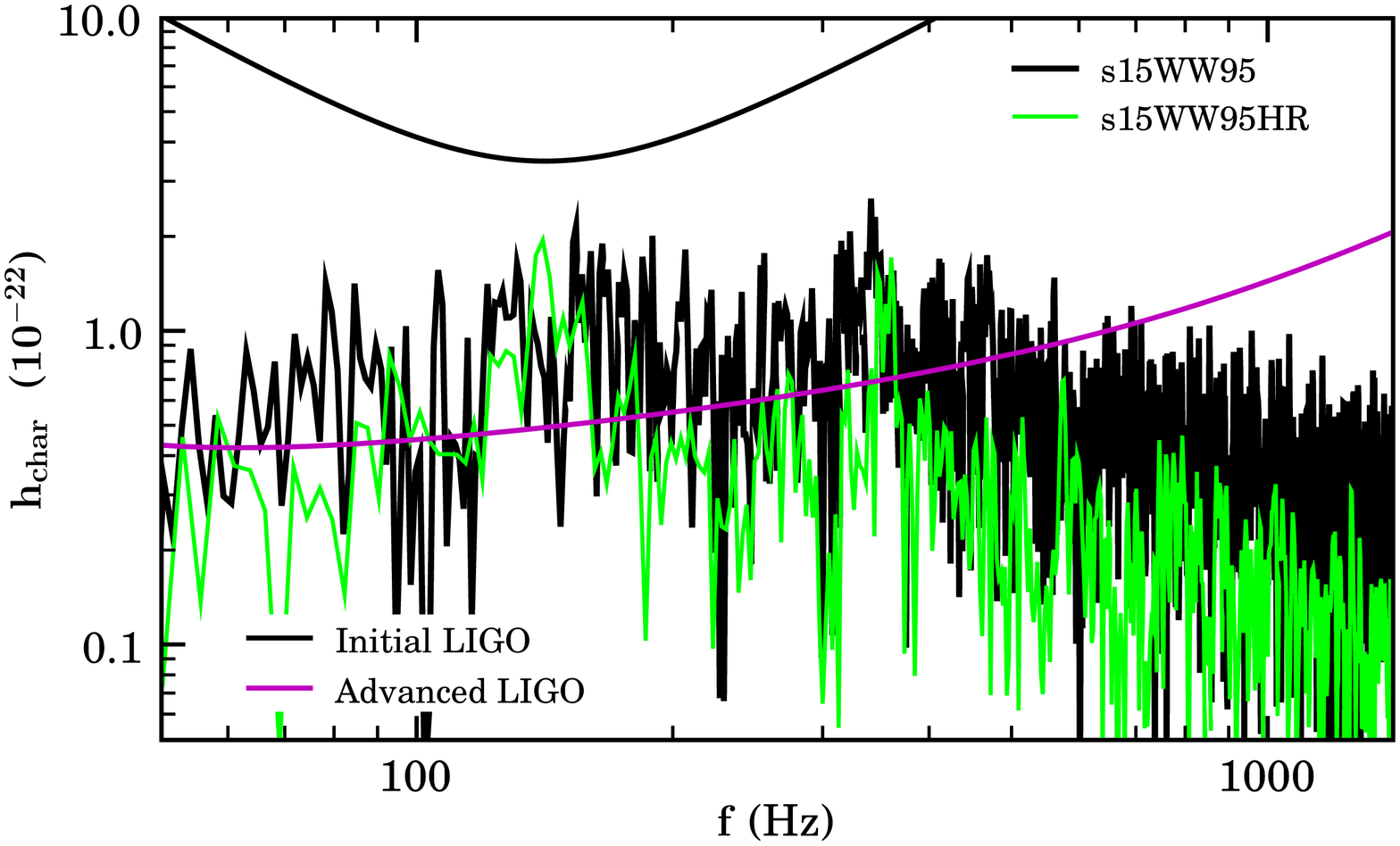}
\caption{{\bf Left:} GW signal ($h_+\,D$ in units of cm) due to
  postbounce neutrino-driven convection in the core of a nonrotating
  $15$-$\mmsun$ presupernova model of~\cite{ww:95}.  The first
  $20\;\mathrm{ms}$ of postbounce evolution are cut in order to
  exclude the initial burst due to prompt convection. The
  standard-resolution calculation is run to $\sim 850\;\mathrm{ms}$
  after bounce. The development of the SASI to non-linear amplitudes
  around $300\;\mathrm{ms}$ after bounce is reflected in
  increased amplitudes and more pronounced, higher-frequency
  variations.
In order to study the resolution
dependence of the convective GW signal, the high-resolution variant
is run to $\sim 400\;\mathrm{ms}$ after bounce. Because of the 
stochastic nature of turbulent convection, one cannot
expect pointwise convergence of the GW signal, but rather rough agreement
in signal amplitudes and characteristic frequencies. The latter is true
for s15WW95 and s15WW95HR.
 The waveforms are available for
download from \cite{ottcatalog}.
{\bf Right:} $h_{\mathrm{char}}(f)$ spectra (equation~\ref{eq:hchar})
 obtained from the s15WW95 and s15WW95HR waveforms for $D =
 10\;\mathrm{kpc}$ and compared to initial and burst-mode advanced
 LIGO $h_\mathrm{rms}$ noise curves~\cite{shoemaker:06}. Note the
 broadband character of the convective GW emission and the
 high-frequency tail that is emitted predominantly at late postbounce
 times and, hence, is particularly pronounced in the
 standard-resolution variant that is run to $\sim 850\;\mathrm{ms}$
 after bounce. s15WW95HR has overall smaller $h_\mathrm{char}$ simply
 because it is run to only $\sim 400\;\mathrm{ms}$ after bounce.
}
\label{fig:conv}
\end{figure}

The left panel of figure~\ref{fig:conv} shows the GW signal of
neutrino-driven convection and SASI in the VULCAN/2D postbounce
calculations s15WW95 and s15WW95HR. The former calculation is carried
out to $\sim 850\;\mathrm{ms}$ after bounce, while the computationally
more demanding higher-resolution latter is carried out to $\sim
400\;\mathrm{ms}$. In order to exclude PNS convection and PNS
$g$-modes, only regions outside a spherical radius of
$60\;\mathrm{km}$ are taken into account in computing the GW emission.
Since both convection and SASI occur in the same spatial domain and
since the SASI-related distortions of the postshock region dynamically
modify convection, the GW signals of these two hydrodynamic
instabilities cannot be separated cleanly in a SN simulation.
Also note that the GW emissions of PNS and neutrino-driven
convection/SASI generally interact and in linear theory (and certainly
in the Newtonian quadrupole approximation) superpose linearly.

In the calculations considered here, the SASI reaches the non-linear
regime around $\sim 300\;\mathrm{ms}$ and the shock and postshock flow
exhibit strong temporally varying distortions at small and large
scales (see, e.g., \cite{burrows:06}).  This is reflected by the GW
signal in the form of more rapid time variation and a more pronounced
modulation of the late-postbounce emission.  For simulation s15WW95,
we find maximum GW signal amplitudes around $6\times10^{-23}$ (at
$10\;\mathrm{kpc}$; typical amplitudes are about a factor of $2$
smaller), $E_\mathrm{GW} \approx 7.5\times10^{-12}\;\mmsun\,c^2$ (for
the entire $850\;\mathrm{ms}$).  The GW emission is rather broadband
with most of the energy emitted between $100$ and $500\;\mathrm{Hz}$
(though with a significant high-frequency tail which is emitted
predominantly at postbounce times greater than $\sim
400\;\mathrm{ms}$). This is also reflected in the characteristic GW
strain spectrum \cite{flanhughes:98},
\begin{equation}
h_\mathrm{char}(f) = \frac{1}{D}\sqrt{\frac{2}{\pi^{2}} 
  \frac{G}{c^{3}} \frac{dE_{\mathrm{GW}}}{df}}\,\,,
\label{eq:hchar}
\end{equation}
shown in the right panel of figure~\ref{fig:conv} and contrasted there
with initial and advanced LIGO rms noise levels.  Note that
$h_\mathrm{char}$ of the higher-resolution calculation is globally
lower due to the shorter period of postbounce time covered by this
calculation. Such differences are not limited to simulations: After
the onset of explosion, neutrino-driven convection is expected to
cease quickly (within $50$--$100\;\mathrm{ms}$ \cite{mueller:04}; note
that PNS convection continues).  Hence, a real-life SN that explodes
at earlier postbounce times will have a smaller integral and
frequency-differential $E_\mathrm{GW}$ (and $h_\mathrm{char}$) from
neutrino-driven convection/SASI than a counterpart that experiences a
later or no explosion at all.  The GW signals of s15WW95 and s15WW95HR
agree well in average amplitudes and temporal variations in the first
$\sim 400\;\mathrm{ms}$ shown in the left panel of
figure~\ref{fig:conv}. For the long-term evolution of a turbulent
system one generally cannot expect pointwise convergence for different
resolutions. However, agreement ``on average'' of the GW signals is to
be expected.

M\"uller~et~al.~\cite{mueller:04} analyzed the GW emission in 2D
simulations of a nonrotating $11.2$-$\mmsun$ and a slowly rotating
$15$-$\mmsun$ progenitor. They followed the postbounce evolution of
their models for $200$--$250\;\mathrm{ms}$, included an approximate GR
potential (see~\S\ref{section:snmodel}), Boltzmann neutrino transport
in a ray-by-ray fashion \cite{buras:06a}, a relatively soft variant of
the Lattimer-Swesty EOS \cite{lseos:91} and comparable resolution to
the aforementioned VULCAN/2D simulations.

 The average GW signal amplitudes and time variations found in the
VULCAN/2D simulations agree roughly with those of M\"uller~et~al.\ up
to $\sim 150\;\mathrm{ms}$ after bounce, but do not show the large
enhancement (by a factor of up to $\sim 20$ and with rapid time
variation around $500-800\;\mathrm{Hz}$) of the GW emission that
appear at later times in their simulations.  M\"uller~et~al.\
attributed this enhancement to the increased strength of convection,
but there may also be a connection to the SASI excursions that become
strong around these times in their models.  Owing largely to that
enhancement, M\"uller~et~al.\ found that most of the GW energy is
emitted in the interval of $500-800\;\mathrm{Hz}$ and published total
$E_\mathrm{GW}$ in the range of
$10^{-10}-\mathrm{few}\times10^{-9}\;\mmsun c^2$, roughly two orders
of magnitude larger than seen in the VULCAN/2D simulations.

The cause of the discrepancy between the M\"uller~et~al.\ simulations
and GW signals and those presented here is most likely related to differences
in the employed EOS, the neutrino transport scheme and resulting
neutrino luminosities and heating efficiencies, stalled shock radii,
and in the development of neutrino-driven convection and the SASI.
Furthermore, the enhancement of the GW signal may in part be due to
PNS convection or PNS pulsations, since they did not exclude the
entire PNS in their analysis.

Kotake~et~al.~\cite{kotake:07a} focussed on the SASI and employed a
simplified approach for modeling postshock flow and neutrino
transport. They assumed constant neutrino luminosities and starting
from analytic postbounce configurations with a fixed accretion rate of
$1\;\mmsun s^{-1}$ that appear initially stable or only marginally unstable
to convection, they studied the dependence of the SASI and
SASI/convective GW emission on the assumed neutrino luminosity. The
results of Kotake~et~al.\ confirm the expectation that higher neutrino
luminosities lead to a more pronounced SASI, more SASI-induced
turbulence in the postshock region, and greater GW signal amplitudes
and emitted energies.  For the lowest luminosities considered ($L_\nu
= 55\;\mathrm{B\, s^{-1}}$ each for $\nu_e$ and $\bar{\nu}_e$), they found
maximum amplitudes of $\sim 2.5\times10^{-22}$ and maximum
$h_\mathrm{char}$ of $\sim 10^{-21}$ (dependent, of course, on the
duration of the simulation) at $f_\mathrm{peak}$ of
$100$--$200\;\mathrm{Hz}$. The sum of the time-dependent $\nu_e$ and
$\bar{\nu}_e$ luminosities in the s15WW95 postbounce simulations with
VULCAN/2D varies from $85\;\mathrm{B\, s^{-1}}$ at $\sim 100\;\mathrm{ms}$ to
$50\;\mathrm{B\, s^{-1}}$ at $\gtrsim 400\;\mathrm{ms}$ after bounce. It is at
all postbounce times smaller than the $110\;\mathrm{B\, s^{-1}}$ assumed by
Kotake~et~al.  Considering the dependence of the SASI and convective
dynamics on the neutrino luminosity, it is not surprising that
Kotake~et~al.\ found significantly stronger GW emission than the
analysis of the VULCAN/2D simulations suggests.

\renewcommand\arraystretch{1.2}
\begin{table*}
\caption{Semi-quantitative summary of the GW emission by aspherical
fluid motions associated with convection and SASI. The numbers for
prompt convection are based on VULCAN/2D simulations presented here
and in \cite{ott:06prl} and on simulations carried out by
Marek~et~al.~\cite{marek:08b}. The GW emission characteristics of PNS
convection and neutrino-driven convection/SASI are based on the
results of \cite{mueller:04,ott:06prl,marek:08b} and the VULCAN/2D simulations
presented in this article. We provide estimates for the typical GW
strain at $10\;\mathrm{kpc}$, the typical emission frequency $f$,
the duration of the emission $\Delta t$, and the emitted energy
$E_\mathrm{GW}$ in GWs. In addition, we list processes and factors
that may limit the duration of the GW emission.
All numbers, including various upper and
lower limits, should be regarded as estimates that require
confirmation by future studies and that, at best, may guide
expectations.} {\footnotesize
\begin{tabular}{llllll}
\multicolumn{1}{c}{Process}
&\multicolumn{1}{c}{Typical $|h|$}
&\multicolumn{1}{c}{Typical $f$}
& \multicolumn{1}{c}{Duration $\Delta t$}
& \multicolumn{1}{c}{$E_\mathrm{GW}$}
& \multicolumn{1}{c}{Limiting Factors}\\
&\multicolumn{1}{c}{(at $10\;\mathrm{kpc}$)}
&\multicolumn{1}{c}{(Hz)}
&\multicolumn{1}{c}{(ms)}
&\multicolumn{1}{c}{($10^{-10} \mmsun c^2$)}
&\multicolumn{1}{c}{or Processes}\\
\hline
Prompt
&$10^{-23} - 10^{-21}$
&$50 - 1000$
&$0-\sim30$
&$\lesssim 0.01 - 10$
&Seed perturbations,  \\
Convection
&\multicolumn{4}{c}{(Emission characteristics depend on seed perturbations.)}
&entropy/lepton gradient,\\
&&&&&rotation\\
\hline
PNS 
&$2-5\times10^{-23}$
&$300-1500$
&$500- \mathrm{several}\, 1000$
&$\lesssim 1.3 (\frac{\Delta t}{1 s})$
&rotation, \\
Convection&&&&&BH formation, \\
&&&&&strong PNS $g$-modes\\
\hline
Neutrino-
&$10^{-23}-10^{-22}$
&$100-800$
&$100-\gtrsim1000$
&$\gtrsim 0.01 (\frac{\Delta t}{100 ms})$
&rotation, \\
driven&(peaks up
&&
&$\lesssim 15 (\frac{\Delta t}{100 ms})$
&explosion,\\
Convection&$\;\,$to $10^{-21}$)&&&&BH formation\\
and SASI\\
\hline
\end{tabular}}
\label{table:conv}
\end{table*}
\renewcommand\arraystretch{1.0}

Marek~et~al.~\cite{marek:08b} extended the previously
discussed work of M\"uller~et~al.~\cite{mueller:04} and considered for
a nonrotating $15$-$\mmsun$ model variations of the convective/SASI
dynamics and the associated GW emission with the stiffness of the
nuclear EOS.  They followed their calculations to $400\;\mathrm{ms}$
after bounce and analyzed in detail the GW signal due to matter
dynamics and anisotropic neutrino radiation fields (for a summary of the
latter, see \sref{section:nugw_conv}). The qualitative features of the
matter GW signals their models are qualitatively
fairly similar to those obtained with VULCAN/2D (shown in
figure~\ref{fig:conv}). Marek~et~al.\ found that the GW signal
amplitudes and typical emission frequencies increase with postbounce
time and with the growing strength of the SASI-driven shock
distortions. Their models produce peak amplitudes around
$5\times10^{-22}$ (at $10\;\mathrm{kpc}$) and average amplitudes about
a factor of $2$ smaller.  Furthermore, Marek~et~al.\ reported that the
SASI-moderated aspherical convective motions in the
low-density postshock gain layer lead to GW emission at frequencies up
to $\sim 200\;\mathrm{Hz}$. Those GWs, however, make up only the
subdominant, low-frequency part of the total emission, which in their
models is dominated by the GWs emitted from deeper, higher-density
regions where aspherical accretion downstreams are decelerated and
perturb the PNS core.

Comparing a model run with a soft variant of the Lattimer-Swesty
EOS~\cite{lseos:91} with a counterpart model run with the stiff Wolff
EOS~\cite{hillewolff:85}, Marek~et~al.\ found that the Lattimer-Swesty
EOS model yields higher-amplitude (by up to a factor of 2),
higher-frequency (spectral peak at $600-800\;\mathrm{Hz}$
vs. $300-600\;\mathrm{Hz}$) GW emission.  These differences are a
consequence of the more compact PNS and resulting harder neutrino
spectra and increased neutrino luminosities which, in turn, result in
increased neutrino heating and more vigorous convection/SASI.  This
quantitative result is of particular importance, since it demonstrates
that -- despite the stochastic, untemplateable nature of the
convective/SASI GW signal -- key microphysical aspects can still be
constrained via the observation of GWs emitted by convection/SASI.

To conclude this section, we point out that in consideration of the
still small set of mostly axisymmetric studies that have addressed GW
emission from neutrino-driven convection/SASI, the currently available
GW signal estimates should be regarded only as examples that may guide
expectations. They are not yet robust and many more simulations will
be needed to systematically (qualitatively and quantitatively) study
the range of possible postbounce evolutions and corresponding
incarnations of convective overturn. In particular, it will be
important to quantify the dependence on progenitor structure and
explore the dynamics of convection and SASI in 3D.

Table~\ref{table:conv} provides an overview of the GW emission from
prompt, PNS, and neutrino-driven convection / SASI. We provide rough
values for typical $|h|$, typical emission frequencies, emission
durations, emitted energies, and list factors and processes that may
inhibit the development or shorten the duration of the individual
emission processes.

\section{Non-radial PNS Pulsations}
\label{section:puls}
PNSs and cold NSs can pulsate in a multitude of ways (e.g.,
\cite{stergioulas:03,andersson:03}) and their non-radial pulsation
modes of quadrupole and higher order emit GWs.  Polar (even-parity)
fluid modes are grouped into the lower-frequency $g$-modes (buoyancy
modes, restoring force gravity, require compositional or thermal
gradients) and the higher-frequency $p$-modes (restoring force
pressure), separated in frequency by the $f$-mode (fundamental mode
without nodes inside the star). Axial (odd-parity) fluid modes are
modes whose restoring force is the Coriolis force, hence they require
rotation and are degenerate at zero frequency in the nonrotating
limit. Modes of mixed axial and polar character, so-called hybrid
modes, exist as well~(e.g., \cite{lockitch:99,stergioulas:03}).
$r$-modes (Rossby waves), a particular group of axial
inertial modes,  are strongly unstable to the GW-back-reaction
Chandrasekhar-Friedman-Schutz (CFS) instability
\cite{chandrasekhar:70, friedman:78} at all rotation rates
\cite{andersson:98,andersson:03}.  In PNSs, $r$-modes have growth
times of the order of seconds \cite{lindblom:98} and appear to
saturate at low amplitudes~\cite{arras:03}. Hence, they are unlikely
to be relevant sources of GWs in the first second of a PNS's life.

In addition to fluid modes there exist polar and axial modes of 
spacetime curvature, low-frequency $s$-modes and high-frequency
$w$-modes, that are weakly coupled to matter (e.g.,
\cite{kokkotas:99,stergioulas:03} and references therein).

Studies of non-radial pulsations of cold NSs as emission processes of
GWs have a long history (see, e.g., \cite{andersson:03, kokkotas:99,
  dimmelmeier:06, bernuzzi:08} and references therein). In the
core-collapse SN context, non-radial PNS oscillation modes and damping
by GW emission and other processes were first considered by
McDermott~et~al.~\cite{mcdermott:83,mcdermott:88},
Finn~\cite{finn:87}, and Reisenegger and
Goldreich~\cite{reisenegger:94}.

Recently, in a series of papers, Ferrari and collaborators
\cite{ferrari:03,ferrari:03b,ferrari:04,ferrari:07}
studied in GR perturbation theory the PNS non-radial mode spectrum and
the GW emission from PNS pulsations on the basis of detailed
spherically-symmetric PNS cooling models. They considered mode
amplification by the CFS instability in rotating PNSs, but did not
study the excitation of non-radial modes in the nonrotating limit.
Assuming the presence of pulsational energy, they provided estimates
in \cite{ferrari:03} for the typical $|h|$ and $E_\mathrm{GW}$
necessary for a particular mode to be detectable by current and future
GW observatories. According to their estimates for a nonrotating or
very slowly rotating PNS, the most efficiently radiating mode, the
lowest-order quadrupole $g$-mode (frequency $f \approx
500-600\;\mathrm{Hz}$), would need to emit GWs at a typical $|h|$ of
$\sim 8 \times 10^{-22}$ (at $10\;\mathrm{kpc}$) and emit a total
$E_\mathrm{GW}$ of $\sim 4\times10^{-8}\,\mmsun\, c^2$ to be
detectable by initial-LIGO-class detectors at a distance of
$10\,\mathrm{kpc}$. In the case of more rapid rotation, the 
$g$-mode frequencies are lower and closer to the maximum detector
sensitivity, leading to lower required $|h|$ and $E_\mathrm{GW}$.

Passamonti~et~al.~\cite{passamonti:05,passamonti:07} explored the
excitation of higher-order non-radial polar modes by non-linear
couplings of radial and non-radial eigenmodes that are excited in the
early postbounce ring-down phase of any core-collapse
event\footnote{In a related work,
  Passamonti~et~al.~\cite{passamonti:06} also studied the coupling of
  radial modes with non-radial axial modes.}.  Such couplings were
first suggested by
Eardley~\cite{eardley:83}. Passamonti~et~al.\ employed a GR
perturbative framework and approximated the PNS by a simple polytropic
stellar structure (which excludes $g$-modes). Their results indicate
that the radial and non-radial fundamental oscillation modes present
in the very early postbounce phase will couple and give rise to a rich
spectrum of daughter modes\footnote{Such radial--non-radial
mode couplings have recently been observed also in numerical models
of phase-transition-induced collapse of NSs~\cite{abdikamalov:08}.}. 
The strongest of these non-linear modes
could be detectable by advanced LIGO-class detectors from a galactic
event and their spectral information would provide additional
insight into the high-density nuclear EOS and the internal structure
of a newborn PNS.

The pulsational mode structure of rapidly rotating NS/PNSs cannot be
determined reliably by perturbative methods and GR hydrodynamic
simulations must be carried out to obtain the spatial and spectral
mode characteristics. Dimmelmeier~et~al.~\cite{dimmelmeier:06}
recently carried out the first extensive study of the axisymmetric
mode structure of rapidly rotating polytropic NS models in
CFC GR~\cite{isenberg:08}, going beyond the Cowling
approximation employed by most previous studies
(e.g.,~\cite{stergioulas:04} and references therein). The mode
structure of rapidly rotating finite-temperature PNSs remains to be
studied.

\subsection{Protoneutron Star $g$-Mode Pulsations in the Context of
the Acoustic Mechanism for Core-Collapse Supernova Explosions}
\label{section:gmodes}

Burrows et~al.~\cite{burrows:06, burrows:07a}, in their simulations
that lead to the proposition of the \emph{acoustic mechanism} for
core-collapse SN explosions, have discovered the excitation of PNS
core $g$-modes by turbulence and by accretion downstreams through the
unstable and highly-deformed stalled shock that experiences the
SASI. The pulsations damp by the emission of strong sound waves and do
not ebb until accretion subsides. As discussed in
\cite{burrows:06,burrows:07a}, these PNS core $g$-mode pulsations
reach nonlinear amplitudes and act as transducers for the conversion
of accretion gravitational energy into acoustic power that is
deposited in the postshock region and may be sufficient to drive an
explosion. Although promising, this acoustic SN mechanism remains
controversial (see, e.g., \cite{yoshida:07,weinberg:08,marek:07}) and
has yet to be confirmed by other groups. In particular,
Weinberg~et~al.~\cite{weinberg:08} pointed out that the saturation
amplitude of the PNS $g$-modes may be limited by non-linear model
coupling via a parametric instability that cannot be resolved in
present simulations.  In addition, there are indications
\cite{burrows:07b,ott:08} that rapid rotation significantly weakens
convection and modifies the SASI and thus may weaken the driving
mechanism for the PNS core $g$-modes.

\begin{figure*}
\centering
\includegraphics[width=7.0cm]{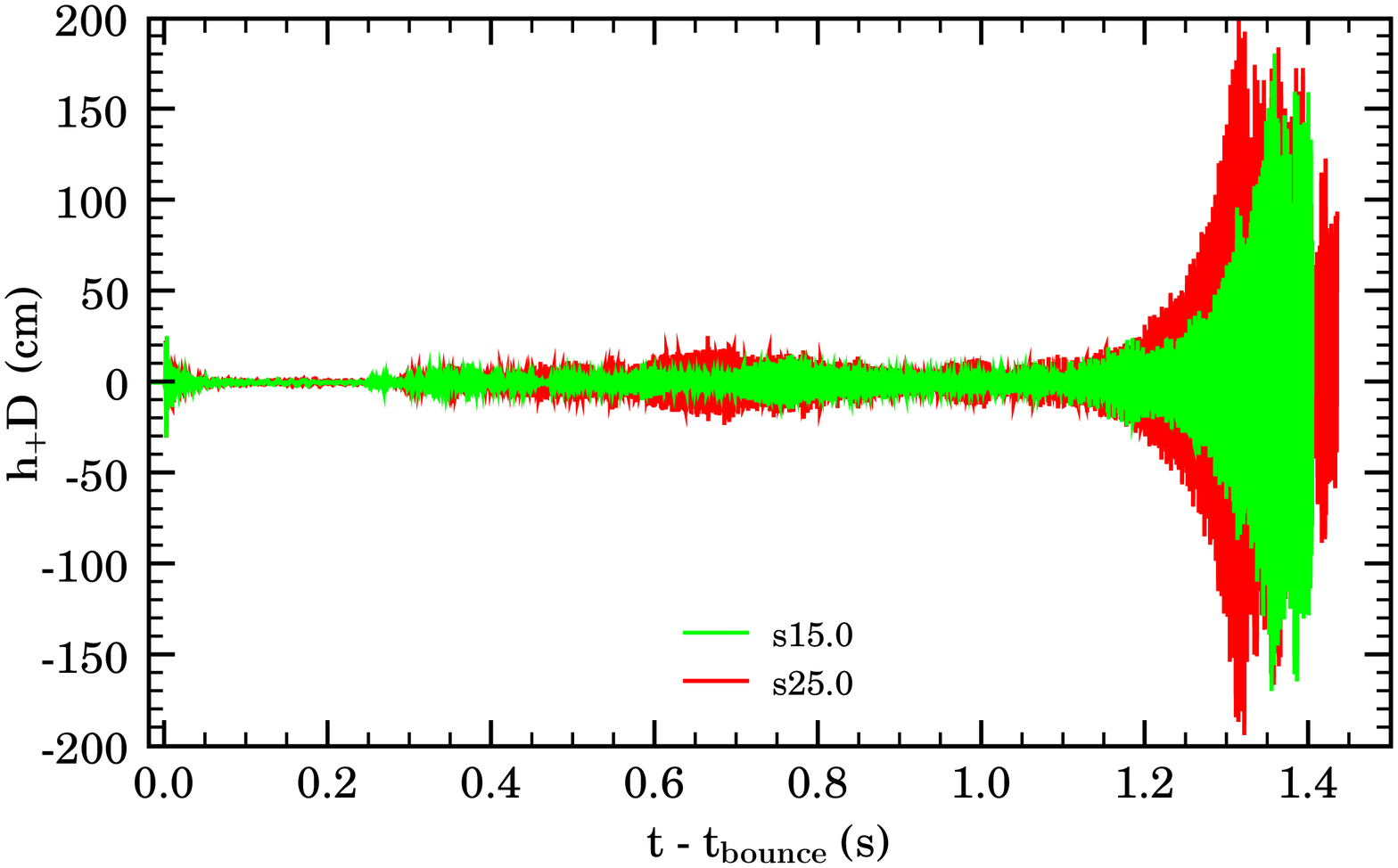}\includegraphics[width=7.0cm]{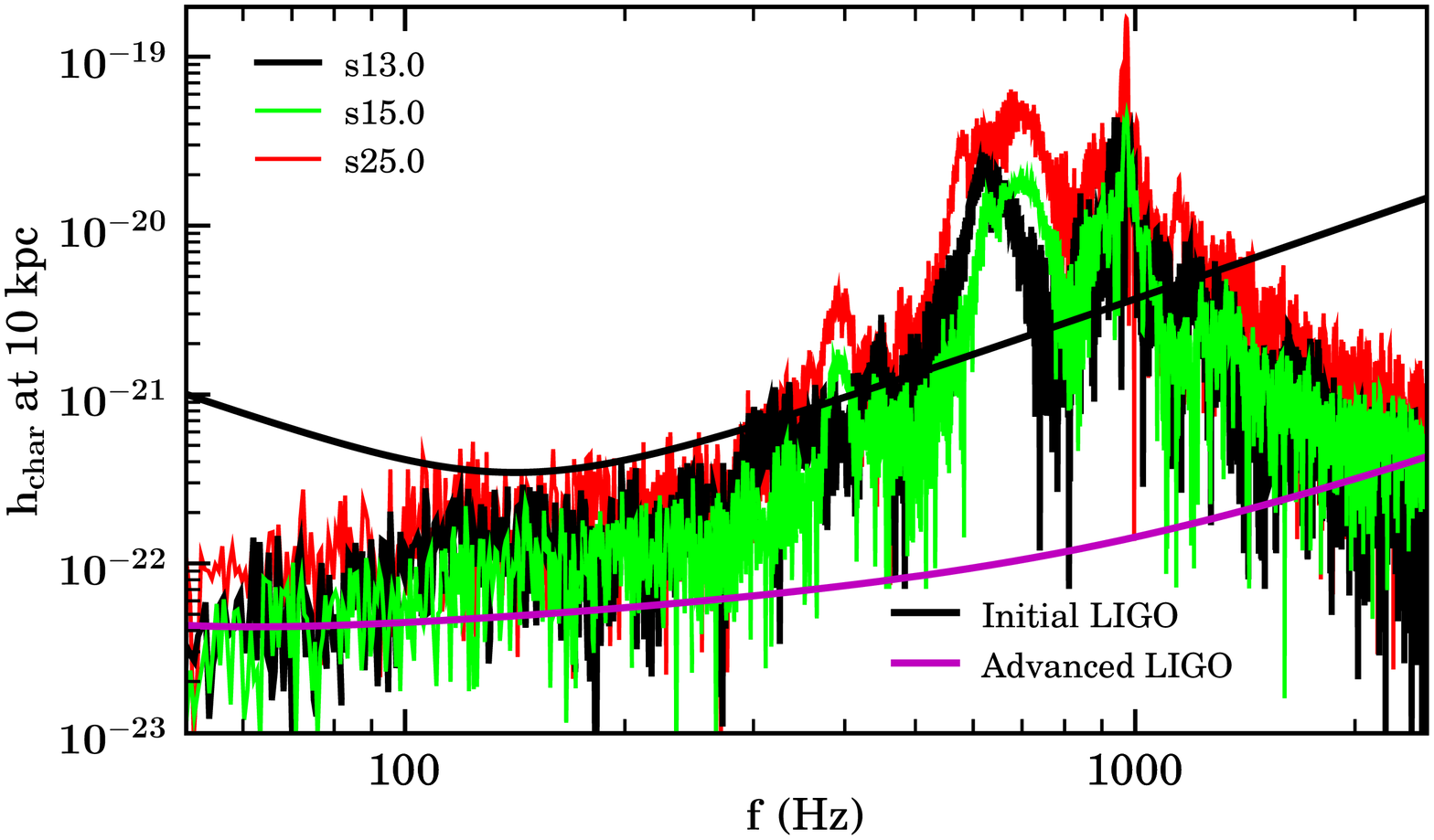}
\caption{GW emission in a set of representative models from
  \cite{burrows:07a}. {\bf Left}: GW wave signal ($h_+ D$ in units of
  cm) as a function of postbounce time in the nonrotating models s25.0
  (red) and s15.0 (green). The burst shortly after core bounce is due
  to prompt convection (see section~\ref{section:convsasi}).  Around
  $300$--$400$~ms, the GW emission becomes dominated by the growing
  $g$-mode amplitudes, but stays at relatively small amplitudes
  initially.  The core pulsations become more vigorous at later times,
  leading to the emission of large-amplitude GWs. Explosion sets in
  around $\sim 1.0$--$1.4\;\mathrm{s}$ after bounce and in some cases
  is unipolar (see \cite{burrows:07a,burrows:06}) -- the star explodes
  on one side while accretion continues on the other. All waveforms
  are available from \cite{ottcatalog}. {\bf Right}: Characteristic GW
  strain spectra (equation~\ref{eq:hchar} and \cite{flanhughes:98}) of
  various models of \cite{burrows:07a} contrasted with the initial and
  advanced LIGO rms noise curves~\cite{shoemaker:06} for a source
  distance of $10\;\mathrm{kpc}$. Note that the spectra of all models
  exhibit a narrow primary peak near $950\;\mathrm{Hz}$ and a broader
  secondary peak around $600$--$700\;\mathrm{Hz}$.  }
\label{fig:gw_gmodes}
\end{figure*}

Burrows~et~al.\ found that the fundamental $\ell = 1$ $g$-mode of the
PNS core is most easily and first excited in their models.
Higher-order eigenmodes and, through non-linear couplings, harmonics
of eigenmodes and other modes with complicated spatial structures
appear at later times.  Ott~et~al.~\cite{ott:06prl} analyzed the GW
emission from the quadrupole components of the PNS core pulsations in
the original nonrotating $11$-$\mmsun$ model presented in
\cite{burrows:06} and, in addition, for a slowly-rotating
$15$-$\mmsun$ model and a nonrotating rather extreme $25$-$\mmsun$
model with a very extended and massive iron core.  For these models
they found maximum GW amplitudes $|h_\mathrm{max}|$ of $\sim 1.3
\times 10^{-21}$ (at $10\;\mathrm{kpc}$) for their $11$-$\mmsun$ and
$15$-$\mmsun$ models and $\sim 5\times10^{-20}$ (at
$10\;\mathrm{kpc}$) for the $25$-$\mmsun$ model. The total emitted
$E_\mathrm{GW}$s were around $1.5\times10^{-8}\;\mmsun\,c^2$ in the
$11$-$\mmsun$ and $15$-$\mmsun$ model, while the $25$-$\mmsun$ model
emitted an amazing $8\times10^{-5}\;\mmsun\, c^2$. GW back-reaction
effects, though neglected in \cite{ott:06prl}, are probably relevant
for this latter model~\cite{ott:06phd}.  Most of the emission in these
models took place in frequency space at $\sim650\pm100\;\mathrm{Hz}$
and $\sim 900\pm100\;\mathrm{Hz}$.

\setcounter{footnote}{1}
In \cite{burrows:07a}, Burrows~et~al.\ considered an extended set of
nonrotating progenitors, covering the mass range from $11.2\,\mmsun$
to $25.0\,\mmsun$ solar masses, and using presupernova models of
\cite{whw:02,nomoto:88}\footnote{Note that the precollapse models used
in \cite{burrows:06,ott:06prl} were taken from \cite{ww:95}
and \cite{heger:05}. Model s25.0 of \cite{burrows:07a} should not
be confused with model s25WW of \cite{ott:06prl}.}. In the following,
we present their previously unpublished GW signals and summarize the
key signal characteristics in table~\ref{table:gmodes}. The waveforms
and $E_\mathrm{GW}$ spectra of all models can be obtained
from~\cite{ottcatalog}.

\begin{table*}[t]
\begin{footnotesize}
\begin{center}
\caption{\small GW data summary for models calculated by
  Burrows~et~al.~\cite{burrows:07a} whose GW signals have not
  previously been published. In addition, the GW emission
  characteristics of the models of \cite{burrows:06,ott:06prl} are
  listed for completeness.  $\Delta t$ is the amount of postbounce
  time covered in each model, $|h_{+,\mathrm{max}}|$ is the overall
  maximum GW amplitude scaled to 10 kpc distance,
  $h_\mathrm{char,max}$ is the global maximum of the characteristic
  strain spectrum \cite{flanhughes:98}, $f_\mathrm{peak}$ is the
  frequency of the global maximum and $\delta f$ is the FWHM.  Data
  for the secondary pronounced peak in $h_\mathrm{char}(f)$ are given
  and all related quantities have the subscript $2$. $E_\mathrm{GW}$
  is the total energy radiated in GWs. Note that the values given for
  $|h_\mathrm{max}|$ in \cite{ott:06prl} are slightly different from
  those presented here for models of \cite{ott:06prl}. This is due to
  a small error in the analysis routines used by
  \cite{ott:06prl}. Also note that model s25WW of \cite{ott:06prl}
  uses a rather extreme $25$-$\mmsun$ precollapse model of
  \cite{ww:95} with a very extended and massive iron core and a very
  shallow density gradient.}
\label{table:gmodes}
\vspace{0.1cm}
\begin{tabular*}{14.0cm}{lr|c|ccc|ccc|r@{~~~~~~}}
\hline
\hline
&&&\multicolumn{3}{|c|}{Primary $h_\mathrm{char}$ peak}
&\multicolumn{3}{|c|}{Secondary $h_\mathrm{char}$ peak}
&\\
Model & \multicolumn{1}{c|}{$\Delta t$}&$|h_{+,\mathrm{max}}|$ 
                   & $h_{\mathrm{char,max}}$ 
                   & $f_\mathrm{peak}$ 
&$\delta f$
&$h_{\mathrm{char,max 2}}$
&$f_\mathrm{peak 2}$
&$\delta f_2$
&\multicolumn{1}{c}{$E_\mathrm{GW}$}\\

&(ms)&(10$^{-21}$ at 
&(10$^{-21}$ at 
&(Hz)
&(Hz)
&(10$^{-21}$ at
&(Hz)
&(Hz)
&\multicolumn{1}{c}{(10$^{-7}$}\\
&&10~kpc)&10~kpc)&&&10~kpc)&&&\multicolumn{1}{r}{$\mmsun\,c^2$)} \\
\hline
\multicolumn{2}{l|}{\color{blue}\bf Models of \cite{burrows:07a}\color{black}}&&&&&&&\\
s11.2&1496&\phantom{0}1.26&\phantom{00}40.3&910&$\sim$100&\phantom{00}5.5&605&$\sim$100&0.60\\
s13.0&1447&\phantom{0}4.00&\phantom{00}44.3&934&\phantom{0}$\sim$75&\phantom{0}27.4&613&\phantom{0}$\sim$90&1.03\\
s15.0&1404&\phantom{0}3.75&\phantom{00}45.8&970&\phantom{0}$\sim$35&\phantom{0}22.0&690&$\sim$130&4.30\\
s20.0&1715&\phantom{0}3.61&\phantom{00}61.6&992&\phantom{0}$\sim$20&\phantom{0}33.4&630&$\sim$200&2.36\\
s25.0&1434&\phantom{0}6.93&\phantom{00}70.9&969&\phantom{0}$\sim$10&\phantom{0}64.7&672&$\sim$200&7.25\\
nomoto13&1237&\phantom{0}0.77&\phantom{00}22.1&907&\phantom{0}$\sim$40&\phantom{0}41.8&602&\phantom{0}$\sim$80&0.12\\
nomoto15&1725&\phantom{0}1.04&\phantom{00}43.7&997&$\sim$100&\phantom{0}44.5&604&$\sim$100&0.70\\
\hline
\hline
\multicolumn{2}{l|}{\color{blue}\bf Models of \cite{ott:06prl}\color{black}}&&&&&&&\\
s11WW&1045&\phantom{0}1.58&\phantom{00}22.8&654&\phantom{0}$\sim$50&\phantom{00}8.5&895&$\sim$40&0.16\\
m15b6&\phantom{0}927&\phantom{0}0.98&\phantom{00}19.3&660&\phantom{0}$\sim$20&\phantom{00}7.9&822&$\sim$30&0.14\\
s25WW&1110&49.91&2514.3&937&\phantom{0}$\sim$10&707.2&790&$\sim$20&824.28\\
\hline
\end{tabular*}
\end{center}
\end{footnotesize}
\end{table*}

The left panel of \fref{fig:gw_gmodes} depicts the GW signal
obtained from models s15.0 ($15\;\mmsun$) and s25.0 ($25\;\mmsun$) as
a function of postbounce time. There is little variation in the
qualitative features of the GW signal throughout the model set
considered by \cite{burrows:07a}. Hence, models s15.0 and s25.0 are
quite representative picks. Core bounce in these
models is marked by a GW burst that is due to prompt convective overturn
setting in very shortly after bounce 
(see section~\ref{section:promptconv}). Typical peak GW amplitudes are around
$\sim 1\times10^{-21}$ at $10\;\mathrm{kpc}$ and most of the energy
is emitted in a frequency interval of $500$--$800\;\mathrm{Hz}$.  The
prompt convection subsides quickly, and during the interval from $\sim
20\;\mathrm{ms}$ to $\sim 300\;\mathrm{ms}$ the GW signal is dominated
by convection in the PNS and in the postshock region.  The PNS
core pulsations reach significant amplitudes at
$300$--$400\;\mathrm{ms}$ after bounce and the GW emission from their
quadrupole components starts to dominate the overall GW signal. From
$\sim 300\;\mathrm{ms}$ to $\sim1.1\;\mathrm{s}$ the models emit
quasi-continuous GWs with maximum amplitudes in the range of $\sim
3\times10^{-22}$--$8\times10^{-22}$ at $10\;\mathrm{kpc}$, modulated
on a timescale of $\sim 100\;\mathrm{ms}$ by variations in the SASI
and accretion-dependent excitation (and damping) of the core
pulsations. At postbounce times greater than $\sim 1.1\;\mathrm{s}$, 
the quadrupole components of the core pulsations reach
very large amplitudes (possibly through the excitation of an $\ell =
2$ eigenmode \cite{ott:06prl}), leading to a significant increase in
the GW amplitudes up to $\sim 7\times10^{-21}$ ($\sim
4\times10^{-21}$) at $10\;\mathrm{kpc}$ in model s25.0 (s15.0). The
other models studied by Burrows~et~al.~\cite{burrows:07a} reach comparable
$|h_\mathrm{max}|$ (see table~\ref{table:gmodes}).

All models of Burrows~et~al.~\cite{burrows:07a} explode between $\sim
1.0\;\mathrm{s}$ and $1.4\;\mathrm{s}$ after bounce and the
simulations were stopped when the shock reached the outer boundary of
the computational grid. In models that exploded globally, the PNS core
pulsations and the corresponding GW signal generally subsided after
the onset of explosion.  However, in a few models, the explosion was
unipolar, setting in along one of the poles while accretion continued
on the opposite side, sustaining to some extent the core pulsations
and prolonging the GW emission.

In the right panel of \fref{fig:gw_gmodes}, characteristic strain
spectra $h_\mathrm{char}(f)$ (equation~\ref{eq:hchar}; for a source
distance of $10\;\mathrm{kpc}$) are shown for models s13.0, s15.0, and
s25.0.  More massive progenitors generally tend to have higher
accretion rates\footnote{Note that the presupernova models of
\cite{whw:02} behave non-monotonically in the scaling of (iron) core
mass and density profile in the mass range from $\sim 15$ to $\sim
20\;\mmsun$.  See also \cite{burrows:07a,dimmelmeier:08}.}, tend to
explode later, reach higher $|h_\mathrm{max}|$, have higher total
$E_\mathrm{GW}$, and more narrow peaks in $dE_\mathrm{GW}/df$ (see
table~\ref{table:gmodes}).  As a consequence, the maximum values of
$h_\mathrm{char}$ generally increase with progenitor mass.

\begin{figure}
\centering
\includegraphics[width=8.7cm]{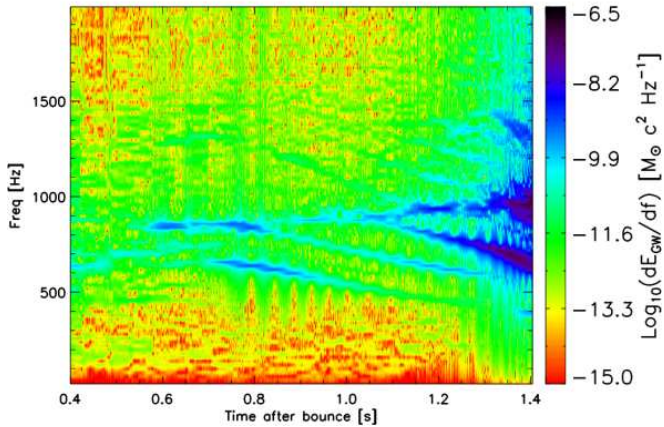}
\caption{Frequency-time evolution of the GW energy
spectrum $dE_\mathrm{GW}/df$ in model s15.0 of \cite{burrows:07a},
computed with a $50\;\mathrm{ms}$ sampling interval and a step width
of $1\;\mathrm{ms}$. Shown is the interval from $0.4\;\mathrm{s}$
after core bounce to the end of the simulation. Given the complex
thermodynamic structure of a hot PNS, the mode spectrum reflected in
$dE_\mathrm{GW}/df$ is rather complicated and exhibits significant
temporal variations in mode frequencies and preferred modes.  Note
that the two strong emission regions around $900$--$1000\;\mathrm{Hz}$
and $600$--$800\;\mathrm{Hz}$ correspond to the the primary and
secondary peak in $h_\mathrm{char}(f)$ shown in the right panel of
\fref{fig:gw_gmodes}.}
\label{fig:timefreq}
\end{figure}

It is interesting to note that $h_\mathrm{char}$ exhibits two
pronounced peaks in all considered models. The primary, quite narrow
peak is centered in frequency around $900$--$1000\;\mathrm{Hz}$ while 
the secondary, smaller and broader peak is located in the
frequency range $500$--$700\;\mathrm{Hz}$. This is due to the complicated
time-dependent mode structure present in the pulsating PNS core and is
also highlighted by \fref{fig:timefreq} which provides a
frequency-time analysis of the GW energy spectrum $dE_\mathrm{GW}/df$
of model s15.0. Due to changes in the hydrodynamic and thermodynamic
structure of the PNS, the $g$-mode frequencies vary (as do the
frequencies of other mode families) with time \cite{ferrari:03}. 
In addition, higher-order harmonics and modes
with complicated spatial structure are excited and their
quadrupole parts emit GWs. At early to intermediate postbounce times,
the emission occurs mainly from quadrupole components with frequencies
around $\sim 600\;\mathrm{Hz}$ and $\sim 850\;\mathrm{Hz}$. The
frequencies initially show a weak upward trend, but decrease beginning
at around $\sim700\;\mathrm{ms}$. At late times, most of the energy is
being emitted around $\sim 850\;\mathrm{Hz}$ and $\sim
600\;\mathrm{Hz}$--$700\;\mathrm{Hz}$, but this time by quadrupole
components of modes who have descended from higher frequencies.

The PNS core $g$-mode pulsations may arguably be so far
the strongest proposed emission process for GWs in core-collapse SNe.
Observing GWs due to PNS $g$-modes and capturing the time-evolution of
the GW spectrum would provide us with invaluable information that
could be used, via comparison with model calculations, to reconstruct the
hydrodynamic and thermodynamic structure and evolution of a very young
PNS. In addition, and since the acoustic mechanism for SN explosions
and the strong PNS core $g$-modes are almost invariably linked, the
observation or the non-observation of the $g$-mode GW signal from a
nearby core-collapse SN may be used to test the acoustic SN 
mechanism (see also \sref{section:conclusions}).

Nevertheless, it is important to keep in mind that the viability of
the acoustic SN mechanism is still unclear and that the dependence on
dimensionality (2D vs.\ 3D), rotation, equation of state, description
of gravity (Newtonian vs.\ GR) and the possibility of mode saturation
at low amplitudes~\cite{weinberg:08} remain to be explored in detail.

\section{Anisotropic Neutrino Emission}
\label{section:neutrinos}

Not only aspherical fluid motion, but in more general terms, any
accelerated transport of energy with a non-zero quadrupole and/or
higher-order component emits GWs. For the case of
anisotropic radiation of neutrinos from a distant point source, this
has been first realized in linear theory via the direct solution of
the inhomogeneous wave equation by Epstein \cite{epstein:78} (but, see
also Turner's independent derivation in the zero-frequency
limit~\cite{turner:78}). 

Burrows \& Hayes \cite{bh:96} and M\"uller \& Janka \cite{jm:97} were
the first authors to implement the formalism. It has since been
employed in a number of other studies
\cite{fryer:04,mueller:04,kotake:06a,kotake:07a,ott:06prl,dessart:06aic,
marek:08b}.
In the following, we present a short overview on the formalism as used
by \cite{bh:96,jm:97} in axisymmetry. More details and the
generalization to 3D can be found in \cite{jm:97,kotake:06a}.

For axial symmetry,~\cite{jm:97,bh:96} write the
dimensionless GW strain for an observer positioned in the
equatorial plane as
\begin{equation}
h^{TT}_{+,\mathrm{eq}}(t) = \frac{2 G}{c^4 D} \int_{-\infty}^{t-D/c}
\alpha(t')L_\nu(t') dt'\,\,,
\label{eq:vulcan_nugw}
\end{equation}
where $L_\nu(t)$ is the total neutrino luminosity and
$\alpha(t)$ is the instantaneous neutrino radiation
anisotropy that includes the transverse-traceless
projections~\cite{jm:97}. It is defined as
\begin{equation}
\alpha(t) = \frac{1}{L_\nu(t)} \int_{4\pi} \Psi(\vartheta',\varphi')
\frac{dL_\nu(\vec{\Omega}',t)}{d\Omega'} d\Omega'\,\,,
\label{eq:vulcan_alpha}
\end{equation}
where $dL_\nu(\vec{\Omega},t) / d\Omega$ is the energy radiated
at time $t$ per unit of time and per unit of solid angle
into direction $\vec{\Omega}$ with
\begin{equation}
L_\nu(t) = \int_{4\pi} \frac{dL_\nu(\vec{\Omega}',t)}{d\Omega'} d\Omega'\,\,.
\end{equation}
$\Psi(\vartheta,\varphi)$ represents the angle dependent factors in
terms of source coordinate system angles $\vartheta$ and $\varphi$ and
depends on the particular GW polarization and the
observer position relative to the source. In axisymmetry,
$h^{TT}_\times = 0$ everywhere, and $h^{TT}_+ = 0$ along the axis of
symmetry. For an observer located in the equatorial plane, observing
the $+$ GW polarization, $\Psi(\vartheta,\varphi)$ is
given~\cite{jm:97,kotake:06a} by
\begin{equation}
\Psi(\vartheta,\varphi) = (1+\sin\vartheta \cos\varphi)
\frac{\cos^2\vartheta - \sin^2\vartheta \sin^2\varphi}
{\cos^2\vartheta + \sin^2\vartheta\sin^2\varphi}\,\,.
\label{eq:vulcan_psi}
\end{equation}
In axisymmetry, there is no $\varphi$ dependence of the
luminosity. By integrating $\Psi(\vartheta,\varphi)$ over
$\varphi$, equations \ref{eq:vulcan_alpha} and
\ref{eq:vulcan_psi} combine to \cite{kotake:06a,kotake:07a},
\begin{equation}
\alpha(t) = \frac{2\pi}{L_\nu(t)} \int_{0}^\pi
\sin\theta'(-1 + 2|\cos\theta'|) 
\frac{dL_\nu(\theta',t)}{d\theta'} d\theta'\,\,.
\label{eq:nu_gw_simple}
\end{equation}

Note that the GW signal due to neutrinos observed at time $t+D/c$
contains contributions from anisotropies in the neutrino radiation
field at all times prior $t$.  This leads to a memory effect in the GW
signal, leaving behind a constant (``DC'') offset after the
anisotropic neutrino emission subsides.  Largely-aspherical mass
ejection can lead to a similar GW memory 
(see section~\ref{section:others}).  The implications and detectability of
such GW bursts with memory were discussed in~\cite{braginskii:87,thorne:87}.

In the context of massive star collapse and core-collapse SNe,
anisotropic neutrino radiation and the associated emission of GWs may
arise (a) from rotationally-deformed PNSs
\cite{mueller:04,kotake:06a,dessart:06aic,ott:06phd}, (b) from
convective overturn and SASI \cite{jm:97,mueller:04,
ott:06prl,ott:06phd,kotake:07a,marek:08b}, and (c) from global asymmetries in
the (precollapse) matter distribution \cite{bh:96,fryer:04}.

The extraction of GWs due to neutrino radiation field anisotropies
requires that the underlying simulation was carried out with some form
of neutrino transport. Unfortunately, the set of presently employed
approaches to neutrino transport generally yield different results
for the degree of anisotropy of the neutrino radiation field for the
same hydrodynamic configuration. For example, any ray-by-ray scheme
(Boltzmann transport or MGFLD) that implements 2D neutrino transport
via many solutions of 1D transport problems along radial rays, even if
neighboring rays are coupled, tends to overestimate local and global
anisotropies. On the other hand, a full-2D MGFLD approach tends
to smooth out anisotropies in the radiation field, in particular at
low optical depths. Ideally, neutrino transport should be carried out
in 2D (or 3D) with energy and momentum-space
angle-dependence. Ott~et~al.~\cite{ott:08} have recently carried out
such simulations using the $S_n$ method (e.g., \cite{castor:04}), but
due to limited number of momentum-space angles that could be used in
long-term evolutions, anisotropies in the neutrino radiation field are
dominated by poor numerical resolution.

When interpreting any results on the GW signal emitted by anisotropic
neutrino radiation obtained with current codes, the dependence on the
neutrino transport formulation and numerics should be kept in mind.

\subsection{Rapid Rotation}
\label{section:nugw_rapid}

Because of the action of centrifugal forces, a rapidly-rotating PNS
has an oblate shape. The neutrino radiation field and its
energy-dependent neutrinospheres follow the matter distribution.
Because of the more compact polar and more extended equatorial density
distribution, neutrinos decouple from matter at smaller radii along
the poles than on the equator. Depending on the degree of rotational
deformation, this can lead to very large pole-equator asymmetries
in the neutrino radiation fields, resulting in the emission of GWs
(see, e.g., \cite{ott:08,walder:05,dessart:06aic,kotake:03aniso}).
Since the qualitative shape of the asymmetry does not vary with time
(viz.\ the PNS stays oblate), the GW signal is monotonically
growing in amplitude and exhibits slow time variation.

\begin{figure}
\centering
\includegraphics[width=8.7cm]{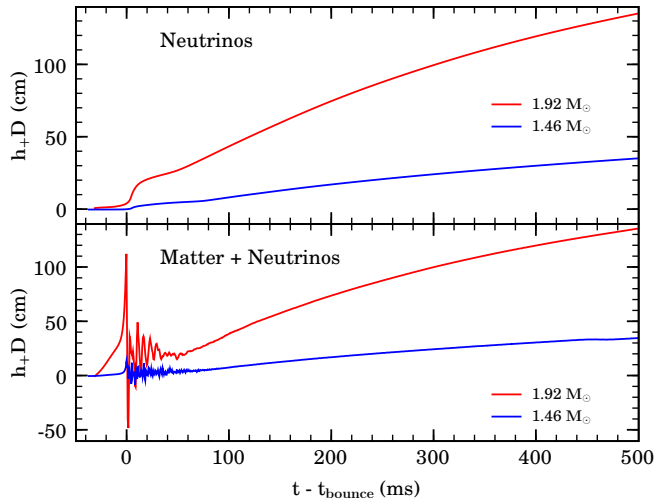}
\caption{GW signal from the rapidly-rotating accretion-induced
collapse (AIC) simulations of Dessart~et~al.~\cite{dessart:06aic} as
analyzed in \cite{ott:06phd}. The $1.92$-$\mmsun$ model has a
postbounce $\beta$ of $\sim 0.25$ while the $1.46$-$\mmsun$ model is
rotating much more slowly with postbounce $\beta \lesssim 0.05$.  {\bf
Top}: GW signal ($h_+ D$ in units of cm) emitted by anisotropic
neutrino radiation. The time is given with respect to the time of
bounce and the first $500\;\mathrm{ms}$ after bounce are shown. Note
the very slow time variation of the signal which is due to the time
integral in equation~\ref{eq:vulcan_nugw} and to the slowly varying
anisotropy of the radiation field owing to the secular contraction of
the PNS.  The GW signal due to neutrinos was extracted at a spherical
radius of $300\;\mathrm{km}$ ($200\;\mathrm{km}$) in the
$1.92$-$\mmsun$ ($1.46$-$\mmsun$) model.
{\bf Bottom}: Combined matter and
neutrino GW signal. The GW signal data are available for download
from \cite{ottcatalog}.}
\label{fig:gw_nu_aic}
\end{figure}

Dessart~et~al.~\cite{dessart:06aic} and Ott~\cite{ott:06phd} analyzed
the GW emission due to aspherical mass motions and asymmetric neutrino
radiation fields in the axisymmetric Newtonian accretion-induced
collapse (AIC) simulations of Dessart~et~al.~\cite{dessart:06aic}.  An
AIC event in a massive ONeMg white dwarf is expected to leave behind
a PNS that is similar in many ways to a PNS formed in standard iron core
collapse.  The two AIC models considered by
Dessart~et~al.~\cite{dessart:06aic} yielded rapidly rotating PNSs with
postbounce $\beta$s of $\sim 0.05$ and $\sim 0.25$ for their
$1.46$-$\mmsun$ model and $1.92$-$\mmsun$ model,
respectively. Dessart~et~al.\ employed a 2D MGFLD approach to neutrino
transport and, hence, the numbers given below for the GW emission due
to neutrinos may be underestimating the true signal strength.

Figure~\ref{fig:gw_nu_aic} shows the contribution of anisotropic
neutrino emission to the GW signal in the AIC models.  At bounce, the
signal shows the greatest variation due to the strong
electron-neutrino burst.  After bounce, the amplitude slowly
but continuously rises. At the end of the simulations,
$|h_\mathrm{max}|$ at $10\;\mathrm{kpc}$ is $\sim 3.6\times10^{-21}$
($\sim 0.55\times10^{-21}$) in the $1.92$-$\mmsun$ ($1.46$-$\mmsun$)
model. The amplitude continues to grow in the subsequent cooling
phase, but its growth rate decreases continuously, since the neutrino
luminosity $L_\nu$ is dropping at a greater rate than the anisotropy
parameter $\alpha$ is increasing. In the frequency domain, the GW
emission due to neutrinos occurs at low frequencies and
$dE_\mathrm{GW}/df$ peaks in the range of $0$--$10\;\mathrm{Hz}$.  The
total emitted $E_\mathrm{GW}$ is $\sim 2\times10^{-11}\;\mmsun c^2$
($\sim 1\times10^{-12} \;\mmsun c^2$) in the $1.92$-$\mmsun$
($1.46$-$\mmsun$) model\footnote{Note that the numbers for
$E_\mathrm{GW}$ given here differ from those stated in \cite{dessart:06aic}.
This is due to an error in the GW analysis of \cite{dessart:06aic}.}.

Kotake~et~al.~\cite{kotake:06a} showed example results of their
unpublished work on the GW signal from rapidly-rotating iron core
collapse.  Unfortunately, they provided only the GW signal from core
bounce and did not provide overall signal characteristics that could
be compared with the AIC models of Dessart~et~al.

M\"uller~et~al.~\cite{mueller:04} studied the GW emission due to
neutrinos in a relatively slowly rotating $15$-$\mmsun$ model.  They
found an almost monotonically-growing GW signal qualitatively similar
to those shown in \fref{fig:gw_nu_aic} with a maximum amplitude at
$10\;\mathrm{kpc}$ of $\sim 1.3 \times 10^{-21}$ at the end of their
simulation ($\sim 270\;\mathrm{ms}$ after bounce). Since their
$15$-$\mmsun$ model, in contrast to the AIC models of Dessart~et~al.,
exhibited significant convective overturn, its GW signal contains
higher-frequency ($\sim 20-100\;\mathrm{Hz}$) components that are due to
either locally-enhanced neutrino emission from rapid downflows or
variations in neutrino absorption due to convective overturn.

Given the above listed results, the GW signal due to anisotropic
neutrino emission associated with rapid PNS rotation is emitted at too
low frequencies and is not sufficiently energetic to be a good
candidate for detection by current and planned ground-based GW
observatories even if the stellar collapse event occurs within the
Milky Way. The situation may be different in the case of moderate
rotation studied by M\"uller~et~al.~\cite{mueller:04} in which the GW
signal contains higher-frequency contributions from
convection/down-stream-induced neutrino radiation
anisotropies. According to the results of M\"uller~et~al., the
neutrino GW signal of such cores may be detectable with advanced
LIGO-class detectors for a galactic source.

\subsection{Convection and SASI} 
\label{section:nugw_conv}

Convection and SASI introduce asphericity into the postshock
flow, leading, among other things, (1) to spatially varying neutrino
cooling/heating in the region between shock and PNS core and (2) to
accretion funnels that allow for rapid downflow of material and
locally enhanced neutrino emission where an accretion funnel hits the
PNS core. Janka \& M\"uller~\cite{jm:97} first investigated (though
with simplified neutrino transport and an artificial inner core
boundary) the GW emission from anisotropic neutrino radiation fields
owing to postbounce convection and accretion downstreams. Since they
provided only GW signals for which neutrino and baryonic matter
contributions were summed up, it is difficult to extract the
neutrino component. A rough estimate gives modest peak amplitudes in
the range of $5\times 10^{-24}$ to $\lesssim 1\times10^{-22}$ at
$10\;\mathrm{kpc}$ and the typical slow waveform variation.  

More recently, M\"uller~et~al.~\cite{mueller:04}, using a ray-by-ray
Boltzmann neutrino-transport scheme and a gravitational potential
whose monopole term is general relativistic, investigated the emission
of GWs due to neutrinos in the first $\sim 200-270\;\mathrm{ms}$ after
bounce in a slowly-rotating $15$-$\mmsun$ model (discussed above in
section~\ref{section:nugw_rapid}) and a non-rotating $11$-$\mmsun$
model.  In the latter, the GW signal due anisotropic neutrino emission
reflects the stochastic nature of postbounce convection and varies
slowly (typical frequencies around $10-20\;\mathrm{Hz}$) with no clear
long-term trend, reaching maximum amplitudes of $\sim 1.4 \times
10^{-22}$ at $10\;\mathrm{kpc}$. Due to the low-frequency variation,
the emitted $E_\mathrm{GW}$ can be expected to be small, was not
stated explicitly by the authors, and is likely to be in or below the
ball park of the numbers mentioned in section~\ref{section:nugw_rapid}
for rapidly-rotating AIC models. In contrast to their slowly-rotating
$15$-$\mmsun$ model, the neutrino GW signal of the non-rotating
$11$-$\mmsun$ model is unlikely to be detectable even by advanced
LIGO-class observatories and when occurring at $10\;\mathrm{kpc}$
distance.

In addition to the $15$-$\mmsun$ and $11$-$\mmsun$ models,
M\"uller~et~al.\ also considered the GWs emitted by anisotropic
neutrino radiation in a PNS convection model of
Keil~et~al.~\cite{keil:96,keil:97} (see section~\ref{section:pnsconv} and
figure~\ref{fig:mueller2004}). The GW signal is qualitatively and
quantitatively similar to what M\"uller~et~al. observed in their full
$11$-$\mmsun$ progenitor. However, before explosion, the true
asymptotic neutrino anisotropy must be extracted outside the
optically-semi-transparent postshock layer (see, e.g.,
\cite{marek:08b}) which was not included in the PNS model of
Keil~et~al. At post-explosion times, the neutrinos practically
free-stream from the PNS surface and the GW signal from the PNS
simulation becomes more relevant.

Using 2D-MGFLD, Ott et al.~\cite{ott:06prl,ott:06phd,ottcatalog} found
for three long-term postbounce models in the mass range from $11$ to
$25\;\mmsun$ maximum amplitudes $|h_\mathrm{max}|$ of $\sim
1.3\times10^{-23}$ to $\sim\;$~$5.5\times 10^{-23}$ (at
$10\;\mathrm{kpc}$) built up over the $\sim 1\;\mathrm{s}$ of
postbounce time covered by the simulations and extracted at a radius
of $200\;\mathrm{km}$.  The neutrino GW signals in these models
exhibit a systematic trend to negative, in absolute value continuously
growing amplitudes.  The total emitted $E_\mathrm{GW}$ due to
neutrinos were small, below $\sim 1\times10^{-13}\;\mmsun c^2$ and
most of the energy emission took place at frequencies below $\sim
10\;\mathrm{Hz}$. Since MGFLD has the tendency to smooth out
radiation-field anisotropies~\cite{ott:08}, the above numbers
unfortunately depend on the extraction radius and should be regarded
as underestimates.

Kotake et al.~\cite{kotake:07a}, using a simplified approach to
neutrino heating and cooling, studied the GW emission from angular
variations in the neutrino cooling in their models that were focussed
on the non-linear development of the SASI.  They found
quasi-monotonically growing GW signals that showed slow variations
($1$--$50\;\mathrm{Hz}$) and reached maximum values after
$\sim500\;\mathrm{ms}$ of $2.0$--$3.5\times10^{-20}$ (at
$10\;\mathrm{kpc}$). The GW signal amplitudes found by
Kotake~et~al.\ are systematically positive which the authors
attributed to SASI-enhanced polar neutrino emission.  The emitted
$E_\mathrm{GW}$ were in the range of $2-6\times10^{-10}\,\mmsun c^2$.
These large numbers are a most likely a consequence of the neutrino
luminosities they imposed which were a factor of $5$ to $10$ larger
than in more realistic simulations~\cite{mueller:04,ott:06prl} and
lead to very vigorous overturn and SASI.

Marek~et~al.~\cite{marek:08b} studied the GW emission
from anisotropic neutrino radiation fields in a nonrotating
$15$-$\mmsun$ model that was run with two nuclear EOS of different
stiffness (see also the discussion of their study in
section~\ref{section:gwconvsasi}).  Similarly to
Ott~et~al.~\cite{ott:06prl,ott:06phd}, but in disagreement with the
less detailed study of Kotake~et~al.~\cite{kotake:07a}, they found a
continuously (in absolute value) growing negative GW signal. According
to their detailed analysis, this negative signal is a consequence of
enhanced equatorial neutrino emission in the postshock region.  At
$400\;\mathrm{ms}$ after bounce, their Lattimer-Swesty
EOS~\cite{lseos:91} model reaches $|h|$ of $\sim 1.2\times10^{-21}$
(at $10\;\mathrm{kpc}$) while the model run with the stiffer Wolff
EOS~\cite{hillewolff:85} exhibits a smaller $|h|$ of $\sim
5.7\times10^{-22}$. These $|h|$ are larger than the typical amplitudes
of GWs emitted by the matter dynamics associated with convection/SASI
(section~\ref{section:gwconvsasi}). However, in agreement with
previous studies, Marek~et~al.~\cite{marek:08b} provided GW spectra
indicating emission predominantly at frequencies $\lesssim
20\;\mathrm{Hz}$, making the component of the overall GW signal
associated with anisotropic neutrino emission very difficult to detect
for ground-based detectors whose sensitivity is limited by
gravity-gradient and seismic noise at low frequencies~(e.g.,
\cite{whitcomb:08}).

\subsection{Global Asymmetries}

\begin{figure}
\centering
\includegraphics[width=4.8cm]{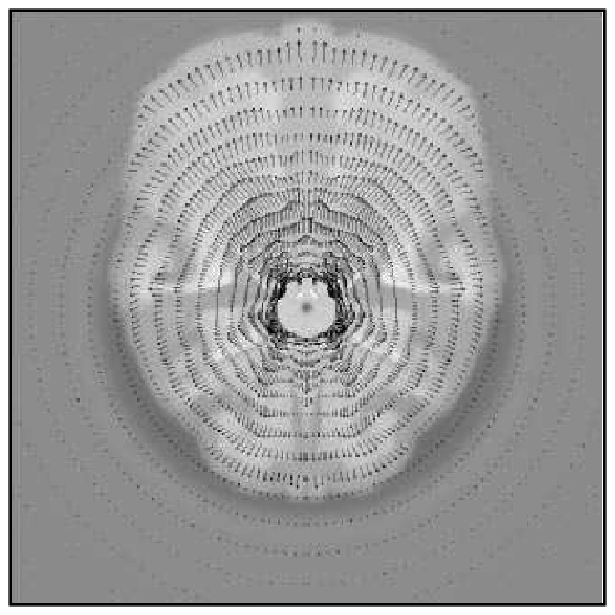}
\includegraphics[width=7.3cm]{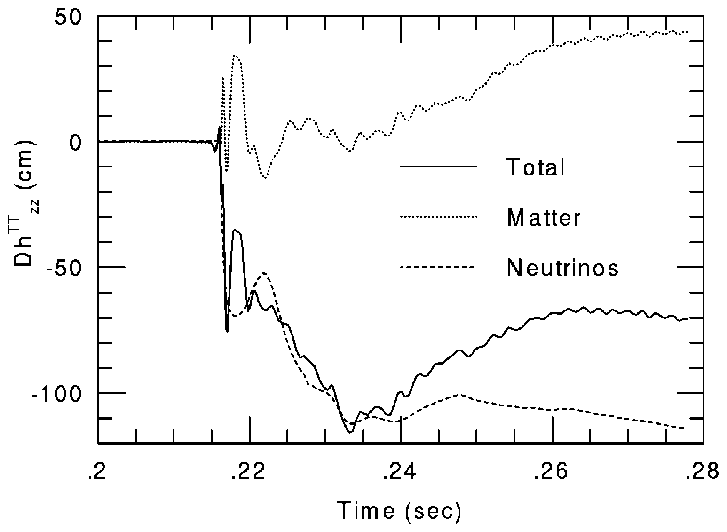}
\caption{{\bf Left}: Grey-scale map showing the entropy distribution
$\sim 50\;\mathrm{ms}$ into the explosion of the 2D asymmetric
collapse model of Burrows and Hayes~\cite{bh:96}. The physical scale
is $4000$ x $4000\;\mathrm{km}$ and velocity vectors are superposed.
This plot corresponds to figure~1 of \cite{bh:96} and is used by kind
permission from the authors.  {\bf Right}: GW signal due to
anisotropic neutrino emission (dashed line), matter (dotted line) and
combined matter+neutrinos (solid line) in the same model of
\cite{bh:96}. This plot corresponds to figure~3 of \cite{bh:96} and is
used by kind permission from the authors.}
\label{fig:bh96}
\end{figure}

Large-scale density perturbations that may be due to inhomogeneities
in the iron core, silicon, and/or oxygen shells lead to globally
asymmetric postbounce mass motions resulting in large angular
variations in the neutrino luminosity (see also
section~\ref{section:global}).  The GWs emission due to neutrinos in
such a globally asymmetric core-collapse event has first been
considered by Burrows and Hayes~\cite{bh:96} in 2D, using a ray-by-ray
gray flux-limited diffusion approach~\cite{bhf:95}.  Their single
model with a 15\% precollapse density perturbation near the north pole
produced a GW signal with a large $|h_\mathrm{max}|$ of $\sim
3.5\times10^{-21}$ (at $10\;\mathrm{kpc}$). The signal, shown in the
right-hand panel of figure~\ref{fig:bh96}, exhibits a strong burst at
core bounce (with $\sim 500\;\mathrm{Hz}$ characteristic frequency)
and a subsequent slow growth to the final amplitude (the model was
tracked to only $\sim 70\;\mathrm{ms}$ after bounce).  The total
emitted $E_\mathrm{GW}$ is around $\sim 3\times10^{-10}\;\mmsun c^2$,
the overall largest $E_\mathrm{GW}$ from anisotropic neutrino emission
found for any presently published model.

Fryer~et~al.~\cite{fryer:04}, also using gray flux-limited diffusion,
but employing a 3D Newtonian smooth-particle hydrodynamics (SPH)
scheme, studied four models with density perturbations of 25\% to 40\%
in 30-degree wedges about the north pole. They found GW signals due to
anisotropic neutrino emission with maximum amplitudes of the same
order of magnitude as those reported by \cite{bh:96}. However, their
signals exhibited much less time variation and a weaker burst
associated with core bounce, suggesting much lower energy emission
(Fryer~et~al.\ did not provide values for $E_\mathrm{GW}$).

\section{Other GW Emission Mechanisms}
\label{section:others}

\subsection{Aspherical Outflows}
\label{section:aspherical}
Core-collapse SN explosions are unlikely to be perfectly spherically
symmetric and all currently considered explosion mechanisms rely on
the breaking of symmetry. Magnetically-driven explosions exhibit
jet-like bipolar outflows (e.g., \cite{burrows:07b} and references
therein), while explosions driven by the neutrino mechanism (e.g.,
\cite{marek:07}) or the acoustic mechanisms (e.g., \cite{burrows:07a})
can range from predominantly unipolar or bipolar outflows to nearly
spherically symmetric explosions.
In addition to the various explosion scenarios, precollapse
large-scale asymmetries in the central regions of the star
may also favor a largely asymmetric explosion.

The GW emission from rapid aspherical outflows has been considered by
Obergaulinger~et~al.~\cite{obergaulinger:06a,obergaulinger:06b} in the
context of 2D magneto-rotational core collapse in Newtonian gravity
and approximate GR (see \sref{section:snmodel}) and by
Shibata~et~al.~\cite{shibata:06} in 2D GRMHD.  Both groups of authors
employed polytropic precollapse models, a simple polytropic + ideal
gas (``hybrid'') EOS, and neglected neutrinos. At this level of
approximation, the shock generally does not stall and a prompt
explosion occurs after bounce.

\begin{figure}
\centering
\includegraphics[width=6.0cm]{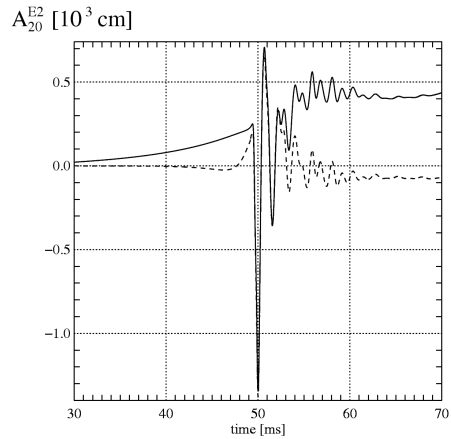}
\caption{GW signal of the
  Obergaulinger~et~al.~\cite{obergaulinger:06a} model A1B3G3-D3M13 in
  terms of the quadrupole pure-spin Tensor harmonic amplitude
  $A_{20}^{E2}$, and $h_+ D = \frac{1}{8}\sqrt{\frac{15}{\pi}}
  \sin^2\theta A^{E2}_{20}$.  The dashed line gives the contributions
  of regions with radius $r \lesssim 60\;\mathrm{km}$ while the solid
  line shows the total signal. The postbounce offset due to
  rapid aspherical outflows in the outer regions is clearly visible in
  the total signal.  This figure corresponds to the lower-left panel
  of figure~8 in \cite{obergaulinger:06a} and is used by kind
  permission from the authors.}
\label{fig:obergau}
\end{figure}

Obergaulinger~et~al.\ and Shibata~et~al.\ both reported a
quasi-monotonically growing contribution to the GW signal from bipolar
outflows, leaving behind a GW memory akin that discussed in the
context of GW emission from anisotropic neutrino radiation
fields~(see section~\ref{section:neutrinos}). An example GW signal taken
from the study of Obergaulinger~et~al.\ is shown in
\fref{fig:obergau}.  Shibata~et~al.~\cite{shibata:06} pointed out that
the GW signal systematics can be understood by considering an outflow
of mass $m$ in the $z$ direction with slowly changing velocity. In
this case, the contribution to the GW signal by the outflow is
$|h_\mathrm{out}| \propto 2 m v_z^2$. Once the outflow has reached a
quasi-steady state, the mass ejection rate can be assumed to stay
roughly constant, hence the mass in the outflow increases roughly
linearly~\cite{shibata:06,burrows:07b}. Using this, the appropriate
factors of $G$ and $c$, and imposing a scaling on $m$ and $v_z$, one
finds \cite{shibata:06}
\begin{equation}
|h_{\mathrm{out},+}| D \approx 300 \bigg(\frac{m}{0.1 \mmsun}\bigg)
\bigg(\frac{v_z}{0.1c}\bigg)^2\; \mathrm{cm}\,\,.
\label{eq:out}
\end{equation}
Similar to the case of anisotropic neutrino emission, the slow
variation of $|h_{\mathrm{out},+}|$ leads to emission of only small
$E_\mathrm{GW}$ that are emitted predominantly at low frequencies.

The approximate $|h_{\mathrm{out},+}|$ given by equation~\ref{eq:out} is in
rough agreement (assuming $m = 0.1\,\mmsun$ and $v_z = 0.1c$) with the
GW memory amplitude of $\sim 6.5\times10^{-21}$ (at
$10\;\mathrm{kpc}$) that Shibata~et~al.\ found after $\sim
15\;\mathrm{ms}$ of postbounce evolution in the most extreme model of
their limited model set. Obergaulinger~et~al.~\cite{obergaulinger:06a,
obergaulinger:06b} performed a more extensive set of calculations, but
did not publish the final GW signal memory amplitudes for all their models.
Those that were shown in \cite{obergaulinger:06a,obergaulinger:06b}
are in rough agreement with equation~\ref{eq:out} and the results of
Shibata~et~al.~\cite{shibata:06}.

Finally, it is important to point out that the MHD-driven bipolar
outflows observed in the microphysically-simple calculations of
\cite{obergaulinger:06a,obergaulinger:06b,shibata:06} occur very early
(around $\sim 5$--$20\;\mathrm{ms}$) in the postbounce evolutions of
their models. More realistic models computed by
Burrows~et~al.~\cite{burrows:07b} (who did not analyze the GW emission
in their models) require of the order of 
$\sim 100\;\mathrm{ms}$ to launch the bipolar outflow. This longer
delay translates to lower accretion and mass expulsion rates and
and may result in smaller GW signal memory amplitudes than
observed by \cite{obergaulinger:06a,obergaulinger:06b,shibata:06}
for their very early explosions.

\subsection{Global Precollapse Asymmetries}
\label{section:global}
Prior to core collapse and after silicon core burning has ceased,
nuclear burning continues in convectively unstable silicon and oxygen
burning shells at the outer edge of the iron core (see, e.g.,
\cite{whw:02}).  The large- and small-scale density, composition and
thermal perturbations induced by convection in these layers and at the
outer edge of the iron core could become large
\cite{bazan:89,meakin:06,meakin:06b,murphy:04}, perhaps up to $\sim
10\%$, and would essentially be frozen in during collapse
\cite{lai:00}.

The effect of these perturbations on the postbounce hydrodynamics
depends on their magnitude and distribution. If they occur primarily
on small scales and reach deep into the iron core, they may act as
seeds for prompt and neutrino-driven convection (see
section~\ref{section:convsasi} for the associated GW signals).  
Larger-scale perturbations, on the other hand, can lead to strongly
aspherical explosions into the direction of lowest density. 
Such largely aspherical ejection of
matter -- as just discussed in the previous
section~\ref{section:aspherical} -- can lead to the burst emission of
GWs with GW memory (e.g., \cite{segalis:01,braginskii:87}) and, in the
case of primarily unipolar outflow, may be responsible for pulsar
birth kicks~(e.g., \cite{hansenphinney:97,lai:01kick} and references
therein).

Burrows \& Hayes \cite{bh:96} were the first to perform a numerical
study of core collapse with a globally asymmetric presupernova
configuration.  They employed an axisymmetric Newtonian code with gray
flux-limited neutrino diffusion along rays.  Their single model was a
$15$-$\mmsun$ presupernova configuration perturbed by a 15\% reduction
of the density of the outer core within 20 degrees of its north pole.
The left panel of figure~\ref{fig:bh96} displays the aspherical
explosion obtained in their model. It erupted preferentially through
the lower-density polar regions and lead to a recoil velocity of the
PNS of $\sim 530\;\mathrm{km/s}$.  The aspherical matter dynamics
resulted in a GW burst signal (shown in figure~\ref{fig:bh96}) that
set in $\sim 5\;\mathrm{ms}$ after bounce and reached maximum
amplitudes of $\sim 1\times10^{-21}$ (at $10\;\mathrm{kpc}$) at early
times with rapid time variation around $\sim 500\;\mathrm{Hz}$ and an
asymptotic memory of $\sim 1.5 \times 10^{-21}$ (at
$10\;\mathrm{kpc}$) built up over $\sim 70\;\mathrm{ms}$.  The
combined neutrino and matter GW signal exhibited a memory of $\sim
2.3\times 10^{-21}$ (at $10\;\mathrm{kpc}$) and the total emitted
energy in GWs was $\sim 1.1\times10^{-9}\;\mmsun\,c^2$.

Fryer~et~al.~\cite{fryer:04,fryer:04kick}, using their 3D SPH scheme
with gray flux-limited diffusion, performed a set of four calculations
of a $15$-$\mmsun$ presupernova star. Three calculations were
perturbed with an initial $30\%$ to $40\%$ density reduction in the
oxygen and silicon layers in a $30$-degree wedge centered about one of
the poles. One calculation was set up with $25\%$ lower density in a
$30$-degree polar wedge throughout iron core, silicon and oxygen
shells. Despite the larger perturbations present in their models,
Fryer~et~al.\ did not find strongly asymmetric explosions and reported
neutron star kick velocities below $\sim 200\;\mathrm{km/s}$, arguing
that momentum transfer by neutrinos was partly counteracting the
hydrodynamic kick mechanism.

The GW emission from aspherical flow in the Fryer~et~al.\ models sets
in within $\sim2$--$5\;\mathrm{ms}$ after bounce and, according to the
authors, is mostly due to asymmetries caused by oscillations in the
PNS core that are excited by the counteracting recoils due to
aspherically ejected matter and neutrino momentum. However, $h_+$ and
$h_\times$ of their models show no periodicity and appear uncorrelated. 
Considering the discussion
on prompt convection in section~\ref{section:convsasi}, it appears not
unlikely that at least part of the GW signal is due to prompt
convection. The GW signals reach their maximum amplitudes of $\sim
2$--$8\times 10^{-22}$ (at 10 kpc) around $10$--$20\;\mathrm{ms}$
after bounce, decay thereafter and contain no large-amplitude GW
memory. The absence of the latter is not discussed by Fryer~et~al.,
but may be due to the relatively small asphericity of their
explosions. The authors provided neither GW spectra nor numbers for
the energy emitted in GWs.

\subsection{Magnetic Fields}

As already pointed out in section~\ref{section:neutrinos}, not only
aspherical matter dynamics, but any kind of accelerated transport of
energy, including magnetic-field energy, may lead to the emission of GWs.
Kotake~et~al.~\cite{kotake:04} extended the Newtonian quadrupole
formalism of wave generation to include the
contribution from magnetic fields and introduced terms associated with
Lorentz force and magnetic-field energy. They focussed on the GW signal of
rotating core collapse and bounce 
and performed a limited set of Newtonian 2D magneto-rotational
core-collapse calculations with a finite-temperature nuclear
EOS and a leakage scheme for neutrinos. They demonstrated that the
magnetic contribution to the GW signal at core bounce remains smaller
than $\sim 1\%$ of the matter contribution for cores with precollapse
magnetization below $\sim 10^{12}\;\mathrm{G}$. The latter is already
an extreme value, considering that garden-variety precollapse iron
cores are likely to be weakly magnetized with $B \lesssim
10^{7}$--$10^{10} \mathrm{G}$~\cite{heger:05}. Since matter and
B-field are strongly coupled, the magnetic component of the GW signal
exhibits time variations similar to those of the matter component. 
Interestingly, it leaves behind a secularly growing GW memory that, at
$20\;\mathrm{ms}$ after bounce, dominates the overall signal with an
amplitude of $\sim 2\times10^{-21}$ (at $10\;\mathrm{kpc}$) in
Kotake~et~al.'s model with precollapse magnetic field of
$5\times10^{12}\;\mathrm{G}$ and central angular velocity
$\Omega_{\mathrm{c,i}} = 4\;\mathrm{rad}\,\mathrm{s}^{-1}$. This GW
memory may be due to the gradual build-up of the magnetic field energy
in polar regions via an $\Omega$ dynamo (e.g.,
\cite{obergaulinger:06a,cerda:07,burrows:07b} and references therein), but was
not studied in detail by Kotake~et~al. Also note that the models of
Kotake~et~al. were not followed through explosion, hence do not
exhibit a contribution to the GW signal and GW memory from aspherical
outflows (see section~\ref{section:aspherical}).

Obergaulinger~et~al.~\cite{obergaulinger:06a,obergaulinger:06b}
performed an extensive set of magneto-rotational core-collapse
simulations with simplified microphysics, leading to prompt explosions
with bipolar outflows (see section~\ref{section:aspherical}).  They
confirmed the previous result of Kotake~et~al.~\cite{kotake:04} that
MHD effects on the dynamics and the magnetic contribution to the GW
signal of core bounce remain small for cores with precollapse magnetic
field strengths below $\sim 10^{12}\;\mathrm{G}$.  For models with
stronger precollapse magnetic fields, Obergaulinger~et~al.\ found
significant magnetic contributions to the GW signal that grow after
bounce due to the continuously increasing magnetic stresses in the
polar regions and inside the PNS.  The time variation in the magnetic
component is much slower than in the matter component, peaking in
frequency below $\sim 100\;\mathrm{Hz}$. In multiple models with
strong magnetic fields (initial $B\gtrsim
10^{12}$--$10^{13}\;\mathrm{G}$), the magnetic component eventually
dominates the GW signal and leaves behind a GW memory that may have
positive or negative sign, varying from model to model.  
The maximum GW signal amplitudes
due to the magnetic component generally scale with the precollapse
magnetic field strength.  For their most extreme model,
Obergaulinger~et~al.\ found a maximum magnetic GW signal amplitude of
$\sim 4.5\times 10^{-21}$ (at $10\;\mathrm{kpc}$) while most other
models show one to two orders of magnitude smaller maximum
amplitudes. Note that the GW memory due to the magnetic contribution
linearly combines with the memory that is due to the bipolar outflow
of matter.  Depending on the sign of the magnetic contribution, the
total amplitude of the GW memory may be increased or decreased.

\subsection{GWs from Collapse to a Black Hole}
\label{section:GWBH}

Ordinary massive stars in the mass range from $\sim 8$ to $\sim
100\;\mmsun$ burn their nuclear fuel all the way to iron-group nuclei
(or a mixture of oxygen, neon, and magnesium nuclei at the lowest
masses) and their cores stay in hydrostatic quasi-equilibrium
throughout their nuclear burning lives. The final iron or ONeMg cores
are supported primarily by the pressure of degenerate electrons
($P_e$), secondarily by the thermal pressure of the heavy ions
($P_\mathrm{ion}$), and tertiarily by radiation pressure
($P_\mathrm{rad}$) such that $P_e \gg P_\mathrm{ion} \gg
P_\mathrm{rad}$ \cite{bethe:90}.  Hence, iron/ONeMg cores resemble
white dwarfs and have effective Chandrasekhar masses in the range of
$\sim 1.3$ to $\sim 2.0\;\mmsun$ \cite{baron:90,whw:02}. When such a
core collapses, it separates into a subsonically contracting
inner core and a supersonically infalling and consequently rarefying
outer core. At bounce, the inner core has a mass of $\sim 0.5$ to
$\sim 0.7\, \mmsun$ (depending on its average electron fraction $Y_e$,
average specific entropy $s$, and rotational
configuration~\cite{dimmelmeier:08,burrows:83}). The solid-core of the
nuclear force and the resulting stiff EOS above nuclear density
\emph{easily} stabilizes the collapse of the inner core. Collapse
\emph{always} results in a PNS and direct collapse to a black hole (BH)
without core bounce and a PNS phase \emph{occurs never} in ordinary
massive stars\footnote{There is considerable confusion in
the astrophysics community about this fact.  This is in part due to
the misleading use of the terms 'direct/prompt BH formation' and
'delayed BH formation' by Fryer, Woosley, and collaborators. In their
context (e.g., \cite{hegeretal:03}), these terms describe 'BH
formation without SN explosion' and 'BH formation after SN explosion
by fallback accretion', respectively.}.

After bounce, outer core material accretes onto the PNS.  The maximum
baryonic NS mass\footnote{Astronomical observations
generally measure the gravitational mass, the baryonic mass minus the
mass-energy equivalent of the NS gravitational binding
energy.} that can be supported against gravity
depends sensitively on the stiffness of the high-density
EOS and ranges between $\sim 1.5\;\mmsun$ and $\sim
2.5-3\;\mmsun$ (e.g.,
\cite{lattimer:07,sumiyoshi:06,sumiyoshi:07}). Rapid differential
rotation can enhance the maximum mass by up to $\sim 50\%$ (e.g.,
\cite{baumgarte:00}), while the thermal structure of the PNS
has less influence on the maximum mass~\cite{burrows:88,sumiyoshi:07}.

In order to leave behind a neutron star, the SN explosion has to set
in before the PNS has reached its maximum mass. Depending on the
latter's actual value and on the accretion rate set by the structure
of the progenitor, an explosion has to occur within $\sim
1-2\;\mathrm{s}$ after bounce.  If the explosion is weak and does not
unbind the entire stellar envelope, fall-back accretion can still push
the PNS over its mass limit at later times (e.g., \cite{zhang:08}).
When a PNS becomes gravitationally unstable before the explosion is
launched, the SN engine is immediately shut off and the star ends its
life as a collapsar, possibly exploding in a gamma-ray burst if it
possesses the needed amount and distribution of angular
momentum~\cite{woosley:93,wb:06,dessart:08a}.

Nonspherical PNS collapse to a BH results in a burst  of GWs
(1) due to the rapidly shrinking mass-quadrupole moment of the PNS
during the collapse and (2) due to the quasi normal mode (QNM) ringing of
the nascent BH that is initially distorted from its stationary Kerr
shape and experiences subsequent distortion by accreting material.

The GW emission from the collapse of a compact star to a BH was first
considered in the late 1970's by Cunningham, Price, and
Moncrief~\cite{cunningham:78,cunningham:79,cunningham:80} whose work was
later improved by Seidel and collaborators~\cite{seidel:90,seidel:91}
and more recently by Harada, Iguchi, and Shibata~\cite{harada:03}.  These
authors employed a perturbative approach and studied the GW emission
from nonspherical perturbations on a spherically-symmetric background.
They found that the overall GW signal of BH formation is dominated by
the QNM ringing of the formed BH and that most of the GWs are emitted
from the fundamental quadrupole mode. If the perturbation is axisymmetric
(i.e., primarily due to rotation) the emission is dominated by the
$\ell = 2, m = 0, n=0$ ($200$) multipolar component, while nonaxisymmetric
perturbations show up as $\ell =2, m \ne 0, n=0$ modes.  These modes
emit at characteristic frequencies that depend only on the BH mass $M$
and its dimensionless angular momentum parameter $j =
(cJ)/(GM^2)$. Berti~et~al.~\cite{berti:06} provide fitting-formulae
and coefficients that describe to better than $\sim 5\%$ accuracy the
BH QNM frequencies $f_{lmn}$. 
Focussing on the $200$ (axisymmetric) and $220$
(nonaxisymmetric, bar-like) QNMs, we find using their tables,
\begin{eqnarray}
\label{eq:f200}
f_{200} &=& 14.4 \bigg(\frac{M}{\mmsun}\bigg)^{\!-1} 
           (1-0.165(1-j)^{0.355})\, \mathrm{kHz}\,\,,\\
\label{eq:f220}
f_{220} &=& 49.4 \bigg(\frac{M}{\mmsun}\bigg)^{\!-1} 
           (1-0.759(1-j)^{0.1292}) \, \mathrm{kHz}\,\,.
\end{eqnarray}
Hence, a nonrotating BH with a mass of $2 \,\mmsun$ rings when
perturbed in its $200$ (or $220$) mode with a frequency of $\sim
6\;\mathrm{kHz}$ and this frequency decreases linearly as more matter
is accreted. In the case of slow rotation (small $j$), one would
expect a 'reverse chirp' from higher to lower frequencies in the GW
signal as a BH is formed and accretes the massive stellar envelope in
a failing core-collapse SN.  With increasing spin, the QNM frequencies
increase at different rates for different azimuthal mode number $m$
and for rapid rotation, the decrease of the QNM frequencies by
increasing BH mass could be partly compensated by increasing $j$.
Initial and advanced LIGO-class detectors have their maximum
sensitivity at frequencies below $\sim 1\;\mathrm{kHz}$, thus will
require large signal amplitudes and emitted energies to detect even a
galactic NS collapse event.

\begin{figure}
\centering
\includegraphics[width=6.0cm]{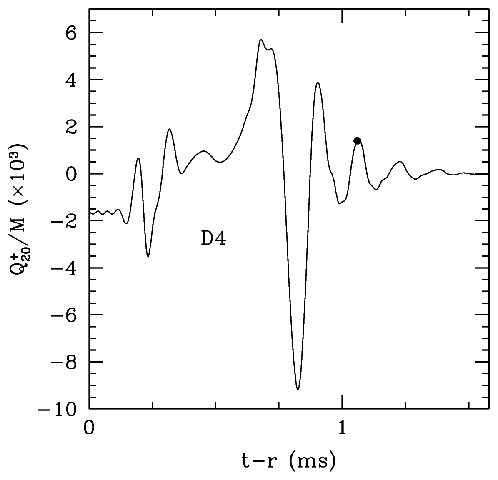}
\includegraphics[width=6.0cm]{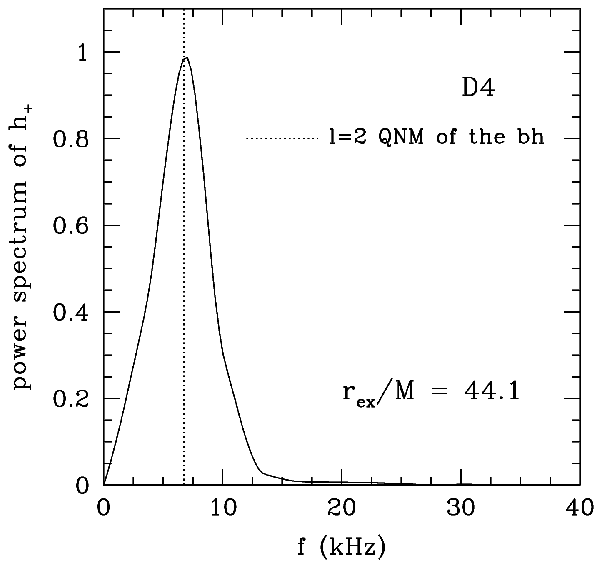}
\caption{{\bf Left}: GW signal of the collapse of a rapidly rigidly
  rotating polytropic NS to a black hole in terms of the dominant
  even-parity metric perturbation $Q^+_{20}$ as computed by
  Baiotti~et~al~\cite{baiotti:05,baiotti:07b}. Even-parity and
  odd-parity metric perturbations are related to the two physical GW
  polarizations according to $h_+ - i h_\times = \frac{1}{2D}
  \sum_{l,m} (Q^+_{lm} - i \int_{-\infty}^t Q^\times_{lm} dt')
  _{-2}Y^{lm}$, where $_{-2}Y^{lm}$ is the $-2$ spin-weighted
  spherical harmonic~\cite{baiotti:05,nagar:05}. This plot corresponds
  to the bottom-right panel of figure 5 of \cite{baiotti:07b} and is
  used by kind permission from the authors. {\bf Right}: Power
  spectrum of the BH ring-down part of the GW signal for the same
  model the units are chosen in such a way that the peak of the power
  spectrum is approximately 1. The dotted line marks the BH QNM
  frequency obtained by perturbation theory. This plot corresponds to
  the bottom-right panel of figure 6 of \cite{baiotti:07b} and is used
  by kind permission from the authors.}
\label{fig:baiotti}
\end{figure}

By making estimates of the energy in the perturbations and its
azimuthal $m$ distribution, one can use equations~\ref{eq:f200} and
\ref{eq:f220} and analogous expressions for other $m$ to estimate GW
signal amplitudes. This can be done as outlined in \cite{fryer:02} or
by assuming that the GW signal is of sine-Gaussian shape as discussed,
e.g., in \cite{abbott:07burst}. However, this shall not be repeated
here, since such estimates for spatial distribution and energy of
perturbations cannot be made reliably. The latter highlights a
fundamental problem of the perturbative approach: While the effects of
a (small) perturbation on a background can be studied with great
accuracy, the perturbation itself cannot be determined and must be put
in by hand or provided as initial data from numerical simulations
(e.g., \cite{berti:06b}).

Direct numerical simulation of nonspherical BH formation using GR
hydrodynamics coupled to spacetime curvature evolution allows for the
self-consistent determination of the GW signal. The first 2D
simulations of BH formation were carried out in 1981 by
Nakamura~\cite{nakamura:81}, but GWs could not be extracted due to
numerical difficulties. The first waveforms and GW energy estimates
from axisymmetric rotating stellar collapse to a BH were provided in
1985 by Stark and Piran~\cite{stark:85} who used highly approximate
initial data that were not in rotational equilibrium. More recently,
Baiotti~et~al.~\cite{baiotti:05,baiotti:07b} presented the first GW
signals from the 3D collapse of rigidly-rotating polytropic NS models
in rotational equilibrium. They found that a collapsing rotating NS
stays essentially axisymmetric and confirmed that the GW emission is
primarily due to axisymmetric BH QNM oscillations. They also identified a
high-frequency, pre-BH formation component that is sensitive to the way
the collapse of the polytropic NS is instigated. In
figure~\ref{fig:baiotti}, we present the GW signal (left panel) and GW
power spectrum of the QNM ringing (right panel) extracted from the
rapidly rigidly rotating collapse model D4 with $M=1.86\,\mmsun$ and
$j=0.54$ of Baiotti~et~al.~\cite{baiotti:07b}. While the early part of
the waveform is dominated by details of the matter dynamics and may be
sensitive to the EOS and the NS angular momentum distribution, the BH
QNM ringing depends only on the mass $M$ and angular momentum
parameter $j$, hence, will be qualitatively and quantitatively
similar in simulations that take into account a more realistic matter
treatment and differential rotation.  For their model suite,
Baiotti~et~al.~\cite{baiotti:05,baiotti:07b} give a maximum energy
emission of $E_\mathrm{GW} = 1.45 \times 10^{-6} (M/\mmsun) \mmsun
c^2$ which obtains in their most rapidly rotating model D4 and scales
roughly as $j^4$ at small $j$. Convolving their GW signals with
detector noise curves as discussed in \cite{thorne:87}, they find an
upper-limit characteristic strain $h_c$ of $\sim 5.5 \times 10^{-22}
(M/\mmsun)$ at $10\;\mathrm{kpc}$ and at a characteristic frequency
$f_c$ of $\sim 530\;\mathrm{Hz}$ for initial LIGO detectors. These
numbers suggest that BH formation in a galactic core-collapse event
with rapid rotation may be marginally detectable already with initial
LIGO-class detectors.  

Simulations that track BH formation in a microphysically detailed
multi-D model of a failing core-collapse SN remain to be
carried out. Such simulations are necessary for the self-consistent
study of the early, pre-QNM ringing GW signal of PNS collapse as well
as for the GW signal of BH QNM oscillations excited by long-term
accretion (but see \cite{nagar:07} for a perturbative treatment of
the latter).

\section{Nearby Core-Collapse SNe, the recent SN 2008bk, and their
Detectability in Gravitational Waves}
\label{section:2008bk}

Optimistic estimates of the core-collapse SN rate in the Milky Way and
the close-by Small and Large Magellanic Clouds predict one
core-collapse SN in $\sim 30-50$ years and even for the entire local
group of galaxies, including M31 at $0.8\,\mathrm{Mpc}$, one
core-collapse SN in $\sim 20$ years is an optimistic estimate (see,
e.g., \cite{vdb:91} and the compilation of rate estimates and references in
\cite{ott:06phd}). This rate stays roughly constant until a distance of
$\sim 3\,\mathrm{Mpc}$. The galaxies of the M81 group and neighboring
groups with high star-formation rates that are located at
$3$--$5\;\mathrm{Mpc}$ from Earth increase the core-collapse SN rate
to an optimistic 1 core-collapse SN in $\sim 2$ years
within $\sim 5\;\mathrm{Mpc}$ \cite{ando:05,arnaud:04}. The next
significant increase in the SN rate occurs when the outskirts of the
Virgo cluster are beginning to contribute at $7$--$10\;\mathrm{Mpc}$
\cite{ando:05}.

Since LIGO/GEO600 \cite{ligo,geo600} science operations began in 2002, five
core-collapse SNe have been discovered optically within
$5\;\mathrm{Mpc}$ from Earth, a number roughly consistent with the
above quoted rate estimate.  Unfortunately, all events occurred outside
of LIGO/GEO600 and VIRGO \cite{virgo} science runs. 
 However, the very recent (as of August 2008) nearby
core-collapse SN, SN 2008bk, occurred while GEO600 and the 2-km LIGO Hanford
interferometer (H2) were taking data in ``astrowatch'' mode
\cite{katsavounidis:08priv}. All other LIGO-class 
detectors, including VIRGO, were offline for upgrades.
\begin{table}
\caption{List of optically discovered core-collapse SNe that occurred
within 5 Mpc from Earth between January 1, 2002 and August 31, 2008. The
table is based on the comprehensive online SN listing at
\texttt{http://www.cfa.harvard.edu/iau/lists/Supernovae.html}. Given
are the SN name, the name of the host galaxy, the date of discovery, the
core-collapse SN spectral subtype (see, e.g., \cite{filippenko:97}), 
and the approximate distance. Note that the date of
astronomical discovery always postdates the actual explosion.  }
\begin{footnotesize}
\begin{tabular}{lllll}
SN & Host Galaxy & Date & Type & Distance \\
\hline
2008bk & NGC 7793 & 20080325 \cite{cbet:1315} &II-P &$\sim 3.9$ \cite{karachentsev:04}\\
2005af & NGC 4945 & 20050208 \cite{iauc:8482} &II-P &$\sim 3.6$ \cite{karachentsev:04}\\
2004dj & NGC 2403 & 20040731 \cite{iauc:8377} &II-P &$\sim 3.3$ \cite{karachentsev:04}\\
2004am & M 82     & 20040305 \cite{iauc:8297} &II-P &$\sim 3.5$ \cite{karachentsev:02}\\
2002kg & NGC 2403 & 20021026 \cite{iauc:8051} &IIn  &$\sim 3.3$ \cite{karachentsev:04}\\
\end{tabular}
\end{footnotesize}
\label{table:sne}
\end{table}

SN 2008bk was found on March 25, 2008 in NGC 7793
\cite{cbet:1315}, a spiral galaxy, approximately $3.9\;\mathrm{Mpc}$
away from Earth. Based on the lightcurve evolution of similar type
II-P (``plateau,'' see, e.g., \cite{filippenko:97}), 2008bk exploded
$\sim 20$--$36$ days before its discovery \cite{cbet:1315}. The progenitor
star appears to have been on the low-mass end of the massive-star
population and may have had a mass between $8.5 \pm 1\;\mmsun$ 
\cite{atel:1464,mattila:08}.

Since LIGO H2 and GEO600 were most likely taking data when SN 2008bk
exploded, it is interesting to estimate the detectability by H2 and
GEO600 for the GW emission mechanisms discussed in
in this review. Given the high rate of core-collapse SNe between
$\sim 3$ and $\sim 5\;\mathrm{Mpc}$ it is also useful to consider
their detectability by other current and future GW observatories.

Since waveforms for some of the GW emission mechanisms are available,
we compute single-detector optimal (assuming perfect orientation)
matched-filtering signal-to-noise ratios $(\mathrm{SNR})$, using
\begin{equation}
(\mathrm{SNR})_\mathrm{optimal}^2 = 4 \int_0^\infty \frac{|\tilde{h}_+(f)|^2 +
|\tilde{h}_\times (f)|^2}{S(f)}\, df\,\,,
\label{eq:snr1}
\end{equation}
where $S(f)$ is the one-sided detector noise power spectral density in
units of $\mathrm{Hz}^{-1/2}$ and $\tilde{h}$ is the Fourier transform
of the wave signal, computed via \cite{thorne:87,flanhughes:98}
\begin{equation}
\tilde{h}_{+,\times}(f) = 
\int_{-\infty}^\infty e^{2\pi i f t} h_{+,\times} (t) dt\,\,.
\end{equation}
Using the characteristic strain $h_\mathrm{char}(f)$ defined by
equation~\ref{eq:hchar}, equation~\ref{eq:snr1} can be rewritten
\cite{flanhughes:98} to
\begin{equation}
(\mathrm{SNR})_\mathrm{optimal}^2 = \int_0^\infty d\ln f \frac{h_\mathrm{char}^2 (f)}
{h_\mathrm{rms}^2 (f)}\,,
\label{eq:snr2}
\end{equation}
where $h_\mathrm{rms}(f)$ is the dimensionless 
detector rms noise given by $h_\mathrm{rms}(f) = \sqrt{fS(f)}$. Note that the
expression for the optimal $\mathrm{SNR}$ given here is a factor of $3/2$
smaller than the expression used in \cite{dimmelmeier:08} and derived in
\cite{thorne:87}. See \cite{flanhughes:98} for details.

\begin{table*}
\caption{Upper-limit $\mathrm{SNR}$ estimates for SN 2008bk at
  $3.9\;\mathrm{Mpc}$ using theoretical noise power spectral densities
  for advanced LIGO 4-km interferometers (LIGO 2 4km), LIGO 1 4-km
  interferometers (LIGO L1/H1), VIRGO, and LIGO S5 data for the H2 and
  GEO600 interferometers. Considered are various representative models
  for GW emission by rotating core collapse and bounce,
  nonaxisymmetric rotational instabilities, and PNS core
  $g$-modes. Behind each model name a paper reference is provided. For
  some models, $\sqrt{n}$-scaled $\mathrm{SNR}$s are given as estimates for
  longer GW emission than tracked by the model calculations. This is
  denoted by $(\times n$) behind the model name.}
\begin{footnotesize}
\begin{tabular}{ll|r|rrrr}
\hline
\hline
\multicolumn{1}{c}{Process}
& \multicolumn{1}{c}{Model}
& \multicolumn{1}{|c|}{LIGO 2} 
& \multicolumn{1}{c}{LIGO L1/H1} 
& \multicolumn{1}{c}{LIGO H2}
& \multicolumn{1}{c}{GEO600}
& \multicolumn{1}{c}{VIRGO}\\
& \multicolumn{1}{c}{}
& \multicolumn{1}{|c|}{4 km} 
& \multicolumn{1}{c}{4 km} 
& \multicolumn{1}{c}{2 km}
& \multicolumn{1}{c}{600 m}
& \multicolumn{1}{c}{3 km}
\\
\hline
Rotating Collapse
&s11A2O13 
&$0.124$
&$0.008$
&$0.005$
&$0.001$
&$0.009$\\
\& Bounce&s20A2O09 
&$0.130$
&$0.008$
&$0.006$
&$< 0.001$
&$0.010$\\
\cite{dimmelmeier:08}& s40A3O12 
&$0.214$
&$0.024$
&$0.013$
&$< 0.001$
&$0.018$\\
\hline
Rotational Instability&
s20A2B4 
&$0.319$
&$0.021$
&$0.014$
&$0.003$
&$0.022$\\
\cite{ott:07cqg,ott:06phd,ott:07prl}
&s20A2B4 ($\times 5$)
&$0.713$
&$0.047$
&$0.031$
&$0.007$
&$0.049$\\
\hline
PNS $g$-modes
&s11.2 
&$0.147$
&$0.006$
&$0.005$
&$0.002$
&$0.009$\\
\cite{ott:06prl,burrows:07a}&s15.0 
&$0.454$
&$0.021$
&$0.015$
&$0.006$
&$0.027$\\
and \sref{section:gmodes}&s25.0 
&$0.612$
&$0.029$
&$0.020$
&$0.007$
&$0.037$\\
&s25.0 ($\times 2$)
&$0.866$
&$0.041$
&$0.029$
&$0.009$
&$0.052$\\
&s25WW 
&$5.331$
&$0.217$
&$0.151$
&$0.057$
&$0.328$\\
\hline
\end{tabular}
\end{footnotesize}
\label{table:snr}
\end{table*}

For computing $\mathrm{SNR}_\mathrm{optimal}$, we employ noise power
spectral densities $S(f)$ for H2 and GEO600 from the LIGO/GEO600 S5
runs. The H2 $S(f)$ data have so far not been published and were
kindly made available to us by M.~Landry~\cite{landry:08}. The GEO600
$S(f)$ data were obtained from the GEO600 website~\cite{geo600}.  In
addition, we compute $\mathrm{SNR}_\mathrm{optimal}$ based on the
design $S(f)$ for VIRGO \cite{virgo}, LIGO Livingston/Hanford 1
\cite{shoemaker:06}, and advanced LIGO $4$-km interferometers (LIGO 2)
in burst mode \cite{shoemaker:06}.

In table~\ref{table:snr}, we list $\mathrm{SNR}_\mathrm{optimal}$
for all considered detectors and for models focussing on GW signals
from rotating collapse and bounce, nonaxisymmetric rotational
instabilities, and PNS $g$-modes at the distance of SN 2008bk. 
GW emission from convective overturn and the
other emission mechanisms considered in this review are not included
in table~\ref{table:snr}, since their
$\mathrm{SNR}_\mathrm{optimal}$ would be of order $10^{-3}$
and below at a distance of $\sim 4\;\mathrm{Mpc}$  even for advanced LIGOs 
(see, e.g., \cite{mueller:04} and sections \ref{section:convsasi} and
\ref{section:others}).

The first model given for each emission process in
table~\ref{table:snr} is an ``average emitter'' and the subsequent
models range from ``strong'' to ``extreme''. Since in many computed
models the simulation was stopped before the GW emission had subsided,
we also provide $\sqrt{n}$-scaled (where $n$ is the factor by which
the emission is prolonged) $\mathrm{SNR}_\mathrm{optimal}$ for a
subset of models.

Based on the $\mathrm{SNR}_\mathrm{optimal}$ listed in
table~\ref{table:snr}, we surmise that LIGO H2 and GEO600 had no
chance of seeing GWs emitted in SN 2008bk. Even the most extreme
models yield optimal $\mathrm{SNR}$s below $\sim 0.2$ at a source
distance of $3.9\;\mathrm{Mpc}$. For a detection, a $\mathrm{SNR}$
significantly greater than $1$ and probably in the range of $5-7$ would
be needed (e.g., \cite{flanhughes:98, mueller:04}). Once in operation
and at design sensitivity, advanced LIGO could constrain the GW
emission from PNS $g$-modes out to the distance of SN 2008bk and
beyond. All other emission processes require a closer core-collapse
event (or more sensitive future detectors) to be observable.

\section{Putting things together: Summary and a Conjecture}
\label{section:conclusions}

The GW signature of core-collapse SNe is rich and
multi-faceted. The aim of this review was to provide an overview and
summary of the current knowledge about the various GW emission
processes that may occur in a core-collapse event.

We have outlined and discussed rotating core collapse and bounce,
nonaxisymmetric rotational instabilities, postbounce convective
overturn, and non-radial PNS pulsations as the prominent candidate
processes whose multi-D dynamics are likely to emit GWs. In addition,
we have summarized the emission characteristics of the GW signals
associated with anisotropic neutrino emission, aspherical outflows,
magnetic stresses, global precollapse asymmetries, and the collapse of
hypermassive PNSs to BHs.

All of the listed processes are burst emitters of GWs, but range in
temporal and spectral characteristics from short, few-ms bursts with
rather narrow-band emission (e.g., rotating collapse and bounce) to
broadband emission lasting, perhaps, for seconds (e.g., convection, in
particular PNS convection).  In addition, a number of the emission
processes produce bursts with GW memory, leaving behind essentially
zero-frequency, permanent distortions of spacetime.

Rotating core collapse and bounce is the most extensively investigated
and arguably the best understood source of GWs in the core collapse
context. 2D and 3D simulations in conformally-flat and full GR were
carried out and included the dominant microphysical aspects, results
of different approaches agree, and waveform catalogs are available
\cite{garchingcat,ottcatalog} to the GW data analysis community which
is beginning to employ them~(e.g., \cite{summerscales:08,brady:04}).
Nevertheless, uncertainties remain. Without doubt, the biggest is the
lack of presupernova models from multi-D stellar evolution
simulations.  Present simulations have to rely upon profiles from 1D
stellar evolution codes that take into account crucial multi-D effects
such as convection and rotation in only highly approximate ways. Core
collapse investigations try to deal with this by large-scale parameter
studies, but until multi-D stellar evolution models become available
and reliable, even extensive parameter studies could, in principle, be
missing the real precollapse configurations that obtain in nature.

In order to produce a burst of GWs at core bounce comfortably
detectable by advanced LIGOs throughout the Milky Way, a precollapse
iron core needs (\cite{dimmelmeier:08} and \sref{section:rotcoll}) to
be spinning very rapidly with central periods $P_o \lesssim
6\;\mathrm{s}$ (corresponding to central angular velocities $\Omega_o
\gtrsim 1\;\mathrm{rad}\,\mathrm{s}^{-1}$). Initial LIGOs might see
cores with $P_0 \lesssim 3\;\mathrm{s}$. Perhaps only $\sim 1\%$ or
less of the massive star population rotates with such short periods at
the onset of core collapse~\cite{woosley:06}, while the vast majority
is likely spinning much more slowly with periods of $30-100\;\mathrm{s}$
as currently predicted~\cite{heger:05,ott:06spin}.

Rapidly spinning iron cores with the above stated short central
periods produce millisecond-period PNSs with rotational energy of 
order $10\, \mathrm{B}$ of which a fraction would be sufficient to
power an energetic SN explosion. Moreover, such PNSs
have strongly differentially-rotating outer cores (e.g.,
\cite{ott:06spin, dimmelmeier:08,wheeler:07}) and are likely to become
subject to nonaxisymmetric rotational instabilities which would lead
to the GW emission characteristics outlined in \sref{section:rotinst}.
As discussed by a number of groups (e.g.,
\cite{burrows:07b,shibata:06,akiyama:03,
  yamada:04,obergaulinger:06a,dessart:08a}), magnetic field
amplification can draw from rotational energy. In particular,
Burrows~et~al.~\cite{burrows:07b} found that cores with $P_0 \lesssim
2-4\;\mathrm{s}$ develop MHD-driven jet-like SN
explosions\footnote{Provided either that the magneto-rotational
  instability (MRI)~\cite{bh:91} occurs as envisioned or that the
  precollapse core magnetic field is at least of the order of
  $10^{11}\; \mathrm{G}$.}.  Hence, there is a close link
between a detectable GW burst from core bounce and MHD-driven
explosions. The same may be true for 3D rotational instabilities and
their GW signal, but it is presently not known in which way 3D
hydrodynamic and MHD instabilities interact in a postbounce SN core.

Convection and the SASI are very likely to be present in the
postbounce pre-explosion phase of the vast majority of core-collapse
events. Rapid rotation damps convection and a very early onset of
explosion, probably relevant only in the case of ONeMg cores in the
lowest-mass massive stars \cite{janka:07,burrows:07c}, can shut off
convection before it is fully developed.

In contrast to the GW signal of rotating collapse and bounce, the GW
emission characteristics of postbounce convective overturn and the
SASI have been investigated by only a few studies and the systematics
of the GW signal with variations in progenitor structure, rotational
configuration etc.\ have not been investigated in detail. Furthermore,
owing to the stochastic nature of both large-eddy and small-scale
turbulent convection, GW templates that could be expected to match
real signals cannot be predicted. It will most likely take an advanced
LIGO-class detector to detect for a galactic event the broadband $\sim
0.5 - 1.5\;\mathrm{s}$ GW signal emitted by convection/SASI before
explosions and the longer-duration, but lower-amplitude GW signal from
PNS convection.

\begin{table*}
\caption{Overview on prominent GW emission processes in core-collapse
  SNe and their possible emission 'levels' in the context of the
  magneto-rotational (MHD; e.g. \cite{burrows:07b}), the neutrino
  (e.g., \cite{marek:07}), and the acoustic SN mechanism (e.g.,
  \cite{burrows:06,burrows:07a}).  For a \emph{galactic} SN, 'strong'
  corresponds to 'probably detectable by initial and advanced LIGO',
  'weak' means 'probably marginally detectable by advanced LIGO', and
  'none' means 'absent or probably not detectable by advanced
  LIGO'. The three considered explosion mechanisms are likely to have
  mutually exclusive GW signatures and could be distinguished by the
  mere detection or non-detection of a GW signal without the solution
  of the full inverse problem.  Note that the GW signal due to
  anisotropic neutrino emission, though present in all three
  mechanism, is not considered, since its low-frequency character
  severely limits its detectability.}
\vskip.1cm
\includegraphics[width=13cm]{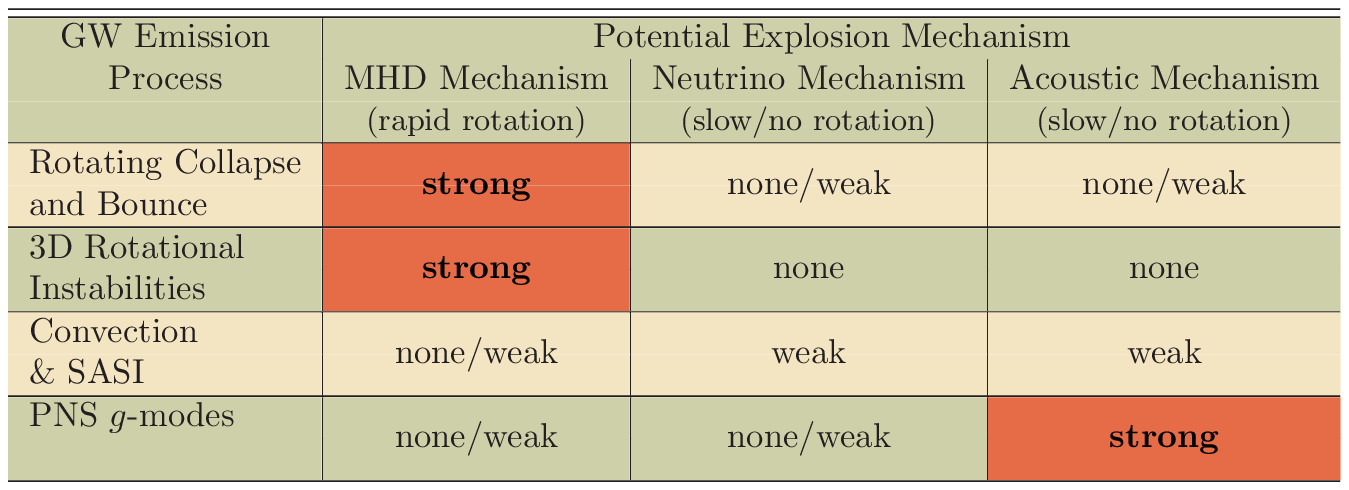}
\label{table:mech}
\end{table*}

The explosion scenario currently favored for slowly rotating and
nonrotating massive stars -- the neutrino mechanism (e.g.,
\cite{bethewilson:85,bethe:90,janka:07}) -- relies on the deposition
of sufficient energy by neutrinos in the postshock region to revive
the stalled shock and unbind the stellar envelope. A multitude of
multi-D simulations have demonstrated that convection and SASI are key
ingredients for neutrino-driven core-collapse SN explosions~(e.g.,
\cite{marek:07,murphy:08} and references therein). The acoustic
mechanism, recently proposed by
Burrows~et~al.~\cite{burrows:06,burrows:07a} and intensely debated in
the community \cite{marek:07,weinberg:08,yoshida:07}, is an
alternative to the neutrino mechanism and relies on the excitation by
accretion and turbulence of large-amplitude PNS core $g$-modes and their
damping by strong sound waves that deposit energy in the postshock
region, eventually leading to explosion. The GW signature of an
acoustically-driven core-collapse SN explosion would be dominated by
the GWs emitted from the quadrupole components of the PNS
core pulsations. Already first-generation LIGOs should be able to see
such a strong signal throughout the Milky Way. Advanced detectors
may be able to put significant constraints on the signal strength
out to $3-5\;\mathrm{Mpc}$ where the SN rate is favorably high.

In table~\ref{table:mech} we summarize the prominent GW emission
processes that are expected to be active in a core-collapse event and
contextualize them with the three core-collapse SN explosion
mechanisms by semi-quantitatively assigning 'emission strengths' for a
galactic event ($D \approx 10\;\mathrm{kpc}$) based on the GW emission
estimates collected in this review.  In the scale we assume, 'none'
refers to 'absent or probably not detectable by advanced LIGOs
($\mathrm{SNR} \ll 1$)', 'weak' means 'probably only marginally
detectable by advanced LIGOs ($\mathrm{SNR} \gtrsim 1$),' 'strong'
reflects 'probably detectable by initial and advanced LIGOs
($\mathrm{SNR} \gtrsim 5-7$).'  The roughness of the scale is intended
to emphasize the independence of the overall argument from variations
in quantitative details.

The MHD mechanism is limited to rapidly rotating cores,
hence will involve strong GW emission from core bounce and,
perhaps, nonaxisymmetric rotational instabilities. The acoustic
mechanism, on the other hand, may work best in nonrotating or slowly
rotating cores and will emit a strong GW signal from PNS
pulsations. The neutrino mechanism, also probably most
relevant in slowly or nonrotating cores, will very likely have
convection/SASI as its strongest source of GWs.

Based on the above discussion and on table~\ref{table:mech}, we
conjecture that \emph{the GW signatures of the neutrino, MHD, and
  acoustic core-collapse SN mechanisms are mutually exclusive}. Hence,
for a galactic SN and even initial LIGOs, the mere detection of a GW
signal and its association with an emission process and, in fact, also
the non-detection of GWs, have the potential of strongly constraining
the way massive stars explode.

Unfortunately, galactic core-collapse SNe are quite rare events,
occurring at a rate of one in a few decades.  Initial LIGO-class
detectors cannot see core-collapse SNe outside the Milky Way.
Advanced detectors, at least as currently planned~\cite{shoemaker:06},
may be able to see most core-collapse SNe throughout the local group
of galaxies ($D\approx 1\;\mathrm{Mpc}$). This, however, would
increase the observable event rate by not more than a factor of $\sim
2$~\cite{vdb:91,ando:05}. Third-generation GW observatories (e.g.\ the
envisioned EURO detector~\cite{euro}) reaching out to at least
$3-5\;\mathrm{Mpc}$ will be necessary for detailed GW astronomy of
core-collapse SNe to become possible.

\section*{Acknowledgements}
The author wishes to thank A.~Burrows, L.~Lehner, and B.~Schutz for
encouraging him to write this topical review. The author is indebted
to his collaborators A.~Burrows, L.~Dessart, E.~Livne, and J.~Murphy
for letting him use previously unpublished theoretical GW signal data
in this article. He also thanks K.~Kotake for providing him with the
derivation of equation~\ref{eq:nu_gw_simple} and for helpful
exchanges.  Furthermore, the author finds it a pleasure to acknowledge
helpful discussions with S.~Akiyama, S.~Ando, E.~Berti, A.~Burrows,
A.~Calder,  P.~{Cerd{\'a}-Dur{\'a}n},
 L.~Dessart, H.~Dimmelmeier, L.~S.~Finn, H.-T.~Janka,
E.~Katsavounidis, M.~Landry, A.~Marek, T.~Marquart, M.~Marschall,
R.~O'Shaugnessy, E.~M\"uller, J.~Murphy, E.~O'Connor, B.~Schutz,
E.~Seidel, E.~Schnetter, K.~S.~Thorne, and L.~Wen. Part of this work
was carried out during the Gravitational Wave Astronomy Workshop at
the Aspen Center of Physics, May-June 2008. This work was supported by
a Joint Institute for Nuclear Astrophysics postdoctoral fellowship,
sub-award no.~61-5292UA of NFS award no.~86-6004791 at the University
of Arizona, by a Sherman Fairchild postdoctoral fellowship at Caltech,
by an Otto Hahn Prize awarded to the author by the Max Planck Society,
and by a research fellowship from the Albert Einstein Institute.

{\scriptsize

}

\end{document}